\documentclass[apj]{emulateapj}

\usepackage{color}
\usepackage{comment}
\usepackage{natbib,aas_macros}
\newcommand{\OII}{[O\,\textsc{ii}]}
\newcommand{\OI}{[O\,\textsc{i}]}
\newcommand{\Ha}{H$\alpha$}
\newcommand{\Hb}{H$\beta$}
\newcommand{\Hc}{H$\gamma$}
\newcommand{\Hd}{H$\delta$}

\newcommand{\CIV}{C\,\textsc{iv}}
\newcommand{\NII}{[N\,\textsc{ii}]}

\newcommand{\OIII}{[O\,\textsc{iii}]}

\newcommand{\SII}{[S\,\textsc{ii}]}
\newcommand{\SIII}{[S\,\textsc{iii}]}
\newcommand{\HeII}{He\,\textsc{ii}}

\newcommand{\ArIII}{[Ar\,\textsc{iii}]}
\newcommand{\ArIV}{[Ar\,\textsc{iv}]}
\newcommand{\NeIII}{[Ne\,\textsc{iii}]}
\newcommand{\FeIII}{[Fe\,\textsc{iii}]}

\newcommand{\TeOIII}{$T_\mathrm{e}$(O\,\textsc{iii})}

\newcommand{\ergscm}{erg\,s$^{-1}$\,cm$^{-2}$}

\newcommand{\kms}{km\,s$^{-1}$}
\newcommand{\metal}{$12$+$\log ({\rm O/H})$}

\newcommand{\HII}{H\,\textsc{ii}}
\newcommand{\mass}{$\log$($M_{\star}$/$M_{\odot}$)}

\newcommand{\EW}{$\textrm{EW}_0$}
\newcommand{\EBV}{$E$($B$$-$$V$)}

\newcommand{\HSCEMPGa}{HSC J1429$-$0110} 
\newcommand{\HSCEMPGb}{HSC J2314$+$0154} 
\newcommand{\HSCEMPGc}{HSC J1142$-$0038} 
\newcommand{\HSCEMPGd}{HSC J1631$+$4426} 
\newcommand{\SDSSEMPGa}{SDSS J0002$+$1715} 
\newcommand{\SDSSEMPGb}{SDSS J1642$+$2233} 
\newcommand{\SDSSEMPGc}{SDSS J2115$-$1734} 
\newcommand{\SDSSEMPGd}{SDSS J2253$+$1116} 
\newcommand{\SDSSEMPGe}{SDSS J2310$-$0211}  
\newcommand{\SDSSEMPGf}{SDSS J2327$-$0200} 
\newcommand{\IzotovEMPG}{J0811$+$4730} 

\slugcomment{Accepted to Astrophysical Journal on 2021 March 2}

\shorttitle{
EMPRESS. II. Fe-Enriched Oxygen-Poor Galaxies
}
\shortauthors{Kojima et al.}

\begin{document}
\title{
EMPRESS. II. \\
Highly F\lowercase{e}-Enriched Metal-poor Galaxies with $\sim 1.0$ (F\lowercase{e}/O)$_\odot$ and $0.02$ (O/H)$_\odot$ :\\ Possible Traces of Super Massive ($>300 M_{\odot}$) Stars in Early Galaxies
}
\author{
Takashi Kojima\altaffilmark{1,2}, 
Masami Ouchi\altaffilmark{3,1,4},
Michael Rauch\altaffilmark{5},
Yoshiaki Ono\altaffilmark{1},
Kimihiko Nakajima\altaffilmark{3},
Yuki Isobe\altaffilmark{1,2},
\\ 
Seiji Fujimoto\altaffilmark{6,7},
Yuichi Harikane\altaffilmark{3,8,1},
Takuya Hashimoto\altaffilmark{9},
Masao Hayashi\altaffilmark{3},
Yutaka Komiyama\altaffilmark{3},\\
Haruka Kusakabe\altaffilmark{10},
Ji Hoon Kim\altaffilmark{11,12},
Chien-Hsiu Lee\altaffilmark{13}, 
Shiro Mukae\altaffilmark{1,14},
Tohru Nagao\altaffilmark{15},\\
Masato Onodera\altaffilmark{11,16},
Takatoshi Shibuya\altaffilmark{17},
Yuma Sugahara\altaffilmark{18,4,1,2},
Masayuki Umemura\altaffilmark{19},
Kiyoto Yabe\altaffilmark{4}
}
\email{E-mail: tkojima@icrr.u-tokyo.ac.jp}
\altaffiltext{1}{%
Institute for Cosmic Ray Research, The University of Tokyo,
5-1-5 Kashiwanoha, Kashiwa, Chiba 277-8582, Japan
}
\altaffiltext{2}{%
Department of Physics, Graduate School of Science, The University of Tokyo, 7-3-1 Hongo, Bunkyo, Tokyo 113-0033, Japan
}
\altaffiltext{3}{%
National Astronomical Observatory of Japan, 2-21-1 Osawa, Mitaka, Tokyo 181-8588, Japan
}
\altaffiltext{4}{%
Kavli Institute for the Physics and Mathematics of the Universe (WPI), 
University of Tokyo, Kashiwa, Chiba 277-8583, Japan
}
\altaffiltext{5}{%
Carnegie Observatories, 813 Santa Barbara Street, Pasadena, CA 91101, USA
}
\altaffiltext{6}{%
Cosmic Dawn Center (DAWN), Copenhagen, Denmark
}
\altaffiltext{7}{%
Niels Bohr Institute, University of Copenhagen, Lyngbyvej 2, DK2100 Copenhagen, Denmark
}
\altaffiltext{8}{%
Department of Physics and Astronomy, University College London, Gower Street, London WC1E 6BT, UK
}
\altaffiltext{9}{%
Tomonaga Center for the History of the Universe (TCHoU), Faculty of Pure and Applied Sciences, University of Tsukuba, Tsukuba, Ibaraki 305-8571, Japan
}
\altaffiltext{10}{%
Observatoire de Gen\`{e}ve, Universit\'e de Gen\`{e}ve, 51 Ch. des Maillettes, 1290 Versoix, Switzerland
}
\altaffiltext{11}{%
Subaru Telescope, National Astronomical Observatory of Japan, National Institutes of Natural Sciences (NINS), 650 North Aohoku Place, Hilo, HI 96720, USA
}
\altaffiltext{12}{%
Metaspace, 36 Nonhyeon-ro, Gangnam-gu, Seoul 06312, Republic of Korea
}
\altaffiltext{13}{%
NSF's National Optical Infrared Astronomy Research Laboratory, Tucson, AZ 85719, USA
}
\altaffiltext{14}{%
Department of Astronomy, Graduate School of Science, The University of Tokyo, 7-3-1 Hongo, Bunkyo, Tokyo, 113-0033, Japan
}
\altaffiltext{15}{%
Research Center for Space and Cosmic Evolution, Ehime University, 2-5 Bunkyo-cho, Matsuyama, Ehime 790-8577, Japan
}
\altaffiltext{16}{%
Department of Astronomical Science, SOKENDAI (The Graduate University for Advanced Studies), Osawa 2-21-1, Mitaka, Tokyo, 181-8588, Japan
}
\altaffiltext{17}{%
Kitami Institute of Technology, 165 Koen-cho, Kitami, Hokkaido 090-8507, Japan
}
\altaffiltext{18}{%
Waseda Research Institute for Science and Engineering, Faculty of Science and Engineering, Waseda University, 3-4-1, Okubo, Shinjuku, Tokyo 169-8555, Japan
}
\altaffiltext{19}{%
Center for Computational Sciences, University of Tsukuba, Tsukuba, Ibaraki 305-8577, Japan.
}

\altaffiltext{\dag}{Partly based on data obtained with the Subaru Telescope. The Subaru Telescope is operated by the National Astronomical Observatory of Japan.}
\altaffiltext{\ddag}{The data presented herein were partly obtained at the W. M. Keck Observatory, which is operated as a scientific partnership among the California Institute of Technology, the University of California and the National Aeronautics and Space Administration. The Observatory was made possible by the generous financial support of the W. M. Keck Foundation.}
\altaffiltext{\#}{This paper includes data gathered with the 6.5 meter Magellan Telescopes located at Las Campanas Observatory, Chile.}

\begin{abstract}
We present element abundance ratios and ionizing radiation of local young low-mass ($\sim 10^{6} M_{\odot}$) extremely metal poor galaxies (EMPGs) with a 2\% solar oxygen abundance (O/H)$_{\odot}$ and a high specific star-formation rate (sSFR$\sim$300\,Gyr$^{-1}$), and other (extremely) metal poor galaxies, which are compiled from Extremely Metal-Poor Representatives Explored by the Subaru Survey (EMPRESS) and the literature.
Weak emission lines such as {\FeIII}4658 and {\HeII}4686 are detected in very deep optical spectra of the EMPGs  taken with 8m-class telescopes including Keck and Subaru \citep{Kojima2020a,Izotov2018a}, enabling us to derive element abundance ratios with photoionization models. 
We find that neon- and argon-to-oxygen ratios are comparable to those of known local dwarf galaxies, and that the nitrogen-to-oxygen abundance ratios (N/O) are lower than 20\% (N/O)$_{\odot}$ consistent with the low oxygen abundance. 
However, the iron-to-oxygen abundance ratios (Fe/O) of the EMPGs are generally high; the EMPGs with the 2\%-solar oxygen abundance show high Fe/O ratios of $\sim 90-140$\% (Fe/O)$_{\odot}$, which are unlikely explained by suggested scenarios of Type Ia supernova iron productions, iron's dust depletion, and metal-poor gas inflow onto previously metal-riched galaxies with solar abundances. 
Moreover, the EMPGs with the 2\%-solar oxygen abundance have very high \HeII4686/\Hb\ ratios of $\sim$1/40, which are not reproduced by existing models of high-mass X-ray binaries with progenitor stellar masses $<$ $120 M_\odot$. 
Comparing stellar-nucleosynthesis and photoionization models with a comprehensive sample of EMPGs identified by this and previous EMPG studies, we propose that both the high Fe/O ratios and the high \HeII4686/\Hb\ ratios are explained by the past existence of super massive ($>300 M_\odot$) stars, which may evolve into intermediate-mass black holes ($\gtrsim100 M_\odot$).
\end{abstract}


\section{INTRODUCTION} \label{sec:intro}
The early universe is dominated by a large number of young, low-mass, metal-poor galaxies.
Theoretical arguments suggest that the first galaxies are formed at $z\sim10$--$20$ from gas already metal-enriched by Pop-III (i.e., metal free) stars. 
According to hydrodynamical simulations \citep[e.g.,][]{Wise2012}, the first galaxies are created in dark matter (DM) mini halos with $\sim10^{8}M_{\odot}$ and have low stellar masses ({\mass}$\sim$4--6), low metallicities ($Z\sim0.1$--$1\% Z_{\odot}$), and high specific star-formation rates (${\rm sSFR}\sim100$ Gyr$^{-1}$) at $z\sim10$.
The typical stellar mass is remarkably small, comparable to those of star clusters.
Such cluster-like galaxies are undergoing early phases of the galaxy formation.
One of critical goals of the modern cosmology is to understand the early-phase galaxy formation by probing the cluster-like, star-forming galaxies (SFGs).

The stellar population and the star-formation history are critical information to understand galaxies undergoing the early-phase star formation.
Element abundances such as iron (Fe) and nitrogen (N) are good tracers of the past star formation and the stellar population because these elements are produced and ejected by the different stellar populations at different ages.
First, iron elements are effectively produced and released into ISM by type-Ia supernovae (SNe) $\sim$1 Gyr after the start of the star formation.
The type-Ia SNe are triggered by gas accretion from a main sequence star onto a white dwarf, whose progenitor weighs $\sim$1--10 $M_{\odot}$ \citep[e.g.,][]{Nomoto2013}, in a binary system.
The type-Ia SNe start contributing to the increase of iron-to-oxygen ratio (Fe/O) at the age of $\sim$1 Gyr \citep[e.g.,][]{Steidel2016}.
As reported in studies of Galactic stars \citep[][]{Bensby2006, Lecureur2007, Bensby2013}, the increasing Fe/O trend is seen in a metallicity range of $Z_{*}/Z_{\odot}\gtrsim0.2$ (corresponding to {\metal}$\gtrsim$8.0).
Below {\metal}$\sim$8.0 (or $\lesssim$1 Gyr), the core-collapse SNe mainly contribute to the production and release of iron and oxygen.
Under the assumption of this mechanism, low-mass, metal-poor, young SFGs are expected to have a low Fe/O ratio due to their young ages. 
Second, nitrogen elements trace activities of massive stars and low- and intermediate-mass stars at low and high metallicities, respectively.
As suggested by previous studies \citep{Perez-Montero2009, Perez-Montero2013, Andrews2013},   nitrogen-to-oxygen ratios (N/O) of SFGs present a plateau in the range of {\metal}$\lesssim$8.0 and a positive slope at higher metallicities as a function of metallicity.
Model calculations of the N/O evolution \citep[e.g.,][]{Vincenzo2016} also support this trend.
The plateau basically results from the primary nucleosynthesis of massive stars, while the positive slope is mainly attributed to the secondary nucleosynthesis of low- and intermediate-mass stars \citep[e.g.,][]{Vincenzo2016}.
In the nitrogen-enrichment mechanism, low-mass, metal-poor, young SFGs may have a low N/O ratio because of their low metallicities and young ages. 

Ionizing radiation is another key to understand the stellar population of galaxies in the early-phase star formation.
Ionizing radiation is produced by massive stars and/or a hot accretion disk around compact objects such as black holes (BHs).
Observational studies \citep{Lopez-Sanchez2010a, Shirazi2012, Senchyna2017, Schaerer2019} have suggested that SFGs show strong {\HeII}4686 emission lines represented by {\HeII}4686/H$\beta$$\sim$1/300--1/30.
Especially, the {\HeII}4686/H$\beta$ shows an increasing trend as metallicity decreases in the range of {\metal}$<$8.0.
The {\HeII}4686 line is sensitive to ionizing photons above 54.4 eV, which are not abundant in radiation of O- and B-type hot stars.
Under the assumption of stellar radiation, \citet{Xiao2018} have created nebular emission models with the combination of the photoionization code {\sc cloudy} \citep[][]{Ferland2013} and the {\sc bpass} (Binary Population and Spectral Synthesis) code \citep[][]{Stanway2016, Eldridge2017}.
The \citet{Xiao2018} models predict {\HeII}4686/{\Hb}$\lesssim$1/1000, well below the observed {\HeII}4686/{\Hb} ratios of $\sim$1/300 to $\sim$1/30 \citep{Schaerer2019}.
This means that the main contributors of {\HeII}4686 are not hot stars. 
\citet{Schaerer2019} have estimated {\HeII}4686/{\Hb} ratios with high mass X-ray binary (HMXB) models of \citet{Fragos2013b, Fragos2013a}, and suggested that high {\HeII}4686/{\Hb} ratios can be partly explained by the HMXB models.
However, the HMXB models still do not explain the high {\HeII}4686/{\Hb} ratios for galaxies with a high {\Hb} equivalent width, {\EW}({\Hb})$>$100 {\AA} (i.e., younger than 5 Myr).
\citet{Schaerer2019} suggest another possible contribution from old stellar population and/or shock-heated gas. 
\citet{Saxena2020b} investigate a connection between the {\HeII}4686 and X-ray emission of 18 {\HeII}-emitting galaxies at $z\sim2$--4, concluding that the observed {\HeII}4686/{\Hb} ratios and X-ray intensities are not explained by the HMXBs simultaneously.
The mass range of the sample is {\mass}$=$8.8--10.7 \citep{Saxena2020a, Saxena2020b}, which corresponds to $\sim$0.2--1.0 $Z_{\odot}$ under the assumption of the mass-metallicity relation at $z\sim2$--4 \citep{Shapley2017}.
The conclusion of \citet{Saxena2020b} may only be applicable to relatively mature galaxies with intermediate masses and metallicities, in contrast to metal-poor galaxies at $z\sim0$ \citep[e.g.,][]{Schaerer2019}. 
\citet{Senchyna2020} also investigate the connection between the {\HeII}4686 and X-ray emission with a sample of 11 local galaxies.
\citet{Senchyna2020} conclude that HMXBs are not dominant sources of the He$^{+}$-ionizing photons, although a possibility of soft X-ray sources such as intermediate-mass black holes (IMBHs) is not ruled out.
In the sample of \citet{Senchyna2020}, 4 galaxies show {\it low} metallicities (0.06--0.1$Z_{\odot}$) and {\it low} {\Hb} equivalent widths ({\EW}({\Hb})$\sim$50--100{\AA}), and 3 galaxies show {\it high} metallicities ($>$0.1$Z_{\odot}$) and {\it high} {\Hb} equivalent widths ({\EW}({\Hb})$\sim$200--400{\AA}).
Galaxies with very {\it low} metallicities ($\sim$0.01$Z_{\odot}$) and {\it high} {\Hb} equivalent widths ({\EW}({\Hb})$\gtrsim$100{\AA}) at the same time are missing in the sample of \citet{Senchyna2020}.
In summary, the main contributor of the {\HeII}4686 emission is still under debate, especially at low metallicities and high {\Hb} equivalent widths, which are expected to be metal-poor galaxies undergoing very early phases of the galaxy evolution.

In the local universe, extremely metal-poor galaxies (EMPGs) have been discovered \citep[e.g.,][]{Izotov1998a,Thuan2005b,Izotov2009,Izotov2018a,Izotov2019a,Ly2014} by exploiting wide-field data such as Sloan Digital Sky Survey \citep[SDSS,][]{York2000}.
These galaxies have low metallicities, {\metal}$\sim$7.0--7.2, low stellar masses, {\mass}$\sim$6--8, and high sSFR, $\sim100$ Gyr$^{-1}$.
Such local EMPGs are regarded as local analogs of high-$z$ galaxies because they have low metallicities, low stellar masses, and large emission line equivalent widths similar to low-mass galaxies with {\mass}$\sim$6--8 at $z\sim2$--$3$ \citep[][]{Christensen2012b, Christensen2012a, Stark2014a, Vanzella2016a} and $z\sim6$--$7$ \citep[][]{Stark2015b, Mainali2017}.
However, the stellar mass ranges of the previous studies, {\mass}$\sim$6--8, are not as low as cluster-like galaxies in the early-phase of the galaxy formation, {\mass}$\sim$4--6, as described above.
To reach a lower mass range than the previous EMPG studies (e.g., SDSS, $i$-band limiting magnitude $\sim$21 mag), deeper, wide-field imaging survey has been expected.

We have initiated a new EMPG survey with wide-field optical imaging data obtained in Subaru/Hyper Suprime-Cam \citep[HSC;][]{Miyazaki2012, Miyazaki2018, Komiyama2018, Kawanomoto2018, Furusawa2018} Subaru Strategic Program \citep[HSC-SSP;][]{Aihara2018a} in \citet[][Paper I hereafter]{Kojima2020a}.
The new EMPG survey has been named ``Extremely Metal-Poor Representatives Explored by the Subaru Survey'' (EMPRESS).
We have created a source sample based on the deep, wide-field HSC-SSP data, which covers $\sim500$ deg$^2$ area with a $5\sigma$ limit of $\sim 26$ mag in Paper I. 
In this EMPRESS project, we try to select low-mass EMPGs with a large {\EW}(H$\alpha$) (e.g., $\gtrsim$800 {\AA}) because our motivation is to discover local counterparts of high-$z$, low-mass galaxies whose specific star-formation rate (sSFR) is expected to be high \citep[$\gtrsim$10 Gyr$^{-1}$, e.g.,][]{Ono2010b, Stark2017, Elmegreen2017b, Harikane2018b}, which are expected to be undergoing very early phases of the galaxy evolution.

This paper is the second paper from our EMPRESS project.
The detailed sample selection and results of the first spectroscopic observations have been reported in Paper I.
These paper will be followed by other papers in which we investigate details of size and morphology, and kinematics of our EMPG sample \citep[e.g.,][Paper III hereafter]{Isobe2020}.
The outline of this paper is as follows.
In Section \ref{sec:sample}, we briefly explain our samples selected from the Subaru HSC-SSP data and the SDSS data.
In Section \ref{sec:spectroscopy}, we describe our optical spectroscopy carried out for our EMPG candidates and explain the reduction and calibration processes of our spectroscopy data.
In Section \ref{sec:analysis}, we measure emission line fluxes and estimate galaxy properties such as stellar mass, star-formation rate, metallicity, and element abundance. 
Section \ref{sec:result_discussion} shows results and discussions of element abundance ratios and ionizing radiation.
Then Section \ref{sec:summary} summarizes this paper.
Throughout this paper, magnitudes are on the AB system \citep{Oke1983}. 
We adopt the following cosmological parameters, $(h, \Omega_m, \Omega_{\Lambda}) = (0.7, 0.3, 0.7)$.
The definition of the solar metallicity is given by {\metal}=8.69 \citep{Asplund2009}.
We also define an EMPG as a galaxy with {\metal}$<$7.69 (i.e., $Z/Z_{\odot}$$<$0.1) in this paper, which is almost the same as in previous metal-poor galaxy studies \citep[e.g.,][]{Kunth2000, Izotov2012, Guseva2017}.

\section{SAMPLE} \label{sec:sample}
This paper uses samples obtained by Paper I.
In Paper I, we select the EMPG candidates from HSC-SSP and SDSS data with our machine learning (ML) classifier.
We briefly describe selections of EMPG candidates in this section.
Hereafter, these candidates chosen from the HSC-SSP and SDSS source catalogs are called ``HSC-EMPG candidates''  and ``SDSS-EMPG candidates'', respectively.

\subsection{HSC-EMPG candidates} \label{subsec:HSCcandidate}
%
We use the HSC-SSP internal data of the S17A and S18A data releases, which are explained in the second data release (DR2) paper of HSC-SSP \citep[][]{Aihara2019}.
Although the HSC-SSP survey data are taken in three layers of Wide, Deep, and UltraDeep, we only use the Wide field layer in this study.
In the HSC-SSP S17A and S18A data releases, images were reduced with the HSC pipeline, {\tt hscPipe} v5.4 and v6.7 \citep[][]{Bosch2018}, respectively, with codes of the Large Synoptic Survey Telescope (LSST) software pipeline \citep[][]{Ivezic2008, Axelrod2010, Juric2015, Ivezic2019}.
The pipeline conducts the bias subtraction, flat fielding, image stacking, astrometry and zero-point magnitude calibration, source detection, and magnitude measurement.
As reported in Paper I, there are slight differences in our results between S17A and S18A data due to the different pipeline versions.
Thus, although part of the S17A and S18A data are duplicated, we use both S17A and S18A data in this study to maximize the size of our EMPG sample.
The details of the observations, data reduction, detection, photometric catalog, and pipeline are described in \citet{Aihara2019} and \citet{Bosch2018}.
We use {\tt cmodel} magnitudes \citep{Bosch2018} corrected for Milky-Way dust extinction \citep{Schlegel1998} in the estimation of the total magnitudes of a source.

Below we explain how we construct an HSC source catalog, from which we select EMPG candidates.
We use isolated or cleanly deblended sources that fall within $griz$-band images.
We also require that none of the pixels in their footprints are interpolated, none of the central 3 $\times$ 3 pixels are saturated, none of the central 3 $\times$ 3 pixels suffer from cosmic rays, and there are no bad pixels in their footprints. 
Then we exclude sources whose {\tt cmodel} magnitude or centroid position measurements have a problem. 
We require a detection in the $griz$-band images.
We mask sources close to a bright star \citep{Coupon2018, Aihara2019} in the S18A data.
Here we select objects whose photometric measurements are brighter than 5$\sigma$ limiting magnitudes, $g<26.5$, $r<26.0$, $i<25.8$, and $z<25.2$ mag, which are estimated by \citet[][]{Ono2018} with 1.5-arcsec diameter circular apertures.
We also require that the photometric measurement errors are less than 0.1 mag in $griz$ bands.
Finally, we obtain 17,912,612 and 40,407,765 sources in total from the HSC-SSP S17A and S18A data, respectively, with the selection criteria explained above.
The effective area is 205.82 and 508.84 deg$^2$ in the HSC-SSP S17A and S18A data, respectively.
See Paper I for details of our HSC source sample.

We select EMPG candidates from the HSC-SSP source catalog in four steps: 
i) An initial rough selection based on colors, extendedness, and blending. 
ii) the ML classifier selection.
iii) Transient object removal by measuring the flux variance in multi-epoch images.
iv) Visual inspection of the $gri$-composite images.
Refer to Paper I for the selection details.
Eventually, we obtain 12 and 21 HSC-EMPG candidates from the S17A and S18A catalogs, respectively.
We find that 6 out of the HSC-EMPG candidates are duplicated between the S17A and S18A catalogs.
Thus the number of our independent HSC-EMPG candidates is 27 ($=$12$+$21$-$6).
A magnitude range of the 27 HSC-EMPG candidates is $i=19.3$--$24.3$ mag.

\subsection{SDSS-EMPG candidates} \label{subsec:SDSScandidate}
We construct a SDSS source catalog from the 13th release \citep[DR13;][]{Albareti2017} of the SDSS photometry data. 
Although the SDSS data are $\sim$5 mag shallower ($i_{\rm lim}$$\sim$21 mag) than HSC-SSP data ($i_{\rm lim}$$\sim$26 mag), we also select EMPG candidates from the SDSS data to complement brighter EMPGs. 
Here we select objects whose photometric measurements are brighter than SDSS limiting magnitudes, $u<22.0$, $g<22.2$, $r<22.2$, $i<21.3$, and $z<21.3$ mag\footnote{Magnitudes reaching 95\% completeness, which are listed in \texttt{https://www.sdss.org/dr13/scope/}}.
We only obtain objects whose magnitude measurement errors are $<$0.1 mag in $ugriz$ bands.
Note that we use {\tt Modelmag} for the SDSS data.
Among flags in the {\tt PhotoObjALL} catalog, we require that a {\tt clean} flag is ``1'' (i.e., {\it True}) to remove objects with photometry measurement issues.
The {\tt clean} flag\footnote{Details are described in\\ \texttt{http://www.sdss.org/dr13/algorithms/photo\_flags\_recommend/}} eliminates the duplication, deblending/interpolation problems, suspicious detections, and detections at the edge of an image.
We also remove objects with a {\it True} cosmic-ray flag and/or a {\it True} blended flag, which often mimics a broad-band excess in photometry. 
We reject relatively large objects with a ninety-percent petrosian radius greater than 10 arcsec to eliminate contamination by HII regions in nearby spiral galaxies. 
Finally, we derive 31,658,307 sources in total from the SDSS DR13 photometry data.
The total unique area of SDSS DR13 data is 14,555 deg$^2$.

We select EMPG candidates from the SDSS source catalog similarly to the HSC source catalog in Section \ref{subsec:HSCcandidate}.
After the selection, we derive 86 SDSS-EMPG candidates from the SDSS source catalog, whose $i$-band magnitudes range $i=14.8$--$20.9$ mag. 
One out of the 86 candidates ({\HSCEMPGa}) is also selected as an HSC-EMPG candidate in Section \ref{subsec:HSCcandidate}. 
Details are described in Paper I.

\section{SPECTROSCOPIC DATA} \label{sec:spectroscopy}
In this section, we explain our spectroscopic data of 10 galaxies described in Paper I, which are selected from our HSC and SDSS source catalogs and confirmed to be metal-poor galaxies.
We have identified that the 2 out of the 10 metal-poor galaxies ({\HSCEMPGd} and {\SDSSEMPGc}) satisfy the EMPG condition of $Z<0.10Z_{\odot}$ with metallicity estimates based on the electron temperature measurement.
{\HSCEMPGd} shows a metallicity of 0.016 $Z_{\odot}$, which is the lowest metallicity reported to date. 
In addition to the 10 metal-poor galaxies, we include another EMPG from the literature \citep[{\IzotovEMPG},][]{Izotov2018a} in the sample of this paper.
{\IzotovEMPG} has the second lowest metallicity of 0.019 $Z_{\odot}$ reported to date.

In paper I, we report on our spectroscopy of the 10 metal-poor galaxies performed with 4 spectrographs of the Low Dispersion Survey Spectrograph 3 (LDSS-3) and the Magellan Echellette Spectrograph (MagE) on Magellan telescope, the Deep Imaging Multi-Object Spectrograph (DEIMOS) on Keck-II telescope, and the Faint Object Camera And Spectrograph (FOCAS) on Subaru telescope. 
Although the spectroscopy and reduction are detailed in Paper I, we briefly summarize them in Sections \ref{subsec:ldss3}--\ref{subsec:focas}.
In Section \ref{subsec:detection}, we newly report very faint emission lines detected in our spectroscopy, such as {\OIII}4363, {\ArIV}4711, {\FeIII}4658, {\HeII}4686, {\NII}6584, and {\ArIII}7136, which are required to estimate element abundance ratios and constrain the FUV spectral hardness (Section \ref{sec:intro}).
Signal to noise ratios of our spectra depend on observational conditions (e.g., telescope, instrument, and integration time) and object brightnesses. 
Thus, the detection of such faint emission lines also depends on the observational conditions and object brightnesses.

\begin{figure*}[!htbp]
\epsscale{0.5}
\plotone{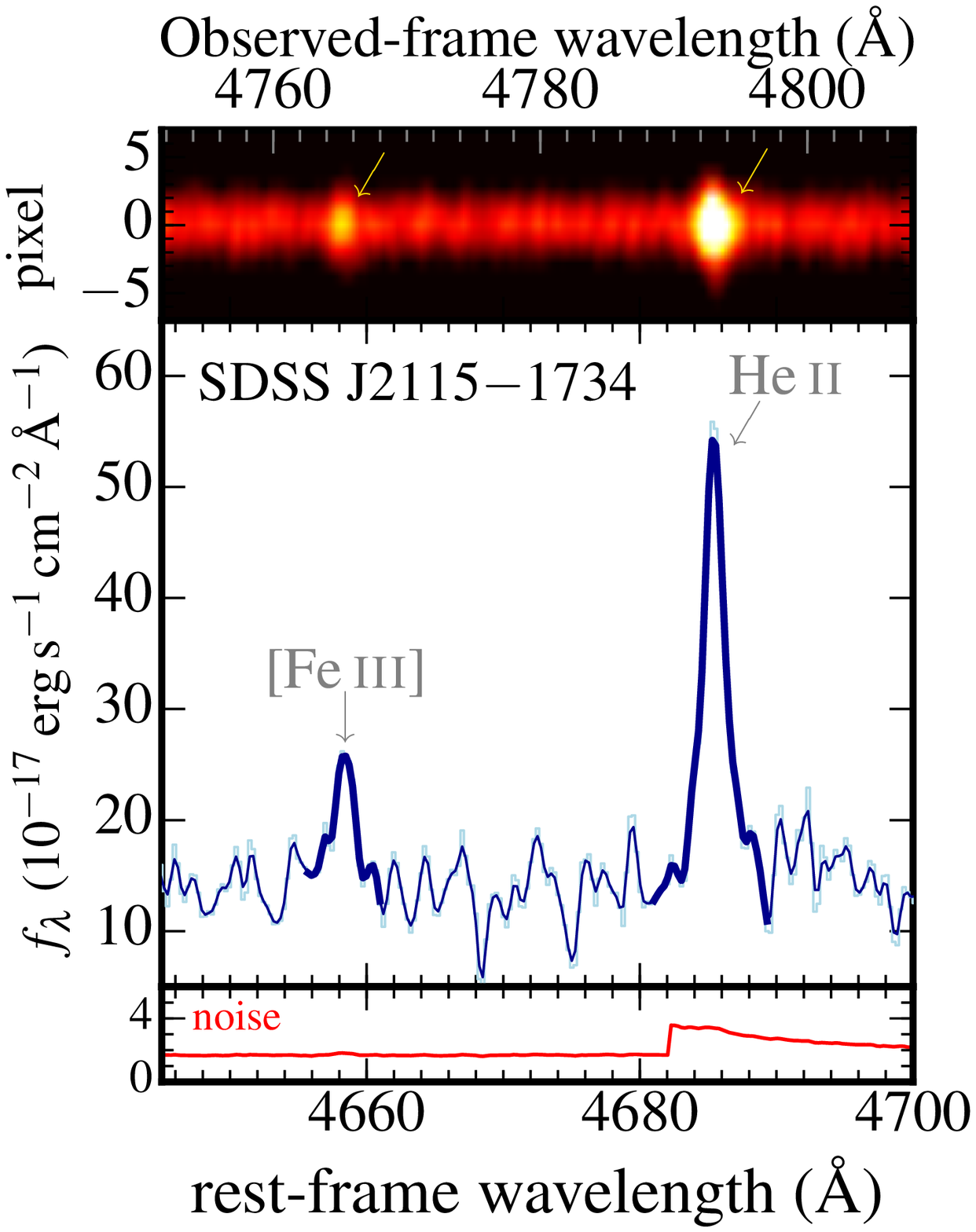}
\plotone{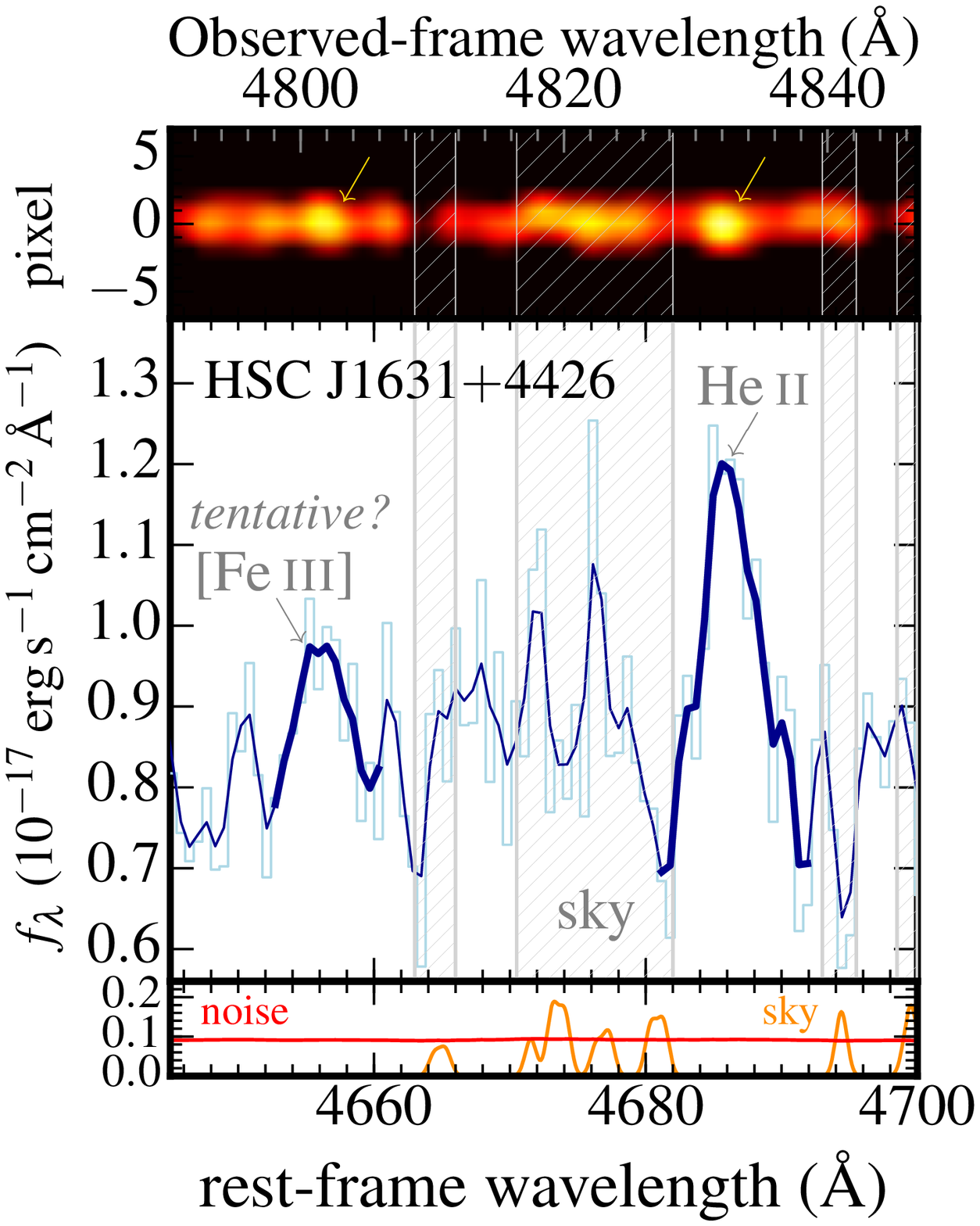}
\caption{Spectra of {\SDSSEMPGc} (left) and {\HSCEMPGd} (right) taken in our MagE and FOCAS spectroscopy, respectively, around the {\FeIII}4658 and {\HeII}4686 emission lines.
The top, center, and bottom panels show 2D spectra, signal spectra, and noise$+$sky spectra, respectively.
In the center panel, the light blue line is the un-smoothed background subtracted spectrum, while the dark blue line is spectra smoothed with a Gaussian profile. 
The shaded regions indicate positions of the sky emission lines.
In the bottom panel, we exhibit a noise spectrum with the red lines. 
The orange line show a sky emission spectrum (in arbitrary units) modeled with the sky emission data of \citet{Hanuschik2003}. 
For {\SDSSEMPGc}, no sky emission line falls on the wavelength range of this panel.
\label{fig:spectra}}
\end{figure*}

\subsection{Magellan/LDSS-3} \label{subsec:ldss3}
We conducted spectroscopy for one galaxy selected from our HSC catalog ({\HSCEMPGa}) with LDSS-3 at Magellan telescope. 
We used the VPH-ALL grism with the $0.\!\!^{\prime\prime}75$$\times$4$^{\prime}$ long-slit, which was placed at the offset position two-arcmin away from the center of the long-slit mask so that the spectroscopy could cover the bluer side.
The exposure time was 3,600 seconds. 
The spectroscopy covered $\lambda\sim$3,700--9,500 {\AA} with a spectral resolution of $R \equiv \lambda/\Delta\lambda \sim$ 860.

We used the {\sc iraf} package to reduce and calibrate the LDSS-3 data. 
The reduction and calibration processes include the bias subtraction, flat fielding, one-dimensional (1D) spectrum subtraction, sky subtraction, wavelength calibration, flux calibration, and atmospheric-absorption correction.
A one-dimensional spectrum was derived from an aperture centered on the blue compact component of our galaxies.
A standard star, CD-32\,\,9972 was used in the flux calibration.
The wavelengths were calibrated with the HeNeAr lamp.
Atmospheric absorption was corrected with the extinction curve at Cerro Tololo Inter-American Observatory (CTIO).
Our LDSS-3 spectroscopy may have been affected by the atmospheric refraction because a slit was not necessarily placed perpendicular to the horizon (i.e., at a parallactic angle) in our spectroscopy, which may lead to the wavelength-dependent slit loss.
The slit angles of each target are determined so that we can simultaneously observe multiple emission regions.
To estimate the wavelength-dependent slit loss $SL(\lambda)$ carefully, we made a model of the atmospheric refraction.

\subsection{Magellan/MagE} \label{subsec:mage}
We carried out spectroscopy for 8 galaxies selected from our HSC and SDSS catalogs ({\HSCEMPGb}, {\HSCEMPGc}, {\SDSSEMPGa}, {\SDSSEMPGb}, {\SDSSEMPGc}, {\SDSSEMPGd}, {\SDSSEMPGe}, and {\SDSSEMPGd}) with MagE of Magellan telescope. 
We used the echellette grating with the $0.\!\!^{\prime\prime}85$$\times$10$^{\prime\prime}$ or $1.\!\!^{\prime\prime}2$$\times$10$^{\prime\prime}$ longslits.
The exposure time was 1,800 or 3,600 seconds, depending on luminosities of the galaxies. 
The MagE spectroscopy covered $\lambda\sim$3,100--10,000 {\AA} with a spectral resolution of $R \equiv \lambda/\Delta\lambda \sim$ 4,000.

To reduce the raw data taken with MagE, we used the MagE pipeline from Carnegie Observatories Software Repository\footnote{\texttt{https://code.obs.carnegiescience.edu}}. 
The MagE pipeline has been developed on the basis of the {\tt Carpy} package \citep{Kelson2000, Kelson2003}.
The bias subtraction, flat fielding, scattered light subtraction, two-dimensional (2D) spectrum subtraction, sky subtraction, wavelength calibration, cosmic-ray removal, 1D-spectrum subtraction were conducted with the MagE pipeline.
Details of these pipeline processes are described on the web site of Carnegie Observatories Software Repository mentioned above. 
One-dimensional spectra were subtracted by summing pixels along the slit-length direction on a 2D spectrum.

We conducted the flux calibration with the standard star, Feige\,110, using {\sc iraf} routines.
Wavelengths were calibrated with emission lines of the ThAr lamp.
Spectra of each order were calibrated separately and combined with the weight of electron counts to generate a single 1D spectrum.
Atmospheric absorption was corrected in the same way as in Section \ref{subsec:ldss3}.
Our MagE spectroscopy may have been also affected by the atmospheric refraction for the same reason as the LDSS-3 spectroscopy.
Thus, we corrected the wavelength-dependent slit loss carefully in the same manner as the LDSS-3 spectroscopy described in Section \ref{subsec:ldss3}.

\subsection{Keck/DEIMOS} \label{subsec:deimos}
We conducted spectroscopy for one galaxy selected from our HSC catalog ({\HSCEMPGd}) with DEIMOS of the Keck-II telescope. 
We used the multi-object mode with the $0.\!\!^{\prime\prime}8$ slit width.
The exposure time was 2,400 seconds. 
We used the 600ZD grating and the BAL12 filter with a blaze wavelength at 5,500 {\AA}. 
The DEIMOS spectroscopy covered $\lambda\sim$3,800--8,000 {\AA} with a spectral resolution of $R \equiv \lambda/\Delta\lambda \sim$1,500. 

We used the {\sc iraf} package to reduce and calibrate the DEIMOS data. 
The reduction and calibration processes were the same as the LDSS-3 data explained in Section \ref{subsec:ldss3}.
A standard star, G191B2B was used in the flux calibration.
Wavelengths were calibrated with the NeArKrXe lamp.
Atmospheric absorption was corrected under the assumption of the extinction curve at Mauna Kea Observatories.
We only used a spectrum within the wavelength range of $\lambda>$4,900 {\AA}, which was free from the stray light (see Paper I for its detail).
We ignore the effect of the atmospheric refraction here because we only use a red side ($\lambda>$4,900 {\AA}) of DEIMOS data, which is insensitive to the atmospheric refraction. 
We also confirm that the effect of the atmospheric refraction is negligible with the models described in Section \ref{subsec:ldss3}.
In the DEIMOS data, we only used line flux ratios normalized to an {\Hb} flux.
Emission line fluxes were scaled with an {\Hb} flux by matching an {\Hb} flux obtained with DEIMOS to one obtained with FOCAS (see Section \ref{subsec:focas}).

\subsection{Subaru/FOCAS} \label{subsec:focas}
We carried out deep spectroscopy for one galaxy selected from our HSC catalog ({\HSCEMPGd}) with FOCAS installed on the Subaru telescope (PI: T. Kojima). 
This object was the same object as in the Keck/DEIMOS Spectroscopy (Section \ref{subsec:deimos}) and observed again with FOCAS with a longer integration time of 10,800 sec.
We used the long slit mode with the $2.\!\!^{\prime\prime}0$ slit width.
The exposure time was 10,800 seconds ($=$3 hours). 
We used the 300R grism and the L550 filter with a blaze wavelength at 7,500 {\AA} in a 2nd order. 
The FOCAS spectroscopy covered $\lambda\sim$3,400--5,250 {\AA} with a spectral resolution of $R$$\equiv$$\lambda/\Delta\lambda$$=$400 with the $2.\!\!^{\prime\prime}0$ slit width.

We used the {\sc iraf} package to reduce and calibrate the FOCAS data. 
The reduction and calibration processes were the same as the LDSS-3 data explained in Section \ref{subsec:ldss3}.
A standard star, BD$+$28\,4211 was used in the flux calibration.
Wavelengths were calibrated with the ThAr lamp.
Atmospheric absorption was corrected in the same way as in Section \ref{subsec:deimos}.
Our FOCAS spectroscopy covered $\lambda\sim$3,800--5,250 {\AA}, which was complementary to the DEIMOS spectroscopy described in Section \ref{subsec:deimos}, whose spectrum was reliable only in the range of  $\lambda>$4900 {\AA}.
We ignore the atmospheric refraction here because FOCAS is equipped with the atmospheric dispersion corrector.
Because an {\Hb} line was over-wrapped in FOCAS and DEIMOS spectroscopy, we used an {\Hb} line flux to scale the emission line fluxes obtained in the DEIMOS observation (see Section \ref{subsec:deimos}).

\subsection{Weak Emission Lines in the Spectra} \label{subsec:detection}
In our spectroscopy, we have detected many emission lines including very faint emission lines such as {\OIII}4363, {\ArIV}4711, {\FeIII}4658, {\HeII}4686, {\NII}6584, and {\ArIII}7136.
These faint emission lines are required to estimate element abundance ratios and constrain the FUV spectral hardness.
Especially, the {\FeIII}4658 and {\HeII}4686 lines are key in this paper, which enable us to investigate the Fe/O abundance ratios and the very hard EUV radiation, as described in Section \ref{sec:intro}.

In Figure \ref{fig:spectra}, we show two spectra of {\SDSSEMPGc} and {\HSCEMPGd} (classified as EMPGs in Paper I) around the {\FeIII}4658 and {\HeII}4686 emission lines. 
In the left panel of Figure \ref{fig:spectra}, the spectrum of {\SDSSEMPGc} clearly exhibits the significant detection of the {\FeIII}4658 and {\HeII}4686 emission lines.
As shown in the right panel, the {\HSCEMPGd} spectrum shows the significant detection of {\HeII}4686 (S/N$=$6.1) as well as the tentative detection of {\FeIII}4658 (S/N$=$2.4).
The flux measurements will be described in Section \ref{subsec:line}.

\begin{figure}[htbp]
\epsscale{1.0}
\plotone{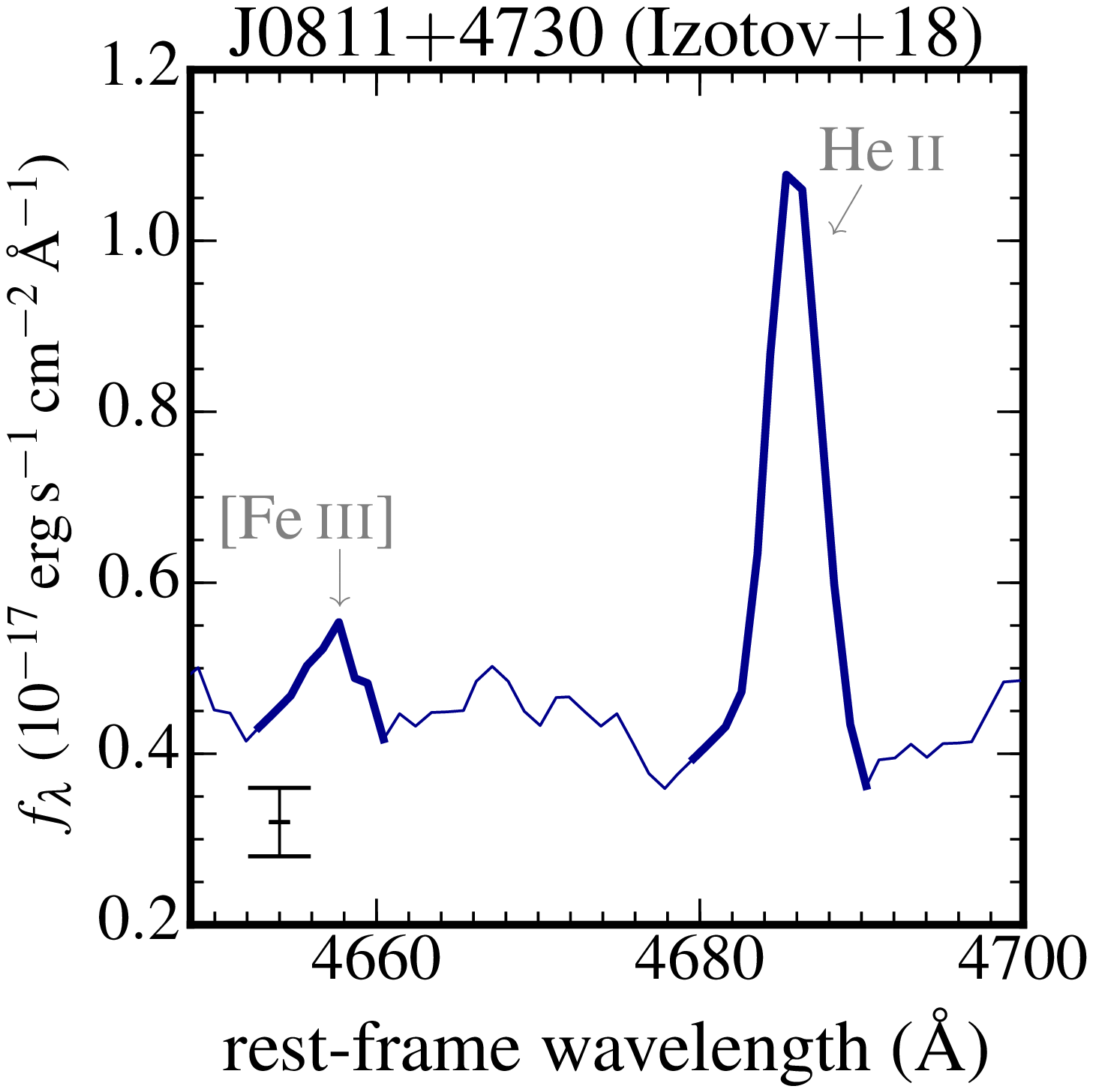}
\caption{Spectrum of J0811$+$4730 \citep{Izotov2018a} around the {\FeIII}4658 and {\HeII}4686 emission lines.
This spectrum was taken with Multi-Object Double Spectrographs (MODS) installed on Large Binocular Telescope (LBT).
This spectrum is adapted from \citet{Izotov2018a} with permission.
The emission lines of {\FeIII}4658 and {\HeII}4686 are detected with S/N$=$5.1 and S/N$=$14.9, respectively.
Continuum is detected at S/N$\sim$10 (in private communication).
The bar at the bottom left represents a typical error of the spectrum.
\label{fig:spectraIzotov}}
\end{figure}

As described above, we also include the EMPG, {\IzotovEMPG} of 0.019 $Z_{\odot}$ in the sample of this paper.
In Figure \ref{fig:spectraIzotov}, we show a spectrum of {\IzotovEMPG} derived from \citet{Izotov2018a}, showing the detection of the two key emission lines of {\FeIII}4658 (S/N$=$5.1) and {\HeII}4686 (S/N$=$14.9) in this paper.

\section{ANALYSIS} \label{sec:analysis}
In this section, we explain the emission line measurement (Section \ref{subsec:line}) and the estimation of galaxy properties (Section \ref{subsec:prop}) for our 10 metal-poor galaxies.
Here we estimate stellar masses, star-formation rates, emission-line equivalent widths, sizes, and metallicities of our 10 metal-poor galaxies.

\subsection{Emission Line Measurements} \label{subsec:line}
We measure central wavelengths and emission-line fluxes with a best-fit gaussian profile using the {\sc iraf} routine, {\tt splot}.
We also estimate flux errors, which originate from read-out noise and photon noise of sky$+$object emission.
As described in Section \ref{subsec:mage}, we correct fluxes of the LDSS-3/MagE spectra assuming the wavelength-dependent slit-loss with the model of the atmospheric refraction.
We measure observed-frame equivalent widths (EWs) of emission lines with the same {\sc iraf} routine, {\tt splot} and convert them into the rest-frame equivalent widths (EW$_0$). 
Redshifts are estimated by comparing the observed central wavelengths and the rest-frame wavelengths in the air of strong emission lines.

Color excesses, {\EBV} are estimated with the Balmer decrement of {\Ha}, {\Hb}, {\Hc}, {\Hd},..., and {H13} lines under the assumptions of the dust extinction curve given by \citet{Cardelli1989} and the case B recombination.
We do not use Balmer emission lines affected by a systematic error such as cosmic rays and other emission lines blending with the Balmer line.
In the case B recombination, we carefully assume electron temperatures ($T_\mathrm{e}$) so that the assumed electron temperatures become consistent with electron temperature measurements of O$^{2+}$, {\TeOIII}, which will be obtained in Section \ref{subsec:prop}.
We estimate the best {\EBV} values and their errors with the $\chi^2$ method \citep{Press2007}.
The {\EBV} estimation process is detailed in Paper I.
We eventually assume $T_{e}$=10,000 K ({\SDSSEMPGa} and {\SDSSEMPGb}), 15,000 K ({\HSCEMPGa}, {\HSCEMPGb}, {\SDSSEMPGd}, {\SDSSEMPGe}, and {\SDSSEMPGf}), 20,000 K ({\HSCEMPGc} and  {\SDSSEMPGc}), 25,000 K ({\HSCEMPGd}), which are roughly consistent with {\TeOIII} measurements.
We summarize the dust-corrected fluxes in Table \ref{table:flux}.

\begin{turnpage}
\begin{deluxetable*}{cccccccccc}
\tablecolumns{10}
\tabletypesize{\scriptsize}
\tablecaption{Flux measurements 
\label{table:flux}}
\tablehead{%
\colhead{\#} &
\colhead{ID} &
\colhead{{\OII}3727} &
\colhead{{\OII}3729} &
\colhead{{\OII}$_\mathrm{tot}$} &
\colhead{H13} &
\colhead{H12} &
\colhead{H11} &
\colhead{H10} &
\colhead{H9} \\
\colhead{(1)} &
\colhead{(2)} &
\colhead{(3)} &
\colhead{(4)} &
\colhead{(5)} &
\colhead{(6)} &
\colhead{(7)} &
\colhead{(8)} &
\colhead{(9)} &
\colhead{(10)} 
}
\startdata
1 & {\HSCEMPGa} & --- & --- & 166.62$\pm$1.99 & --- & --- & --- & --- & 5.45$\pm$0.68 \\
2 & {\HSCEMPGb} & $<$14.21 & $<$13.50 & $<$19.60 & --- & --- & --- & --- & --- \\
3 & {\HSCEMPGc} & 68.06$\pm$0.81 & 91.82$\pm$0.84 & 159.88$\pm$1.16 & 3.26$\pm$0.68 & 3.63$\pm$0.67 & 6.65$\pm$0.79 & 4.39$\pm$0.66 & 6.13$\pm$0.63 \\
4 & {\HSCEMPGd} & --- & --- & 50.12$\pm$2.66 & --- & --- & --- & --- & 4.68$\pm$1.17 \\
5 & {\SDSSEMPGa} & 74.28$\pm$0.50 & 103.08$\pm$0.51 & 177.36$\pm$0.71 & --- & --- & --- & 5.99$\pm$0.30 & 8.42$\pm$0.25 \\
6 & {\SDSSEMPGb} & 103.35$\pm$0.72 & 149.41$\pm$0.76 & 252.75$\pm$1.05 & --- & --- & --- & --- & 6.74$\pm$0.33 \\
7 & {\SDSSEMPGc} & 34.05$\pm$0.27 & 46.85$\pm$0.30 & 80.91$\pm$0.41 & 2.80$\pm$0.20 & 3.42$\pm$0.32 & 4.23$\pm$0.20 & 4.40$\pm$0.19 & 5.88$\pm$0.18 \\
8 & {\SDSSEMPGd} & 39.44$\pm$0.11 & 55.06$\pm$0.12 & 94.50$\pm$0.16 & 2.36$\pm$0.06 & 3.40$\pm$0.05 & 4.08$\pm$0.05 & 5.59$\pm$0.05 & 7.59$\pm$0.05 \\
9 & {\SDSSEMPGe} & 43.70$\pm$0.10 & 59.12$\pm$0.12 & 102.82$\pm$0.16 & 2.38$\pm$0.06 & 2.75$\pm$0.06 & 3.78$\pm$0.06 & 5.72$\pm$0.07 & 7.49$\pm$0.06 \\
10 & {\SDSSEMPGf} & 45.07$\pm$0.11 & 59.96$\pm$0.12 & 105.03$\pm$0.16 & 2.48$\pm$0.06 & 3.36$\pm$0.06 & 3.99$\pm$0.08 & 5.29$\pm$0.06 & 7.19$\pm$0.06 \\
\hline\hline 
\# & {\NeIII}3869 & {\NeIII}3967 & H7 & H{$\delta$} & H{$\gamma$} & {\OIII}4363 & {\FeIII}4658 & HeII4686 & {\ArIV}4711 \\ 
 (1) & (11) & (12) & (13) & (14) & (15) & (16) & (17) & (18) & (19) \\ \hline 
1 & 63.97$\pm$0.70 & --- & 32.48$\pm$0.48 $^\mathrm{a}$& 21.64$\pm$0.33 & 45.55$\pm$0.28 & 9.06$\pm$0.23 & 0.94$\pm$0.17 & 1.36$\pm$0.17 & 2.02$\pm$0.16 \\
2 & $<$5.87 & $<$3.90 & --- & 26.15$\pm$3.11 & 46.56$\pm$1.67 & $<$1.61 & $<$1.15 & $<$1.04 & $<$0.94 \\
3 & 29.33$\pm$0.63 & 7.74$\pm$0.54 & 17.16$\pm$0.54 & 27.36$\pm$0.51 & 48.20$\pm$0.43 & 5.96$\pm$0.39 & $<$0.29 & $<$0.30 & $<$0.41 \\
4 & 21.73$\pm$1.11 & --- & 20.03$\pm$0.87 $^\mathrm{a}$& 27.53$\pm$0.65 & 46.88$\pm$0.50 & 8.18$\pm$0.48 & 0.92$\pm$0.38 & 2.32$\pm$0.38 & $<$0.35 \\
5 & 46.64$\pm$0.30 & 15.00$\pm$0.21 & 16.27$\pm$0.19 & 26.04$\pm$0.17 & 46.92$\pm$0.14 & 6.38$\pm$0.09 & 0.66$\pm$0.05 & 0.86$\pm$0.05 & 0.79$\pm$0.06 \\
6 & 44.64$\pm$0.37 & 12.19$\pm$0.25 & 15.27$\pm$0.24 & 26.44$\pm$0.20 & 44.90$\pm$0.16 & 6.59$\pm$0.10 & 0.65$\pm$0.06 & 1.20$\pm$0.06 & 0.44$\pm$0.11 \\
7 & 43.59$\pm$0.24 & 13.97$\pm$0.17 & 15.54$\pm$0.16 & 26.12$\pm$0.16 & 46.99$\pm$0.16 & 13.94$\pm$0.11 & 0.87$\pm$0.06 & 2.67$\pm$0.11 & 1.89$\pm$0.08 \\
8 & 64.12$\pm$0.11 & 18.82$\pm$0.06 & 16.12$\pm$0.06 & 25.67$\pm$0.05 & 46.52$\pm$0.06 & 14.13$\pm$0.04 & 0.36$\pm$0.01 & 0.30$\pm$0.02 & 2.12$\pm$0.02 \\
9 & 53.54$\pm$0.10 & 15.86$\pm$0.06 & 16.38$\pm$0.06 & 26.66$\pm$0.06 & 48.57$\pm$0.07 & 14.85$\pm$0.04 & 0.45$\pm$0.02 & 0.48$\pm$0.02 & 1.70$\pm$0.02 \\
10 & 50.72$\pm$0.11 & 15.14$\pm$0.07 & 17.45$\pm$0.07 & 25.93$\pm$0.07 & 47.39$\pm$0.08 & 12.81$\pm$0.05 & 0.63$\pm$0.03 & 0.78$\pm$0.03 & 1.51$\pm$0.03 
\enddata

\tablecomments{
(1): Number.
(2): ID.
(3)--(36): Dust-corrected emission-line fluxes normalized to an {\Hb} line flux in the unit of {\ergscm}. 
Upper limits are given with a 1$\sigma$ level. 
Lines suffering from saturation or affected by sky emission lines are shown as no data here.
{\OII}$_\mathrm{tot}$ represents a sum of {\OII}3727 and {\OII}3729 fluxes. If the spectral resolution is not high enough to resolve {\OII}3727 and {\OII}3729 lines, we only show {\OII}$_\mathrm{tot}$ fluxes.
\tablenotetext{a}{A sum of {\NeIII}3867 and H7 fluxes because they are blended due to the low spectral resolution.}
}
\end{deluxetable*}
\end{turnpage}

\addtocounter{table}{-1}
\begin{turnpage}
\begin{deluxetable*}{cccccccccc}
\tablecolumns{10}
\tabletypesize{\scriptsize}
\tablecaption{Flux measurements --- {\it Continued}}
\tablehead{%
\colhead{\#} &
\colhead{ID} &
\colhead{{\ArIV}4740} &
\colhead{H{$\beta$}} &
\colhead{{\OIII}4959} &
\colhead{{\OIII}5007} &
\colhead{HeI5876} &
\colhead{{\OI}6300} &
\colhead{{\SIII}6312} &
\colhead{{\NII}6548} \\
\colhead{(1)} &
\colhead{(2)} &
\colhead{(20)} &
\colhead{(21)} &
\colhead{(22)} &
\colhead{(23)} &
\colhead{(24)} &
\colhead{(25)} &
\colhead{(26)} &
\colhead{(27)} 
}
\startdata
1 & {\HSCEMPGa} & 0.96$\pm$0.16 & 100.00$\pm$0.24 & 210.62$\pm$0.29 & 626.89$\pm$0.46 & 9.28$\pm$0.09 & 2.94$\pm$0.08 & 1.07$\pm$0.08 & $<$0.19 \\
2 & {\HSCEMPGb} & $<$1.01 & 100.00$\pm$1.00 & 69.57$\pm$0.72 & 207.48$\pm$0.88 & 12.85$\pm$0.41 & 11.88$\pm$0.48 & $<$0.44 & $<$0.32 \\ 
3 & {\HSCEMPGc} & $<$0.36 & 100.00$\pm$0.52 & 102.76$\pm$0.47 & 308.14$\pm$0.65 & 11.11$\pm$0.31 & 5.29$\pm$0.31 & $<$0.30 & 1.99$\pm$0.24 \\ 
4 & {\HSCEMPGd} & $<$0.36 & 100.00$\pm$0.37 & 55.76$\pm$0.34 & 170.92$\pm$0.38 & 9.12$\pm$0.60 & --- & $<$0.55 & $<$0.50 \\ 
5 & {\SDSSEMPGa} & 0.88$\pm$0.06 & 100.00$\pm$0.16 & 196.65$\pm$0.19 & 593.18$\pm$0.32 & 11.09$\pm$0.05 & 2.34$\pm$0.04 & 1.63$\pm$0.03 & 1.92$\pm$0.03 \\ 
6 & {\SDSSEMPGb} & 0.78$\pm$0.07 & 100.00$\pm$0.17 & 183.24$\pm$0.21 & 571.59$\pm$0.34 & 10.74$\pm$0.06 & 2.59$\pm$0.04 & 1.88$\pm$0.04 & 1.68$\pm$0.04 \\ 
7 & {\SDSSEMPGc} & 1.75$\pm$0.07 & 100.00$\pm$0.19 & 165.47$\pm$0.22 & --- & 10.80$\pm$0.06 & 1.70$\pm$0.05 & 1.71$\pm$0.05 & 1.18$\pm$0.04 \\ 
8 & {\SDSSEMPGd} & 1.69$\pm$0.02 & 100.00$\pm$0.07 & 250.60$\pm$0.11 & --- & 11.17$\pm$0.02 & --- & --- & 1.16$\pm$0.01 \\ 
9 & {\SDSSEMPGe} & 1.40$\pm$0.02 & 100.00$\pm$0.09 & 214.32$\pm$0.12 & --- & 10.41$\pm$0.03 & 2.39$\pm$0.02 & 1.23$\pm$0.02 & 1.07$\pm$0.02 \\ 
10 & {\SDSSEMPGf} & 1.12$\pm$0.03 & 100.00$\pm$0.11 & 200.95$\pm$0.14 & --- & 11.28$\pm$0.04 & 2.79$\pm$0.03 & 1.52$\pm$0.02 & 1.31$\pm$0.02 \\ 
\hline\hline
\# & H{$\alpha$} & {\NII}6584 & HeI6678 & {\SII}6716 & {\SII}6731 & HeI7065 & {\ArIII}7136 & {\OII}7320 & {\OII}7330 \\ 
 (1) & (28) & (29) & (30) & (31) & (32) & (33) & (34) & (35) & (36) \\ \hline 
1 & 246.66$\pm$0.20 & 5.08$\pm$0.18 & 2.96$\pm$0.07 & 7.87$\pm$0.08 & 6.07$\pm$0.08 & 2.73$\pm$0.07 & 5.62$\pm$0.08 & 1.45$\pm$0.07 & 0.92$\pm$0.07 \\
2 & 278.45$\pm$0.66 & 2.10$\pm$0.29 & --- & 5.18$\pm$0.33 & $<$0.55 & --- & $<$0.85 & $<$0.46 & $<$0.61 \\ 
3 & 272.09$\pm$0.57 & 8.64$\pm$0.26 & 2.75$\pm$0.21 & 16.24$\pm$0.24 & 8.52$\pm$0.27 & $<$0.30 & 5.52$\pm$0.43 & $<$0.36 & $<$0.37 \\ 
4 & 229.46$\pm$1.00 & $<$0.48 & 2.07$\pm$0.59 & --- & --- & --- & --- & $<$0.48 & $<$0.54 \\ 
5 & 280.48$\pm$0.17 & 5.68$\pm$0.04 & 2.99$\pm$0.03 & 11.19$\pm$0.04 & 5.97$\pm$0.03 & 2.58$\pm$0.03 & --- & --- & --- \\ 
6 & 276.34$\pm$0.20 & 5.28$\pm$0.04 & 2.64$\pm$0.03 & 10.39$\pm$0.06 & 8.14$\pm$0.04 & 1.42$\pm$0.04 & 6.30$\pm$0.05 & 1.84$\pm$0.04 & 1.47$\pm$0.04 \\ 
7 & 278.90$\pm$0.22 & 3.11$\pm$0.04 & --- & 5.23$\pm$0.04 & 4.66$\pm$0.04 & 3.61$\pm$0.05 & 4.61$\pm$0.06 & 1.09$\pm$0.04 & 0.83$\pm$0.04 \\ 
8 & --- & 3.29$\pm$0.01 & 2.97$\pm$0.01 & 6.47$\pm$0.01 & 5.07$\pm$0.01 & 3.56$\pm$0.01 & 5.08$\pm$0.02 & 1.55$\pm$0.01 & 1.03$\pm$0.01 \\ 
9 & --- & 2.76$\pm$0.02 & 2.65$\pm$0.02 & 7.65$\pm$0.02 & 5.56$\pm$0.02 & 3.18$\pm$0.02 & 4.70$\pm$0.02 & 1.26$\pm$0.02 & 0.98$\pm$0.02 \\ 
10 & --- & 3.21$\pm$0.02 & 2.93$\pm$0.02 & 8.78$\pm$0.03 & 6.49$\pm$0.03 & 3.83$\pm$0.03 & 5.36$\pm$0.03 & 1.69$\pm$0.02 & --- \\ 
\enddata
\tablecomments{
Continued.
}
\end{deluxetable*}
\end{turnpage}
\begin{deluxetable*}{lcccccccc}
\tablecolumns{9}
\tabletypesize{\scriptsize}
\tablecaption{Parameters of our metal-poor galaxies %
\label{table:prop}}
\tablehead{%
\colhead{\#} &
\colhead{ID} &
\colhead{EMPG?} &
\colhead{redshift} &
\colhead{{\EW}(H$\beta$)} &
\colhead{{12+log(O/H)}} &
\colhead{$\log$($M_{\star}$)} &
\colhead{$\log$($SFR$)} &
\colhead{{\EBV}} \\
\colhead{} &
\colhead{} &
\colhead{} &
\colhead{} &
\colhead{({\AA})} &
\colhead{} &
\colhead{($M_{\odot}$)} &
\colhead{({$M_{\odot}$\,yr$^{-1}$})} & 
\colhead{(mag)} \\
\colhead{(1)} &
\colhead{(2)} &
\colhead{(3)} &
\colhead{(4)} &
\colhead{(5)} &
\colhead{(6)} &
\colhead{(7)} &
\colhead{(8)} & 
\colhead{(9)}
}
\startdata
1  &  {\HSCEMPGa}   & {\it no} & $0.02980$ & $172.6^{+0.7}_{-0.6}$ & $8.27\pm0.02$ & $6.55^{+0.13}_{-0.09}$ & $0.43\pm0.01$ & $0.35\pm0.02$  \\ 
2  &  {\HSCEMPGb}  & {\bf \it yes}$^\mathrm{a}$ & $0.03265$ & $213.3^{+23.4}_{-17.6}$ & $7.23^{+0.03}_{-0.02}$$^\mathrm{a}$ & $5.17\pm0.01$ & $-0.85\pm0.01$ & $0.28\pm0.03$  \\ 
3  &  {\HSCEMPGc}   & {\it no} & $0.02035$ & $111.9^{+1.4}_{-1.3}$ & $7.72\pm0.03$ & $4.95^{+0.04}_{-0.01}$ & $-1.07\pm0.01$ & $0.00^{+0.02}_{-0.00}$  \\ 
4  &  {\HSCEMPGd}  & {\it yes}  & $0.03125$ & $123.5^{+3.5}_{-2.8}$ & $6.90\pm0.03$ & $5.89^{+0.10}_{-0.09}$ & $-1.28\pm0.01$ & $0.19\pm0.03$  \\ 
5  &  {\SDSSEMPGa}  & {\it no} & $0.02083$ & $103.9\pm0.2$ & $8.22\pm0.01$ & $7.06\pm0.03$ & $0.00\pm0.01$ & $0.00^{+0.01}_{-0.00}$  \\  
6  &  {\SDSSEMPGb}  & {\it no} & $0.01725$ & $153.7^{+0.5}_{-0.4}$ & $8.45\pm0.01$ & $6.06^{+0.03}_{-0.13}$ & $-0.17\pm0.01$ & $0.02\pm0.02$  \\  
7  &  {\SDSSEMPGc}  & {\it yes} & $0.02296$ & $214.0^{+0.9}_{-0.8}$ & $7.68\pm0.01$ & $6.56\pm0.02$ & $0.27\pm0.01$ & $0.19\pm0.03$  \\  
8  &  {\SDSSEMPGd}  & {\it no} & $0.00730$ & $264.7\pm0.3$ & $7.973\pm0.002$ & $5.78\pm0.01$ & $-0.54\pm0.01$ & $0.00^{+0.01}_{-0.00}$  \\  
9  &  {\SDSSEMPGe}  & {\it no} & $0.01245$ & $127.6\pm0.2$ & $7.890^{+0.003}_{-0.004}$ & $6.99\pm0.03$ & $-0.16\pm0.01$ & $0.01^{+0.02}_{-0.01}$  \\  
10 & {\SDSSEMPGf}  & {\it no} & $0.01812$ & $111.0\pm0.2$ & $7.866^{+0.004}_{-0.005}$ & $6.51^{+0.02}_{-0.03}$ & $-0.18\pm0.01$ & $0.00^{+0.02}_{-0.00}$  
\enddata
\tablecomments{
(1): Number.
(2): ID.
(3): Whether or not an object satisfies the EMPG definition, {\metal}$<$7.69. If yes (no), we write {\it yes} ({\it no}) in the column.
(5): Rest-frame equivalent width of an {\Hb} emission line.
(6): Gas-phase metallicity based on the $T_\mathrm{e}$ method except for {\HSCEMPGb}.  
(7): Stellar mass.
(8): Star-formation rate. 
(9): Color excess.
\tablenotetext{a}{The metallicity of {\HSCEMPGb} is obtained with the metallicity calibration of \citet{Skillman1989c}.}
}
\end{deluxetable*}

\subsection{Galaxy Properties} \label{subsec:prop}
In this section, we estimate gas-phase metallicities (O/H), and gas-phase element abundance ratios of our 10  galaxies.
Note that metallicities are already estimated in Paper I, as well as stellar masses, SFRs, and color excesses.

We estimate electron temperatures of O$^{2+}$ ($T_{\mathrm e}$({\sc{Oiii}})) and O$^{+}$ ($T_{\mathrm e}$({\sc{Oii}})), using line ratios of {\OIII}4363/5007 and {\OII}(3727+3729)/(7320+7330), respectively.
We use nebular physics calculation codes of PyNeb \citep[][v1.0.14]{Luridiana2015} to estimate electron temperatures.
If an {\OII}5007 line is saturated, we estimate an {\OII}5007 flux with  
	\begin{equation}
	\textrm{\OIII}5007=2.98\times \textrm{\OIII}4959, \label{eq:Oiii_replace}
	\end{equation}
which is strictly determined by the Einstein {\it A} coefficient.
If either of {\OII}7320 or {\OII}7330 line is detected, we estimate a total flux of {\OII}(7320+7330) with a relation of 
	\begin{equation}
	\textrm{\OII}7330=0.56\times \textrm{\OII}7320. \label{eq:Oii_auroral}
	\end{equation}
We have confirmed that Equation (\ref{eq:Oii_auroral}) holds with very little dependence on $T_\mathrm{e}$ and $n_\mathrm{e}$, using PyNeb.
If none of {\OII}7320,7330 line is detected, we estimate $T_{\mathrm e}$({\sc{Oii}}) from an empirical relation of  
	\begin{equation}
	{T_\mathrm{e}}(\textrm{O\,\textsc{ii}}) = 0.7 \times {T_\mathrm{e}}(\textrm{O\,\textsc{iii}}) + 3000,
	\end{equation}
which has been confirmed by \citet{Campbell1986} and \citet{Garnett1992}.
We also assume 
	\begin{equation}
	{T_\mathrm{e}}(\textrm{S\,\textsc{iii}}) = 0.83 \times {T_\mathrm{e}}(\textrm{O\,\textsc{iii}}) + 1700,
	\end{equation}
to estimate electron temperatures associated with S$^{2+}$ ions \citep{Garnett1992}.
We regard $T_{\mathrm e}$({\sc{Oiii}}), $T_{\mathrm e}$({\sc{Siii}}), and $T_{\mathrm e}$({\sc{Oii}}) as representative electron temperatures associated with ions in high, intermediate, and low ionization states, respectively. 

We estimate gas-phase metallicities, {\metal}, based on electron temperature measurements, which are so-called $T_\mathrm{e}$-metallicities.
Hereafter, we call the $T_\mathrm{e}$-metallicity just ``metallicity'' unless we describe explicitly. 
We also use PyNeb to estimate metallicities.
The latest atomic data are used in the PyNeb codes.
We do not estimate a $T_{\mathrm e}$-based metallicity of {\HSCEMPGb} because none of the $T_{\mathrm e}$({\sc{Oiii}}), $T_{\mathrm e}$({\sc{Oii}}), and $T_{\mathrm e}$({\sc{Siii}}) is estimated due to non-detection of {\OIII}4363 and {\OII}7320,7330 emission lines.
Instead, we estimate the metallicity of {\HSCEMPGb} with a calibrator obtained by \citet{Skillman1989c} as described in Paper I.
The estimates of gas-phase metallicities are summarized in Table \ref{table:prop}.
The estimation of electron temperatures and metallicities are detailed in Paper I.

\begin{deluxetable*}{lccccc}
\tablecolumns{6}
\tabletypesize{\scriptsize}
\tablecaption{Element abundance ratios %
\label{table:abund}}
\tablehead{%
\colhead{\#} &
\colhead{ID} &
\colhead{log(Ne/O)} &
\colhead{log(Ar/O)} &
\colhead{log(N/O)} &
\colhead{log(Fe/O)} \\
\colhead{(1)} &
\colhead{(2)} &
\colhead{(3)} &
\colhead{(4)} &
\colhead{(5)} &
\colhead{(6)} 
}
\startdata
1 & {\HSCEMPGa} & $-0.634^{+0.006}_{-0.007}$  & $-2.634^{+0.028}_{-0.026}$  & $-1.753^{+0.020}_{-0.023}$  & $-1.994^{+0.091}_{-0.075}$$^{}_{-0.223}$$^\mathrm{c}$  \\
2 & {\HSCEMPGb} & --- $^\mathrm{a}$& --- $^\mathrm{a}$& --- $^\mathrm{a}$& --- $^\mathrm{a}$\\
3 & {\HSCEMPGc} & $-0.712^{+0.008}_{-0.010}$  & $<-2.253$  & $-1.297^{+0.019}_{-0.014}$  & $<-2.124$  \\
4 & {\HSCEMPGd} & $-0.641^{+0.022}_{-0.019}$  & --- $^\mathrm{b}$& $<-1.710$  & $-1.246^{+0.174}_{-0.313}$$^{}_{-0.222}$$^\mathrm{c}$  \\
5 & {\SDSSEMPGa} & $-0.701^{+0.003}_{-0.002}$  & --- $^\mathrm{b}$& $-1.644\pm0.004$  & $-2.126^{+0.035}_{-0.027}$$^{}_{-0.217}$$^\mathrm{c}$  \\
6 & {\SDSSEMPGb} & $-0.754\pm0.005$  & $-2.704^{+0.016}_{-0.015}$  & $-1.943^{+0.006}_{-0.009}$  & $-2.335^{+0.027}_{-0.038}$$^{}_{-0.211}$$^\mathrm{c}$  \\
7 & {\SDSSEMPGc} & $-0.757^{+0.005}_{-0.004}$  & $-2.274^{+0.007}_{-0.013}$  & $-1.518^{+0.009}_{-0.011}$  & $-1.639^{+0.026}_{+0.026}$$^{}_{-0.209}$$^\mathrm{c}$  \\
8 & {\SDSSEMPGd} & $-0.707\pm0.001$  & $-2.391\pm0.002$  & $-1.563\pm0.003$  & $-2.078^{+0.020}_{-0.022}$$^{}_{-0.221}$$^\mathrm{c}$  \\
9 & {\SDSSEMPGe} & $-0.761\pm0.001$  & $-2.440\pm0.004$  & $-1.710^{+0.005}_{-0.004}$  & $-2.046^{+0.019}_{-0.023}$$^{}_{-0.210}$$^\mathrm{c}$ \\
10 & {\SDSSEMPGf} & $-0.737^{+0.001}_{-0.002}$  & $-2.386\pm0.006$  & $-1.616^{+0.006}_{-0.004}$  & $-1.890^{+0.022}_{-0.017}$$^{}_{-0.213}$$^\mathrm{c}$  
\enddata
\tablecomments{
(1): Number.
(2): ID.
(3)--(6): Gas-phase element abundance ratios of Ne/O, Ar/O, N/O, and Fe/O.
Upper limits are given with a 2$\sigma$ confidence level.
\tablenotetext{a}{Not estimated due to the lack of electron temperature estimates.}
\tablenotetext{b}{Not estimated because the {\ArIII}7136 emission line is strongly affected by the sky emission line.}
\tablenotetext{c}{We show two kinds of Fe/O lower errors. The first term is the statistical error propagated from spectral noise, and the second one is the systematic error originated from ICF uncertainties explained in Section \ref{subsec:prop}.}
}
\end{deluxetable*}

We estimate gas-phase element abundance ratios of neon-to-oxygen (Ne/O), argon-to-oxygen (Ar/O), nitrogen-to-oxygen (N/O), and iron-to-oxygen (Fe/O) in a similar way to \citet{Izotov2006}.
First, we estimate ion abundance ratios of Ne$^{2+}$/H$^{+}$, Ar$^{3+}$/H$^{+}$, Ar$^{2+}$/H$^{+}$, N$^{+}$/H$^{+}$, and, Fe$^{2+}$/H$^{+}$ with the PyNeb codes.
The atomic data used in the PyNeb calculation are shown in Table \ref{table:atomdata}.
Because different ions reside in different parts of an {\HII} region, we choose one of the $T_{\mathrm e}$({\sc{Oiii}}), $T_{\mathrm e}$({\sc{Siii}}), and $T_{\mathrm e}$({\sc{Oii}}) to estimate abundances of each ion according to their ionization potential.
We use $T_{\mathrm e}$({\sc{Oiii}}) to estimate abundances of O$^{2+}$, Ne$^{2+}$, and Ar$^{3+}$.
We adopt $T_{\mathrm e}$({\sc{Siii}}) in the estimation of Ar$^{2+}$ abundances.
We apply $T_{\mathrm e}$({\sc{Oii}}) for abundances of low-ionizion ions, O$^{+}$, N$^{+}$, and Fe$^{2+}$.
Second, we convert the ion abundances into element abundances with ionization correction factors ($ICF$s) of \citet{Izotov2006} shown below:
	\begin{eqnarray}
	\frac{\mathrm{Ne}} {\mathrm{H}} &=& \frac{\mathrm{Ne}^{+}} {\mathrm{H}^{+}} + ICF (\mathrm{Ne}^{+}), \\
	\frac{\mathrm{Ar}} {\mathrm{H}} &=& \frac{\mathrm{Ar}^{3+}+\mathrm{Ar}^{2+}} {\mathrm{H}^{+}} + ICF (\mathrm{Ar}^{3+}+\mathrm{Ar}^{2+}), \\
	\frac{\mathrm{N}} {\mathrm{H}} &=& \frac{\mathrm{N}^{+}} {\mathrm{H}^{+}} + ICF (\mathrm{N}^{+}), \\
	\frac{\mathrm{Fe}} {\mathrm{H}} &=& \frac{\mathrm{Fe}^{2+}} {\mathrm{H}^{+}} + ICF (\mathrm{Fe}^{2+}).
	\end{eqnarray}
The $ICF$s are based on {\HII} region models \citep{Stasinska2003} and are given as a function of $v=$O$^{+}$/(O$^{2+}$+O$^{+}$) or $w=$O$^{2+}$/(O$^{2+}$+O$^{+}$).
Finally, we obtain Ne/O, Ar/O, N/O, and Fe/O ratios by dividing Ne/H, Ar/H, N/H, and Fe/H by O/H (i.e., metallicity).
We do not estimate Ne/O, Ar/O, N/O, and Fe/O ratios of {\HSCEMPGb} because none of the $T_{\mathrm e}$({\sc{Oiii}}), $T_{\mathrm e}$({\sc{Oii}}), and $T_{\mathrm e}$({\sc{Siii}}) is obtained.
The Ar/O ratios of {\HSCEMPGd} and {\SDSSEMPGa} are not estimated as well because the {\ArIII}7136 emission line is strongly affected by the sky emission line.
\citet{Rodriguez2003} suggests that $ICF$(Fe$^{2+}$) of the \citet{Stasinska2003} models may include systematic errors, which originate from uncertainties of a recombination rate of Fe$^{2+}$ and/or uncertain collision strengths of Fe$^{2+}$ and Fe$^{3+}$.
Thus, we also estimate the $ICF$(Fe$^{2+}$) values with another model of \citet{Rodriguez2005}.
The ICFs obtained by the \citet{Stasinska2003} models are $\sim$0.2 dex higher than the \citet{Rodriguez2005} models.
In this paper, we regard the ICF offsets between the two models as systematic errors of Fe/O, which are included in lower errors of Fe/O (see Table \ref{table:abund} and  Figures \ref{fig:abundances} and \ref{fig:FeO}).
For the literature EMPG, {\IzotovEMPG}, we derive the element abundances from \citet{Izotov2018a}.
The element abundances of {\IzotovEMPG} are obtained in the same manner as in this paper.
We summarize the element abundance ratios in Table \ref{table:abund}.

\begin{deluxetable*}{lccc}
\tablecolumns{6}
\tabletypesize{\scriptsize}
\tablecaption{Atomic Data %
\label{table:atomdata}}
\tablehead{%
\colhead{Ion} &
\colhead{Emission Process} &
\colhead{Atomic Data} &
\colhead{Line Data}\\
\colhead{(1)} &
\colhead{(2)} &
\colhead{(3)} &
\colhead{(4)} 
}
\startdata
H$^{0}$ & Re & \citet{Storey1995} & \citet{Storey1995} \\
O$^{+}$ & CE & \citet{Fischer2004} & \citet{Kisielius2009} \\
O$^{2+}$ & CE & \citet{Storey2000}, \citet{Fischer2004} & \citet{Storey2014} \\
Ne$^{2+}$ & CE & \citet{Galavis1997} & \citet{McLaughlin2000} \\
Ar$^{2+}$ & CE & \citet{MunozBurgos2009} & \citet{MunozBurgos2009} \\
Ar$^{3+}$ & CE & \citet{Mendoza1982} & \citet{Ramsbottom1997} \\
N$^{+}$ & CE & \citet{Fischer2004} & \citet{Tayal2011} \\
Fe$^{2+}$ & CE & \citet{Quinet1996}, \citet{Johansson2000} & \citet{Quinet1996}  
\enddata
\tablecomments{
(1): Ion.
(2): Re and CE represent the recombination and the collisional excitation, respectively. 
(3): References of transition probabilities used in this paper.
(4): References of line emissivities in a 2D temperature-density dependent table (Re) and the temperature-dependent collision strengths (CE) applied in this paper.
}
\end{deluxetable*}

\section{RESULTS AND DISCUSSIONS} \label{sec:result_discussion}

\subsection{Element Abundance Ratios} \label{subsec:abund}
\begin{figure*}[htbp]
\epsscale{1.1}
\plotone{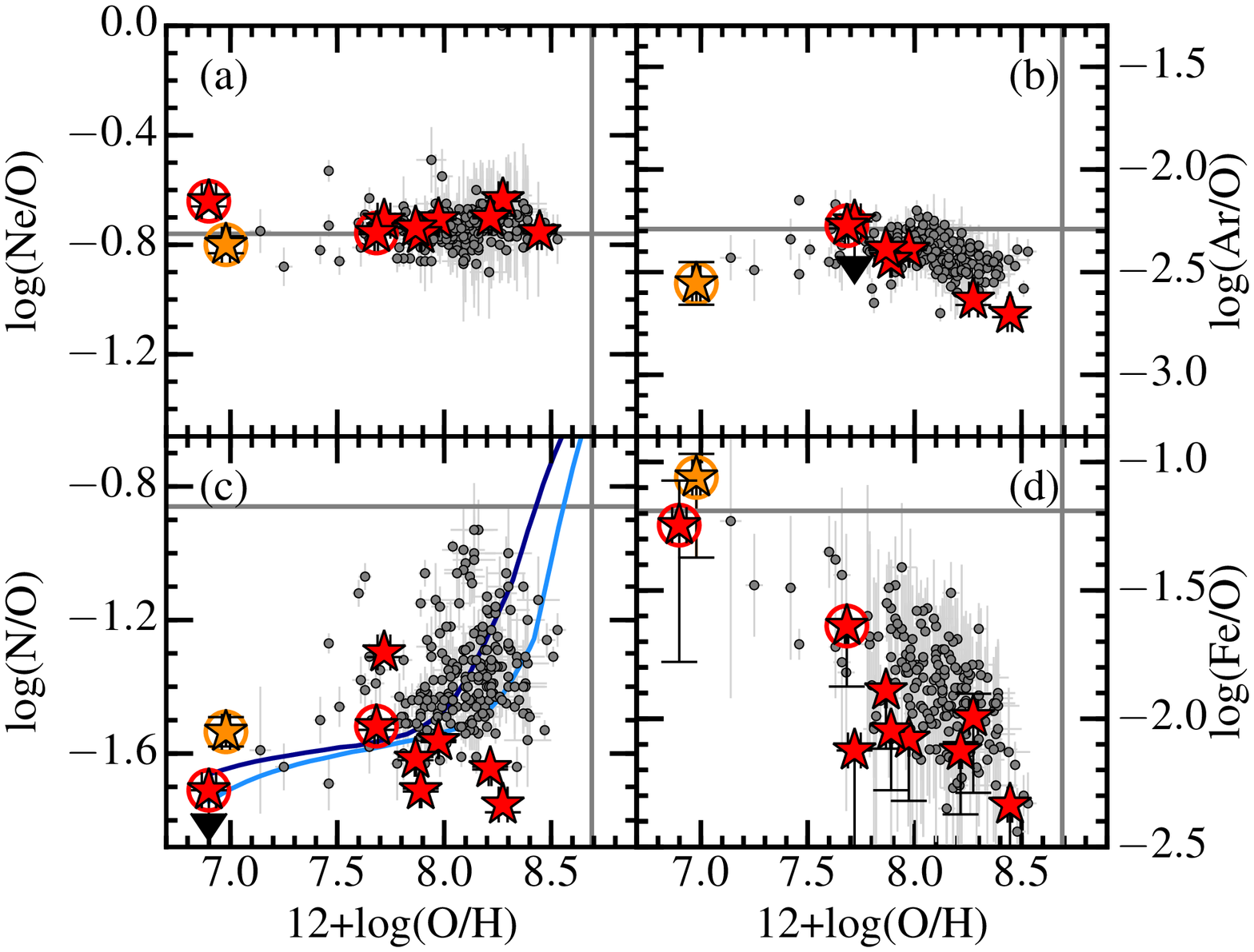}
\caption{Element abundance ratios of neon, argon, nitrogen, and iron to oxygen (Ne/O, Ar/O, N/O, and Fe/O) are shown as a function of metallicity in the panels (a) to (d), respectively.
Our metal-poor galaxies from HSC-EMPG and SDSS-EMPG  source catalogs are shown with red stars.
The EMPG, {\IzotovEMPG}, which is derived from \citet{Izotov2018a} is shown with the orange star.
Metal-poor galaxies that satisfy the EMPG are marked with a large circle.
Here we do not show {\HSCEMPGb}, whose $T_\mathrm{e}$-metallicity and element abundances are not estimated due to the lack of the electron temperature measurement.
Gray dots represent local galaxies of \citet{Izotov2006}.
Gray vertical and horizontal lines indicate solar abundance ratios and metallicity \citep{Asplund2009}.
Solid lines in the panel (c) are the model calculations of the N/O evolution \citep[][]{Vincenzo2016} with star-formation efficiencies of 0.5 (dark blue) and 1.0 (light blue) Gyr$^{-1}$.
The Ar/O ratios of {\HSCEMPGd} and {\SDSSEMPGa} are not shown here because the {\ArIII}7136 emission line is strongly affected by the sky emission line.
\label{fig:abundances}}
\end{figure*}

We show the element abundance ratios of neon, argon, nitrogen, and iron to oxygen (Ne/O, Ar/O, N/O, and Fe/O) of our metal-poor galaxy sample consisting of 10 metal-poor galaxies from Paper I and {\IzotovEMPG} from \citet{Izotov2018a}.
Figure \ref{fig:abundances} shows the Ne/O, Ar/O, N/O, and Fe/O ratios as a function of metallicity, 12+log(O/H).
Thanks to the two representative EMPGs, {\HSCEMPGd} (0.016 $Z_{\odot}$) and {\IzotovEMPG} (0.019 $Z_{\odot}$), we are able to investigate and discuss the low metallicity end (below 0.02 $Z_{\odot}$) of the element abundances for the first time.
We discuss these element abundance ratios in the following subsections.
We compare the element abundance ratios of our metal-poor galaxy sample with a metal-poor galaxy sample of \citet{Izotov2006}, whose typical stellar mass range is larger than our sample galaxies.

\subsubsection{Ne/O and Ar/O ratios} \label{subsubsec:NeO_ArO}
\citet{Izotov2006} report that Ne/O and Ar/O ratios little depend on metallicity because the neon, argon, and oxygen are all $\alpha$ elements, which are produced by the nuclear fusion of $\alpha$ particles inside stars.
As shown in the panels (a) and (b) of Figure \ref{fig:abundances}, we find that our metal-poor galaxy sample shows almost constant values of $\log$(Ne/O) $\sim$$-0.8$ and $\log$(Ar/O)$\sim$$-2.5$ within a scatter of $\pm$0.2 dex, which are almost consistent with the solar abundance ratios.
We also find that the Ne/O and Ar/O ratios are consistent with those of local galaxies reported by \citet{Izotov2006} within the scatter.
The consistency suggests that our metal-poor galaxy sample also shows no metallicity dependence in Ne/O and Ar/O ratios.

Note that the Ar/O ratio might slightly decrease in the range of {\metal}$\gtrsim$8.2 in our sample and the \citet{Izotov2006} sample, in contrast to the Ne/O ratios.
The Ar/O ratio is expected to be constant and consistent with the solar abundance, $\log$(Ar/O)$_\odot$$=$$-2.29$, because there seems to be no physical reason for the Ar/O (i.e., $\alpha$ element ratio) decrease at {\metal}$\gtrsim$8.2.
It may be explained by the underlying unknown systematics in the Ar/O estimation at {\metal}$\gtrsim$8.2.
We do not discuss it further in this paper because the Ar/O abundances at relatively higher metallicities are out of the scope of this paper.

\subsubsection{N/O ratio} \label{subsubsec:NO}
As suggested by previous studies \citep{Perez-Montero2009, Perez-Montero2013, Andrews2013}, N/O ratios of SFGs present a plateau at $\log$(N/O)$\sim$$-1.6$ in the range of {\metal}$\lesssim$8.0 and a positive slope at higher metallicities as a function of metallicity.
The panel (c) of Figure \ref{fig:abundances} presents model calculations of the N/O evolution \citep[][]{Vincenzo2016}, which also show the plateau and positive slope.
The plateau basically results from the primary nucleosynthesis of massive stars, while the positive slope is mainly attributed to the secondary nucleosynthesis of low- and intermediate-mass stars \citep[e.g.,][]{Vincenzo2016}.
We briefly describe the two nitrogen production processes below. 
\begin{itemize}
\item Primary nucleosynthesis: Inside a metal-poor star, protons are burned through the proton-proton ($p$-$p$) chain reaction, and little nitrogen is produced at this stage.
Nitrogen elements are mainly produced after the formation of a heavy-element core (e.g., O and C) and ejected into ISM by SNe, for stars more massive than $\sim$8 $M_{\odot}$.
\item Secondary nucleosynthesis: Metal-rich stars efficiently burn hydrogen through the carbon-nitrogen-oxygen (CNO) cycle, where nitrogen elements accumulate because $^{14}$N fusion ($^{14}$N$+$$p$ $\rightarrow$ $^{15}$O$+$$\gamma$) is the slowest process in the CNO cycle.
Then nitrogen is ejected through stellar winds during the asymptotic giant branch (AGB) phase, $\sim$1 Gyr after the birth of low- and intermediate-mass stars.
\end{itemize}
As shown in the panel (c) of Figure \ref{fig:abundances}, most of our metal-poor galaxies have N/O ratios of $\log$(N/O)$<$$-1.5$ (i.e., less than $\sim$30 percent of the solar N/O ratio).
Especially, {\HSCEMPGd} has a strong, 2$\sigma$ upper limit of $\log$(N/O)$<$$-1.71$, and {\IzotovEMPG} show a low N/O ratio of $\log$(N/O)$=$$-1.53$.
The N/O values of the two EMPGs ({\HSCEMPGd} and {\IzotovEMPG}) will be discussed again in Section \ref{subsubsec:FeO}.
These low N/O ratios suggest that our metal-poor galaxies have not yet started the secondary nucleosynthesis due to their low metallicities and young stellar ages.

We also find several galaxies of our metal-poor galaxy sample have relatively low N/O ratios compared to the model lines of \citet[][]{Vincenzo2016} and \citet{Izotov2006} SFGs at {\metal}$\sim$8.0.
\citet[][]{Vincenzo2016} find that the N/O plateau is lowered under the assumption of high sSFR or the top heavy initial mass function (IMF).
Indeed, our metal-poor galaxy sample have high sSFRs ($\sim$300\,Gyr$^{-1}$) compared to those of the \citet{Izotov2006} SFGs ($\sim$1--10\,Gyr$^{-1}$) because we aim to obtain galaxies with high sSFRs in this study.
Thus, the N/O differences between our metal-poor galaxy sample and \citet{Izotov2006} SFG sample are explained by the sample selection.

\subsubsection{Fe/O ratio} \label{subsubsec:FeO}
In the panel (d) of Figure \ref{fig:abundances}, we find that our metal-poor galaxies show a decreasing trend in Fe/O ratio as metallicity increases.
The same decreasing Fe/O trend is found in a star-forming-galaxy sample of \citet[][]{Izotov2006}.
Most of our metal-poor galaxies have Fe/O ratios comparable to a star-forming-galaxy sample of \citet[][]{Izotov2006}.
Three EMPGs, {\HSCEMPGd}, {\SDSSEMPGc}, and {\IzotovEMPG} (encircled by a red or orange circle) show relatively high Fe/O ratios, log(Fe/O)$>$$-1.7$, among our metal-poor galaxies.
Especially, we find that {\HSCEMPGd} and {\IzotovEMPG}, two of the lowest metallicity galaxies with 0.016 and 0.019 (O/H)$_{\odot}$, have high Fe/O ratios of log(Fe/O)$=$$-1.25^{+0.17}_{-0.53}$ and log(Fe/O)$=$$-1.06^{+0.09}_{-0.31}$, respectively, which are comparable to the solar Fe/O ratio, log(Fe/O)$_{\odot}$$=$$-1.19$.
In this paper, we mainly focus on the two representative EMPGs, {\HSCEMPGd} and {\IzotovEMPG}, which interestingly show high Fe/O ratios.
Note again that {\IzotovEMPG} is an EMPG reported by \citet{Izotov2018a}. 
Table \ref{table:EMPGparam} summarizes the Fe/O ratios and {\HeII}4686/{\Hb} ratios (discussed in Section \ref{subsubsec:strong_Heii}) of the two representative EMPGs ({\HSCEMPGd} and {\IzotovEMPG}), which play an important role in this paper.
We also show the fluxes of the two kew emission lines of {\FeIII}4658 and {\HeII}4686 in Table \ref{table:EMPGparam}.

\begin{deluxetable*}{ccccccc}
\tablecolumns{7}
\tabletypesize{\scriptsize}
\tablecaption{Parameters of two representative EMPGs %
\label{table:EMPGparam}}
\tablehead{%
\colhead{ID} &
\colhead{{12+log(O/H)}} &
\colhead{log(Fe/O)} &
\colhead{log({\HeII}/{\Hb})} &
\colhead{$F$({\FeIII})} &
\colhead{$F$({\HeII})} &
\colhead{Ref.} \\
\colhead{} &
\colhead{} &
\colhead{} &
\colhead{} &
\colhead{(\ergscm)} &
\colhead{(\ergscm)} &
\colhead{} \\
\colhead{(1)} &
\colhead{(2)} &
\colhead{(3)} &
\colhead{(4)} &
\colhead{(5)} &
\colhead{(6)} &
\colhead{(7)} 
}
\startdata
{\HSCEMPGd} & $6.90\pm0.03$ & $-1.25^{+0.17}_{-0.31}$$^{}_{-0.22}$$^\mathrm{a}$ & $-1.58^{+0.07}_{-0.08}$ & $12.0\pm5.02$ & $30.5\pm5.04$ &  {\bf This paper}  \\ 
{J0811$+$4730} & $6.98\pm0.02$ & $-1.06^{+0.09}_{-0.09}$$^{}_{-0.22}$$^\mathrm{a}$ & $-1.64\pm0.03$ & $6.55\pm1.26$ & $28.6\pm1.89$ & I18    
\enddata
\tablecomments{
(1): ID.
(2): Gas-phase metallicity.  
(3): Abundance ratio of log(Fe/O).
(4): Emission line ratio of log({\HeII}/{\Hb}). 
(5)--(6): Emission line fluxes of {\FeIII}4658 and {\HeII}4686 in the unit of 10$^{-18}$ {\ergscm}.
(7): Reference. I18 represents \citet{Izotov2018a}.
\tablenotetext{a}{We show two kinds of Fe/O lower errors. The first term is the statistical error propagated from spectral noise, and the second one is the systematic error originated from ICF uncertainties explained in Section \ref{subsec:prop}.}
}
\end{deluxetable*}

\begin{figure*}[htbp]
\epsscale{0.9}
\plotone{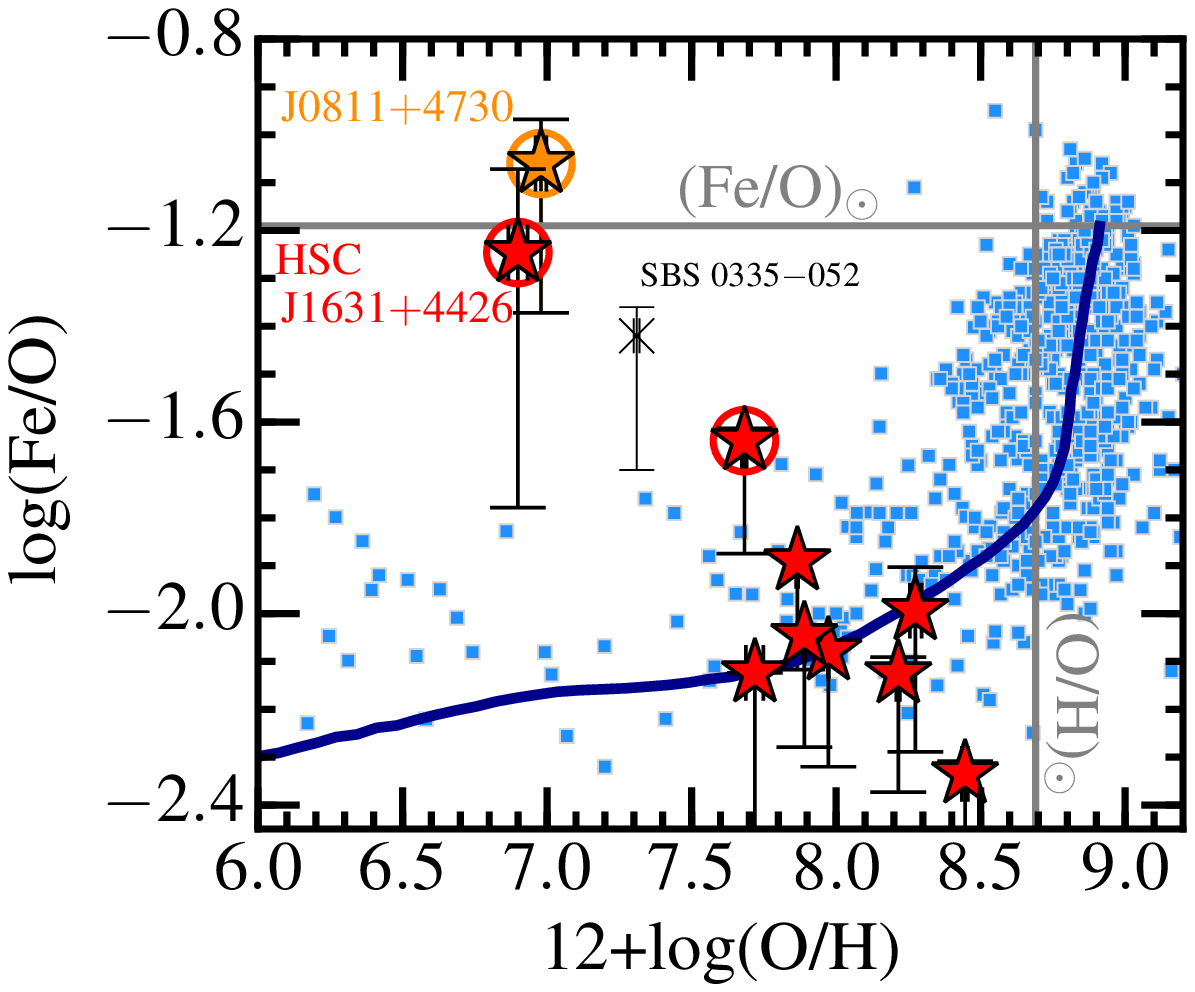}
\caption{Comparison of Fe/O ratios of our metal-poor galaxies (symbols are the same as Figure \ref{fig:abundances}) and Galactic stars (blue squares).
We show observational data of Galactic stars from \citet{Cayrel2004}, \citet{Gratton2003}, and \citet{Bensby2013}.
Blue solid line represents a stellar Fe/O evolution model under the assumption that gas is enriched by massive stars with 9--100 $M_{\odot}$ \citep{Suzuki2018}.
The cross represents another metal-poor galaxy, SBS 0335$-$052 from literature \citep{Izotov1999}. 
SBS 0335$-$052 has 59\% of the solar Fe/O ratio.
Here we do not show {\HSCEMPGb}, whose $T_\mathrm{e}$-metallicity and element abundances are not estimated due to the lack of the electron temperature measurement.
\label{fig:FeO}}
\end{figure*}

To characterize the two EMPGs ({\HSCEMPGd} and {\IzotovEMPG}) with a high Fe/O ratio, we also compare our metal-poor galaxies with Galactic stars \citep{Cayrel2004, Gratton2003, Bensby2013} in Figure \ref{fig:FeO}.
The Galactic star samples are composed of dwarf or subdwarf stars.
The solid line here represents a stellar Fe/O evolution model under the assumption that gas is enriched by massive stars with 9--100 $M_{\odot}$ \citep{Suzuki2018}.
The gaseous abundance ratios at the time of the star formation are imprinted in the stellar abundance patterns because stars are formed from gas.
As explained in Section \ref{sec:intro}, the Fe/O ratio increases at {\metal}$\gtrsim$8.0 due to the contribution of type-Ia SNe $\sim$1 Gyr after the start of the star formation. 
Surprisingly, we find in Figure \ref{fig:FeO} that the two EMPGs ({\HSCEMPGd} and {\IzotovEMPG}) deviate from the observational results of Galactic stars and the Fe/O evolution model.
Below, we mainly focus on the discussion of the two EMPGs of {\HSCEMPGd} and {\IzotovEMPG}, unless we specifically describe explicitly. 
Note that our metal-poor galaxies present gas-phase Fe/O ratios in nebulae, while the Galactic stars show the Fe/O ratio obtained from stellar atmospheric absorption lines.
The nebular abundance ratios are subject to change by the effects of SNe, stellar wind, and galactic inflow in a short time scale (i.g., $\lesssim$ 10 Myr).
By contrast, the abundance ratios of Galactic stars (i.e., dwarf and subdwarf stars) change little across the cosmic time because the element production proceeds very slowly and the heavy elements such as oxygen and iron are produced little in low-mass stars such as dwarf and subdwarf stars.
Thus, the abundance ratios of the dwarf and subdwarf stars are almost fixed at the time of the star formation, and can be regarded as tracers of the past chemical evolution.
The dwarf and subdwarf stars with low metallicities of 0.01--0.1 $Z_{\odot}$ are as old as $\sim$12 Gyr \citep[e.g.,][]{Bensby2013}, which is right after the MW formation.
The two EMPGs ({\HSCEMPGd} and {\IzotovEMPG}), whose Fe/O ratios deviate from the Galactic stars and the Fe/O model, suggest that their Fe/O ratios have increased for some reason after the galaxy formation.

We discuss a possibility that the Fe/O ratios might be overestimated by the contribution of hard EUV radiation (e.g., AGNs) or shock heating (e.g., SNe).
Collisional excitation lines of low-ionization ions such as {\NII}, {\SII}, and {\FeIII} are sensitive to hard EUV radiation or shock heating.
For example, the strong {\NII}6584, {\SII}6717,6731 lines are often used in Baldwin-Phillips-Terlevich (BPT) diagram \citep{Baldwin1981} as indicators of hard EUV radiation from an AGN.
The AGN photoionization models \citep[e.g.,][]{Groves2004a, Groves2004b} actually predict {\NII} and {\SII} line intensities stronger than the stellar photoionization models.
The {\NII} and {\SII} lines are enhanced by the power-law radiation of an AGN because a partially ionized zone is formed at the edge of ionized gas. 
In addition, shock gas models \citep{Allen2008} also demonstrate that the {\NII} and {\SII} lines are boosted when the shock heating contributes to the line emission.
The {\NII} and {\SII} lines are enhanced by the shock heating because low-ionization ions such as N$^{+}$ and S$^{+}$ are abundant in the recombination zone behind a shock front \citep{Allen2008}. 
Thus, abundances of low-ionization ions can be overestimated by the hard EUV radiation or shock heating. 
Especially, because the ionization potentials of N$^{0}$ (14.5 eV) and Fe$^{+}$ (16.2 eV) are very close, the N/O  and Fe/O ratios can be overestimated simultaneously if the hard EUV radiation or shock heating contribute.
However, the N/O ratios of the two EMPGs ({\HSCEMPGd} and {\IzotovEMPG}) are as low as galaxy chemical evolution models at $\log$(N/O)$\sim$$-1.6$.
Only Fe/O ratios of the two EMPGs deviate from the chemical evolution models.
Thus, we rule out the possibility that the Fe/O ratios are overestimated by the hard EUV radiation or shock heating.

We also discuss another possibility that the {\FeIII}\,4658 emission line might be contaminated by a {\CIV}\,4659 emission line, which can lead to the overestimation of Fe/O ratio.
If the {\CIV}\,4659 line exists, the {\FeIII}\,4658 and {\CIV}\,4659 lines may be unresolved due to the low spectral resolutions of the FOCAS and MODS spectroscopy.
The {\CIV}\,4659 line is a C$^{3+}$ recombination line, which is characterized by stellar winds from Wolf-Rayet (WR) stars.
Typical galaxies with WR features (so-called ``WR galaxies'') show broad emission lines such as {\CIV}\,4659, {\HeII}\,4686, and {\CIV}\,5808 with FWHMs $\sim$ 3,000 {\kms} \citep[e.g.,][]{Lopez-Sanchez2010a}.
Stellar wind velocities are expected to decrease with decreasing metallicity \citep[e.g.,][]{Nugis2000, Crowther2006} because atmospheric opacity of metal-poor stars becomes smaller than metal-rich stars.
However, WR galaxies/stars at low metallicities are very rare and have not yet been studied very well. 
One of the most metal-poor galaxies, IZw\,18 ({\metal}$=$7.16, 0.030 $Z_{\odot}$) show prominent broad emission lines of {\CIV}\,4659, {\HeII}\,4686, and {\CIV}\,5808 with FWHMs $\sim$ 2,600--3,600 {\kms} \citep[$\Delta\lambda$$=$50--55{\AA},][]{Legrand1997}, which are as broad as typical metal-rich WR galaxies \citep[$\sim$ 3,000 {\kms},][]{Lopez-Sanchez2010a}.
The result may suggest that line widths of {\CIV}\,4659, {\HeII}\,4686, and {\CIV}\,5808 may be fairly large even when a galaxy has a few percent solar metallicity.
In contrast to IZw\,18, the two EMPGs discussed in this paper ({\HSCEMPGd} and {\IzotovEMPG} with $\sim$0.02 $Z_{\odot}$) show no detection of broad emission lines of {\CIV}\,4659, {\HeII}\,4686, and {\CIV}\,5808.
The detected {\FeIII}\,4658 line of {\HSCEMPGd} and {\IzotovEMPG} (FWHMs$\sim$4 {\AA}, i.e., $\sim$ 260 {\kms}) are much narrower than the broad lines of IZw\,18 (FWHMs $\sim$ 2,600--3,600 {\kms}).
Thus, the {\CIV}\,4659 line originated from stellar winds may not contaminate the {\FeIII}\,4658 line of {\HSCEMPGd} and {\IzotovEMPG} significantly.

We briefly discuss an extreme case of a very narrow {\CIV}\,4659 line ($\lesssim$ 260 {\kms}), although the case is unlikely because the line width of $\lesssim$ 260 {\kms} is one order of magnitude smaller than the the broad {\CIV}\,4659, {\HeII}\,4686, and {\CIV}\,5808 lines of IZw\,18 (FWHMs $\sim$ 2,600--3,600 {\kms}).
Even if the {\CIV}\,4659 line can be extremely narrow, non-detection of the {\CIV}\,5808 line suggests that the {\CIV}\,4659 line is very faint in {\HSCEMPGd} and {\IzotovEMPG}. 
Note that the {\CIV}\,5808 intensity is almost comparable to that of {\CIV}\,4659 \citep[][]{Lopez-Sanchez2010a}.
Thus, we conclude that the {\FeIII}\,4658 line is not contaminated by the {\CIV}\,4659 line significantly even in the extreme case. 
\\

As described above, we have found that the Fe/O ratios of the two EMPGs ({\HSCEMPGd} and {\IzotovEMPG}) deviate from the Galactic stars and the Fe/O chemical evolution models, and ruled out the possibility that the Fe/O ratios are overestimated.
Below, we discuss three scenarios i)--iii) that might be able to explain the Fe/O deviation of the two EMPGs.

{\underline{\bf i) Preferential Dust Depletion:}}
The first scenario is the preferential dust depletion of iron, suggested by \citet{Rodriguez2005} and \citet{Izotov2006}.
\citet{Rodriguez2005} and \citet{Izotov2006} explain that gas-phase Fe/O ratios decrease as a function of metallicity in the range of {\metal}$\lesssim$8.5 because iron elements are depleted into dust more effectively than oxygen.
The depletion becomes dominant in a higher metallicity range, where the dust production becomes more efficient.
For dust-free (i.e., metal-poor) galaxies, gas-phase Fe/O ratios are expected to become comparable to the observational results of Galactic stars and the Fe/O evolution model.
Although the dust depletion may explain the negative Fe/O slope, it does {\it not} explain the fact that the two EMPGs ({\HSCEMPGd} and {\IzotovEMPG}) show higher Fe/O ratios than the Galactic stars and models at fixed metallicity.
In addition, as we have seen in Section \ref{subsec:line} and Table \ref{table:prop}, most of our metal-poor galaxies show {\EBV}$\sim$0 (i.e., less dusty).
At least, we do not find an evidence that galaxies with a larger metallicity show larger color excesses (i.e., dustier).
This means that the Fe/O decrease of our sample is not attributed to the dust depletion.
Based on these facts, we rule out the first scenario.

{\underline{\bf ii) Metal Enrichment and Gas Dilution:}}
The second scenario is a combination of metal enrichment and gas dilution caused by inflow. 
In this scenario, we assume that EMPGs are formed from metal-enriched gas with the solar metallicity and solar Fe/O ratio.
The Fe/O evolution models suggest that the Fe/O ratio increases at {\metal}$\gtrsim$8.0 due to the contribution of type-Ia SNe $\sim$1 Gyr after the start of the star formation. 
Such mature galaxies also tend to have the solar metallicity (i.e, O/H) at the same time. 
If primordial gas (i.e., almost metal free) falls into the metal-enriched galaxies, the metallicity (i.e, O/H) decreases while the Fe/O ratio does not change.
At first glance, this scenario seems to explain the Fe/O deviation of the two EMPGs.
However, if the second scenario is true, both the Fe/O and N/O ratios should match solar abundances because the N/O ratio also reaches the solar N/O ratio, log(N/O)$_{\odot}$$=$$-0.86$, at the solar metallicity (Sections \ref{sec:intro} and \ref{subsubsec:NO}).
As we have seen in the panel (c) of Figure \ref{fig:abundances}, the two deviating EMPGs, {\HSCEMPGd} and {\IzotovEMPG} (encircled by a red or orange circle at {\metal}$\sim$7.0) have a strong 2$\sigma$ upper limit of $<$0.14 (N/O)$_{\odot}$ and a low value of 0.21 (N/O)$_{\odot}$, respectively.
These low N/O ratios suggest that the two deviating EMPGs are experiencing the primary nucleosynthesis, not the secondary nucleosynthesis expected to start $\sim$1 Gyr after the onset of the star formation.
This also means that the high Fe/O ratios are not attributed to the type-Ia SNe, which arise $\sim$1 Gyr after  the onset of the star formation.
This conclusion is also consistent with the fact that the two EMPGs are very young, $\lesssim50$ Myr (Paper I). 
We exclude the second scenario because the second scenario does not explain the observed Fe/O and N/O ratios simultaneously.

{\underline{\bf iii) Super Massive Star Beyond 300 $M_{\odot}$:}}
The third scenario is the contribution of super massive stars beyond 300 $M_{\odot}$.
Super massive stars beyond 300 $M_{\odot}$ eject much iron at the time of core-collapse SN explosion.
\citet{Ohkubo2006} have calculated yields from core-collapse SNe under the assumption of the progenitor stellar mass with 500--1000 $M_{\odot}$, obtaining $\sim$2--40 (Fe/O)$_{\odot}$.
In the super massive stars beyond 300 $M_{\odot}$, an iron core grows until the iron core occupies more than 20 percent of the stellar mass.
Although massive stars with 140--300 $M_{\odot}$ undergo thermonuclear explosions triggered by pair-creation instability \citep[PISNe,][]{Barkat1967}, super massive stars beyond 300 $M_{\odot}$ are too massive to trigger PISNe and thus continue the iron core growth.
The super massive stars beyond 300 $M_{\odot}$ eject a large amount of iron by a jet stream from the massive iron core during the SN explosion.
On the other hand, the core-collapse SNe of typical-mass stars (10--50 $M_{\odot}$) eject gas with an average of $\sim$0.4 (Fe/O)$_{\odot}$ \citep[][IMF integrated in the range of 10--50 $M_{\odot}$]{Tominaga2007b}, which is below the solar Fe/O ratio.
Yields of type-Ia SNe calculated by \citet{Iwamoto1999} show $\sim$40 (Fe/O)$_{\odot}$.
Of the three types of SNe, only the type-Ia SNe and the SNe of super massive stars ($>$300 $M_{\odot}$) contribute to the iron enrichment larger than the solar Fe/O ratio.
As we have discussed in the second scenario above, the low N/O ratios of the two EMPGs suggest that their high Fe/O ratios are not explained by type-Ia SNe. 
Ruling out the type-Ia SNe, we find that the remaining possibility is the contribution from the SNe of super massive stars beyond 300 $M_{\odot}$.
We also confirm that SNe of the super massive stars ($>$300 $M_{\odot}$) do not change N/O ratios in comparison with the core-collapse SNe of typical massive stars \citep{Iwamoto1999, Ohkubo2006}, strengthening the reliability of the super massive star ($>$300 $M_{\odot}$) scenario. \\

In addition to the two EMPGs ({\HSCEMPGd} and {\IzotovEMPG}) with $\sim$2\% solar metallicity, a metal-poor galaxy, SBS 0335$-$052, shows {\metal}$=$7.31$\pm$0.01 (i.e., 4.2\% solar metallicity) and $\log$(Fe/O)$=$$-$1.42$^{+0.06}_{-0.28}$, which is 59\% of the solar Fe/O ratio \citep{Izotov2006b}. 
In Figure \ref{fig:FeO}, SBS 0335$-$052 also deviates from the observational results of Galactic stars and the Fe/O evolution model, which supports the results of this paper.
Note that \citet{Izotov1999} suggest that the Fe/O value of SBS 0335$-$052 can be overestimated by the  {\CIV}\,4659 contamination on the {\FeIII}\,4658 line, based on the low resolution spectroscopy of Multiple Mirror Telescope (MMT) spectrophotometry ($R$ $\sim$ 700).
However, conducting the VLT/GIRAFFE spectroscopy with high spectral resolutions ($R$ $\sim$ 10,000), \citet{Izotov2006b} find that the {\FeIII}\,4658 line is very narrow ($\sim$ 1 {\AA}, i.e., $\sim$ 60 {\kms}), which is not explained by the {\CIV}\,4659 line originated from stellar winds. 
Non-detection of the {\CIV}\,5808 line also suggests that the {\CIV}\,4659 line is not strong enough to contaminate the {\FeIII}\,4658 line significantly.
Thus, SBS 0335$-$052 \citep{Izotov2006b} is another example of a metal-poor galaxy that significantly shows a higher Fe/O ratio than the chemical evolution models.
\citet{Izotov2006b} attribute the higher Fe/O ratio of SBS 0335$-$052 to low dust depletion of iron, which has been ruled out in this paper (i.e., the first scenario in this section).

In summary of this subsection, we have discussed the three scenarios that might be able to explain the high Fe/O ratios of the two EMPGs ({\HSCEMPGd} and {\IzotovEMPG}).
We suggest that the high Fe/O ratios of the two EMPGs are attributed to the contribution from core-collapse SNe of super massive stars beyond 300 $M_{\odot}$.
The contribution of super massive stars beyond 300 $M_{\odot}$ to the iron enhancement has never been discussed by previous studies including \citet{Izotov2006} and \citet{Izotov2018a}.
Many previous studies \citep[e.g.,][]{Fragos2013b, Fragos2013a, Stanway2016, Suzuki2018, Xiao2018} assume the IMF maximum stellar mass ($M_\mathrm{max}$) at $M_\mathrm{max}$$=$100, 120, or 300 $M_\mathrm{\odot}$, ignoring super massive stars beyond 300 $M_{\odot}$, so this paper sheds light on the super massive stars beyond 300 $M_{\odot}$ in metal-poor galaxies undergoing the early-phase galaxy formation.

\subsection{Ionizing Radiation} \label{subsec:radiation}

\subsubsection{Emission Line Ratios} \label{subsubsec:line_ratio}
\begin{figure}[htbp]
\epsscale{1.15}
\plotone{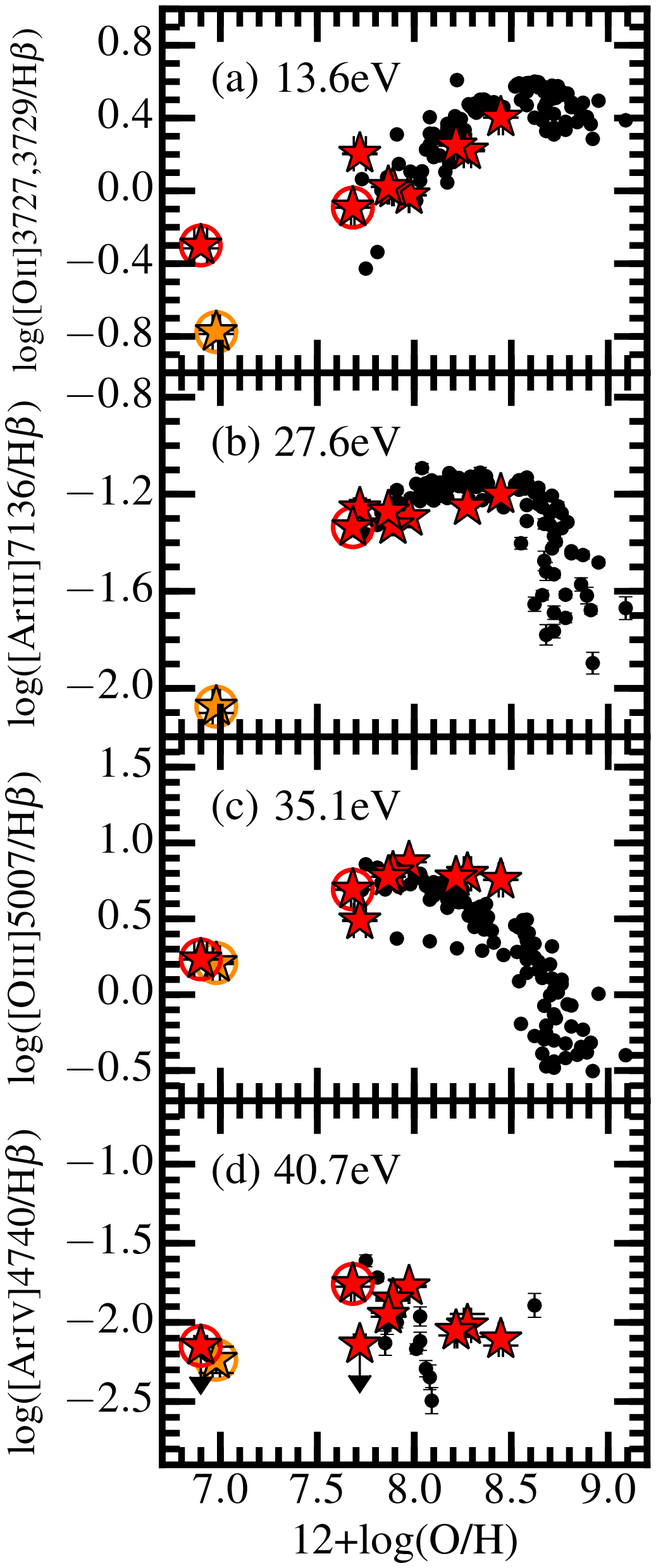}
\caption{Dust-corrected emission line ratios of {\OII}3727,3729, {\ArIII}4740, {\OIII}5007, and {\ArIV}7136 to {\Hb} in panels (a)--(d). 
Symbols are the same as in Figure \ref{fig:abundances}.
Here we do not show {\HSCEMPGb} as well, whose $T_\mathrm{e}$-metallicity and element abundances are not estimated.
The ionization potentials of O$^{0}$, Ar$^{2+}$, O$^{+}$, and Ar$^{3+}$ ions (13.6, 27.6, 35.1, and 40.7 eV, respectively) are presented in panels (a)--(d).
Black circles represent averages of local SFGs obtained with SDSS composite spectra \citep{Andrews2013}.
We do not show galaxies whose nebular emission line is strongly affected by the sky emission line.
\label{fig:lineratio_metal}}
\end{figure}

We investigate ionizing radiation of our metal-poor galaxy sample by comparing emission line ratios of various ions.
Figure \ref{fig:lineratio_metal} shows four emission line ratios of {\OII}3727,3729/{\Hb}, {\ArIII}4740/{\Hb}, {\OIII}5007/{\Hb}, and {\ArIV}7136/{\Hb} as a function of metallicity.
Among many emission lines detected in our spectroscopy, we choose the {\OII}3727,3729, {\ArIII}4740, {\OIII}5007, and {\ArIV}7136 emission lines for two reasons below.
The first reason is that oxygen and argon are both $\alpha$ elements, and thus the Ar/O abundance ratio is almost constant as we confirm in Section \ref{subsec:abund}.
Thus, emission line ratios are simply interpreted by ionizing radiation intensity and/or hardness, free from  variance of element abundance ratio.
The second reason is that the four lines are sensitive to ionization photons in a wide energy range from 13.6 to 40.7 eV.
The {\OII}3727,3729, {\ArIII}4740, {\OIII}5007, and {\ArIV}7136 lines are emitted via spontaneous emission after collisional excitation of O$^{+}$, Ar$^{2+}$, O$^{2+}$, and Ar$^{3+}$, respectively.
Table \ref{tbl:ion_potential} summarizes these emission line processes and corresponding photon energy required to emit these lines.

\begin{deluxetable}{cccc}
\tablecolumns{4}
\tabletypesize{\scriptsize}
\tablecaption{Summary of Emission Line Process, Ionization Process and Ionization Potential \label{tbl:ion_potential}}
\tablehead{%
\colhead{Line} &
\colhead{Emission} &
\colhead{Ionization} &
\colhead{Ionization} \\ 
\colhead{} &
\colhead{Process} &
\colhead{Process} &
\colhead{Potential} \\ 
\colhead{} &
\colhead{} &
\colhead{} &
\colhead{(eV)}
}
\startdata
{\Hb} & 
Re  &
 H$^{0}$+$\gamma$ $\rightarrow$ H$^{+}$ &
13.6 \\{\OII}3727 & 
CE &
O$^{0}$+$\gamma$ $\rightarrow$ O$^{+}$ &
13.6  \\
{\ArIII}4740 & 
CE  &
Ar$^{2+}$+$\gamma$ $\rightarrow$ Ar$^{3+}$  &
27.6 \\
{\OIII}5007 & 
CE  &
O$^{+}$+$\gamma$ $\rightarrow$ O$^{2+}$ &
35.1 \\
{\ArIV}7136 & 
CE  &
Ar$^{3+}$+$\gamma$ $\rightarrow$ Ar$^{4+}$  &
40.7 \\ 
{\HeII}4686 & 
Re  &
He$^{+}$+$\gamma$ $\rightarrow$ He$^{2+}$  &
54.4
\enddata
\tablecomments{
In the column of emission processes, Re and CE represent the recombination and collisional excitation.
}
\end{deluxetable}

In panels of (a)--(d) of Figure \ref{fig:lineratio_metal}, we show local, average SFGs of \citet[][AM13 hereafter]{Andrews2013} with black circles.
We regard the AM13 SFGs as local averages because the AM13 sample is obtained by the SDSS composite spectra in bins of wide SFR and stellar-mass ranges.
In panels of (a)--(d), the AM13 SFGs form sequences as a function of metallicity.
The sequences of {\OII}3727,3729/{\Hb} and {\ArIII}4740/{\Hb} show peaks at around {\metal}$\sim$8.7 and 8.3, respectively.
The {\OIII}5007/{\Hb} and {\ArIV}7136/{\Hb} ratios may also have peaks around {\metal} $\sim$8.0 and {\metal}$\sim$7.2--7.7 by interpolating AM13 SFGs and our metal-poor galaxies.
Recalling that the {\OII}3727,3729, {\OIII}5007, {\ArIII}4740, and {\ArIV}7136 lines are sensitive to ionizing photon above 13.6, 35.1, 27.6, and 40.7 eV, respectively, we find that the peak metallicities decrease with increasing ionizing potentials of the corresponding emission lines.
The peak transition demonstrates that ISM is irradiated by more intense or harder ionizing radiation in lower metallicity, as suggested by previous studies \cite[e.g.,][]{Nakajima2014, Steidel2016, Nakajima2016, Kojima2017}.
We also find that our metal-poor galaxies fall on the sequences of AM13 SFGs within a scatter.
Thus, we infer that our metal-poor galaxies and the AM13 SFGs have a similar spectral shape in the energy range of 13.6--40.7 eV for a given metallicity.

\begin{figure}[htbp]
\epsscale{1.2}
\plotone{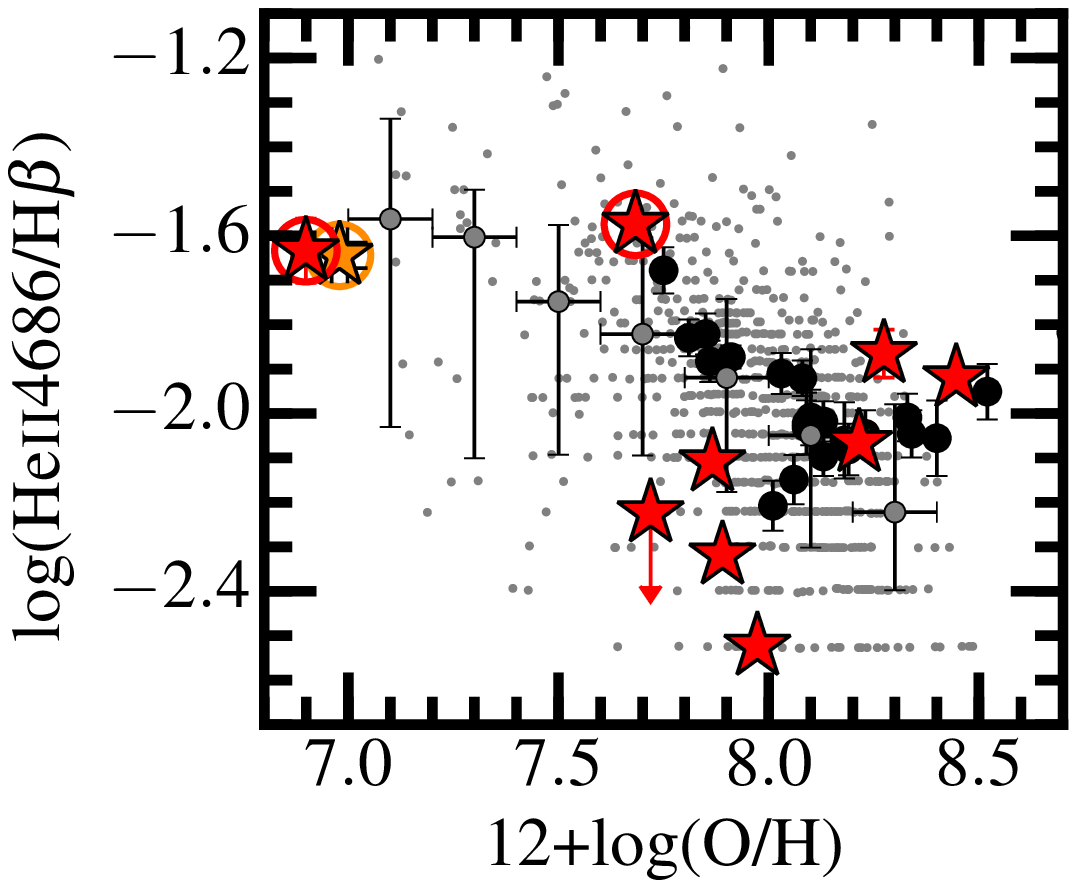}
\caption{Emission line ratios of {\HeII}4686/{\Hb} as functions of metallicity. 
Symbols are the same as in Figure \ref{fig:lineratio_metal}.
Here we do not show {\HSCEMPGb}, whose $T_\mathrm{e}$-metallicity and element abundances are not estimated.
Gray dots are individual local galaxies of S19 with an {\HeII}4686 detection.
Gray circles and error bars show medians and 68\%-percentile scatters of the S19 sample obtained in each metallicity bin, respectively. 
\label{fig:HeIIHb_metal}}
\end{figure}

Figure \ref{fig:HeIIHb_metal} shows {\HeII}4686/{\Hb} ratios of our metal-poor galaxies as a function of metallicity, as well as the AM13 SFGs (black circles).
The AM13 SFGs show almost constant {\HeII}4686/{\Hb} ratios around log({\HeII}4686/{\Hb})$\sim$$-2.0$ in the range of {\metal}$=$8.1--8.6, while the {\HeII}4686/{\Hb} ratios increases with decreasing metallicity below {\metal}$=$8.1.
Our metal-poor galaxies show a wide range of {\HeII}4686/{\Hb} ratios between log({\HeII}4686/{\Hb})$\sim$$-1.6$ and $-2.6$.
The distribution of our metal-poor galaxies are similar to those of SFGs of \citet[][S19 hereafter]{Schaerer2019}.
Among our metal-poor galaxies, three EMPGs (encircled by red and orange circles) show highest {\HeII}4686/{\Hb} ratios around log({\HeII}4686/{\Hb})$\sim$$-1.6$, including the two representative EMPGs, {\HSCEMPGd} and {\IzotovEMPG} \citep{Izotov2018a}.

As described in Section \ref{sec:intro}, previous studies also find large {\HeII}4686/{\Hb} ratios of $\log$({\HeII}4686/H$\beta$)$>$$-1.8$.
In a study of WR galaxies, \citet{Lopez-Sanchez2010a} find 4 galaxies that show large {\HeII}4686/{\Hb} ratios of $\log$({\HeII}4686/H$\beta$)$=$$-1.7$ to $-1.5$ with no WR features in the metallicity range of {\metal}$\sim$7.6--8.1.
\citet{Shirazi2012} construct an SDSS galaxy sample with a {\HeII}4686 detection.
\citet{Shirazi2012} find 68 galaxies that have $\log$({\HeII}4686/H$\beta$)$=$$-2.4$ to $-1.8$ with no WR features in metallicity range of {\metal}$\sim$7.7--8.2.
\citet{Senchyna2017} conduct HST/COS spectroscopy for 5 SDSS galaxies with a nebular {\HeII}4686 detection and no WR features.
The 5 galaxies show $\log$({\HeII}4686/H$\beta$)$=$$-2.1$ to $-1.4$ in metallicity range of {\metal}$\sim$7.8--8.0.
Our EMPGs, {\HSCEMPGd} and {\IzotovEMPG} have {\metal}$\sim$6.90 and 6.98, respectively, which are much lower than those of these samples.
The {\HeII}4686/{\Hb} ratios of {\HSCEMPGd} and {\IzotovEMPG} are $\log$({\HeII}4686/H$\beta$)$\sim$$-1.6$, which is comparable to those of the three samples of \citet{Lopez-Sanchez2010a}, \citet{Shirazi2012}, and \citet{Senchyna2017}.

\subsubsection{Strong {\HeII}4686 Line} \label{subsubsec:strong_Heii}
As described in Section \ref{sec:intro}, the physical mechanism of the {\HeII}4686 emission from SFGs is still under debate.
One of possible sources of He$^{+}$ ionizing photons is the very hot star produced via binary evolution.
\citet{Xiao2018} have created nebular emission models with the combination of the photoionization code {\sc cloudy} \citep[][]{Ferland2013} and the {\sc bpass} code \citep[][]{Stanway2016, Eldridge2017}.
The {\sc bpass} code calculates the stellar binary evolution, including atmosphere stripping, stellar rotation, and stellar mergers.
The binary stellar evolution models of \citet{Xiao2018} predict low values of {\HeII}4686/{\Hb}$\lesssim$1/1000, which are well below the observed {\HeII}4686/{\Hb} ratios of our galaxies and S19 galaxies ({\HeII}4686/{\Hb}$\sim$1/300--1/30).
This result suggests that the main contributors of {\HeII}4686 are not hot stars produced via binary evolution themselves.

Another possible explanation is a high mass X-ray binary (HMXB), where X-ray is emitted from a binary system of a compact object and a star through gas accretion.
S19 aim to explain the {\HeII}4686 emission from local SFGs with HMXB models of \citet{Fragos2013b, Fragos2013a}.
HMXBs are binary systems consisting of a compact object (such as BH) and a companion star.
The companion star provides gas onto the compact object, and creates a hot accretion disk around the compact object.
The hot accretion disk radiates very hard, power-low radiation ranging from UV to X-ray.
\citet{Fragos2013b, Fragos2013a} carefully calculate the HMXB evolution along the star-formation history, and predict total X-ray luminosities ($L_\mathrm{X}$) from a galaxy as functions of metallicity and age.
S19 convert an $L_\mathrm{X}$/SFR ratio to the {\HeII}4686/{\Hb} ratio, under the simple assumptions of {\HeII}4686/{\Hb}=1.74$\times$$Q$(He$^+$)/$Q$(H) \citep[Case B recombination of 20,000K,][]{Stasinska2015}, $Q$(H)/SFR$=$9.26$\times$10$^{52}$ photon\,s$^{-1}$/(M$_{\odot}$\,yr$^{-1}$) \citep[][]{Kennicutt1998}, and hardness of $Q$(He$^+$)/$L_\mathrm{X}$=2$\times$10$^{10}$ photon\,erg$^{-1}$.
Here, the $Q$(He$^+$) and $Q$(H) are defined by ionizing photon production rates above 54.4 and 13.6 eV, respectively.
S19 also use the {\sc bpass} binary stellar synthesis models of \citet{Xiao2018} to associate stellar ages with {\EW}({\Hb}).

\begin{figure*}[htbp]
\epsscale{0.9}
\plotone{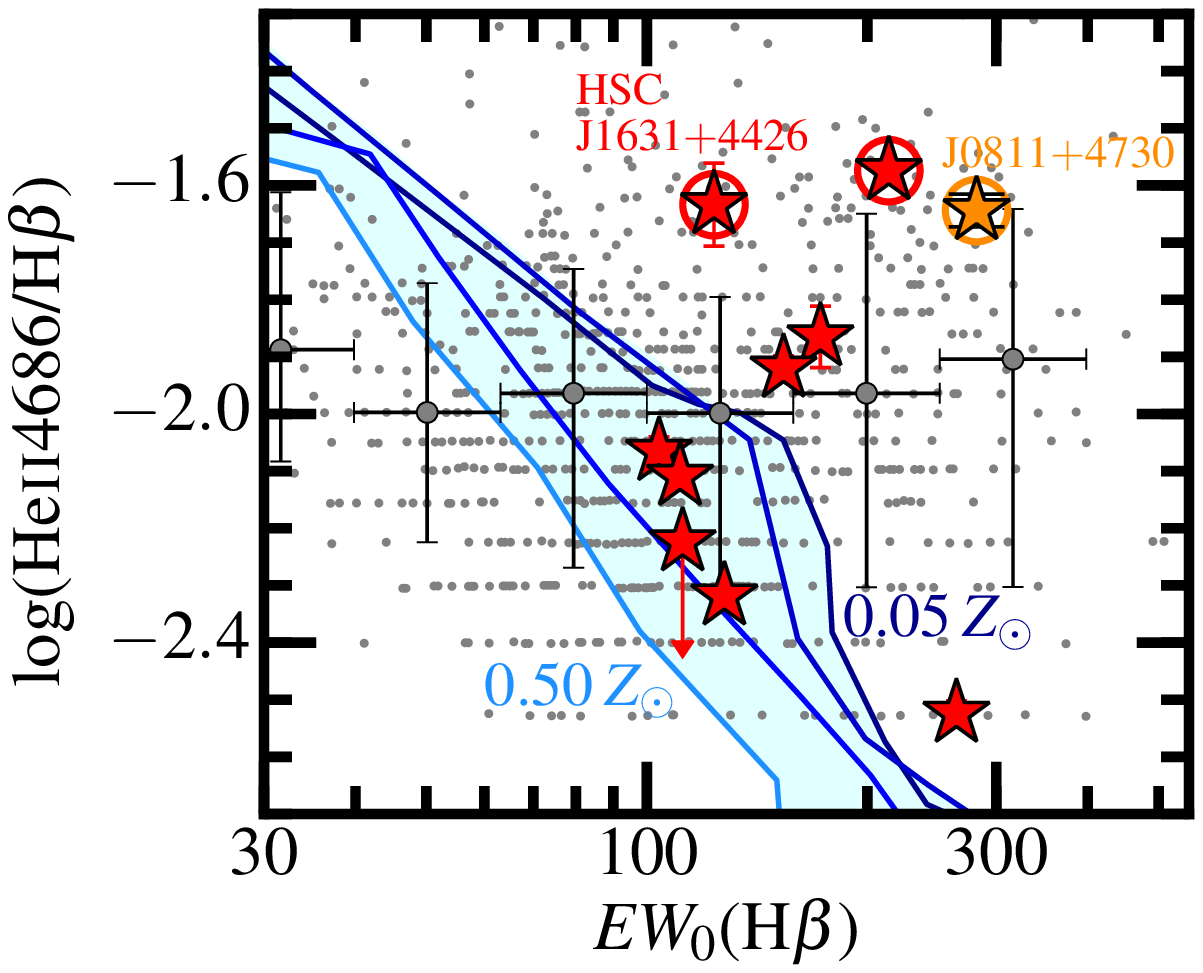}
\caption{Same as Figure \ref{fig:HeIIHb_metal}, but as a function of {\Hb} equivalent width, {\EW}({\Hb}). 
Symbols are the same as in Figures \ref{fig:lineratio_metal} and \ref{fig:HeIIHb_metal}.
Here we do not show {\HSCEMPGb}, whose $T_\mathrm{e}$-metallicity and element abundances are not estimated.
Solid lines represent the S19 HMXB models, tracing time evolution of {\HeII}4686/{\Hb} and {\EW}({\Hb}) with different metallicities of 0.05, 0.10, 0.20, and 0.50 $Z_{\odot}$ (from dark blue to light blue).
\label{fig:HeIIHb_EWHb}}
\end{figure*}

Figure \ref{fig:HeIIHb_EWHb} compares {\HeII}4686/{\Hb} ratios of our metal-poor galaxies and those obtained by the S19 HMXB models (solid lines) as a function of {\EW}({\Hb}).
The solid lines trace time evolution of {\HeII}4686/{\Hb} and {\EW}({\Hb}) with different metallicities of 0.05, 0.10, 0.20, and 0.50 $Z_{\odot}$.
The {\EW}({\Hb}) decreases and the {\HeII}4686/{\Hb} ratio increases as time passes due to the stellar evolution and HMXB evolution, respectively.
The HMXB models (especially 0.05 and 0.10$Z_{\odot}$) show a rapid increase of {\HeII}4686/{\Hb} around {\EW}({\Hb})$\sim$100--300 {\AA}.
The {\EW}({\Hb})$\sim$100--300 {\AA} corresponds to $\sim$5 Myr in the {\sc bpass} binary stellar synthesis models.
The rapid increase is triggered by the first compact object formation (i.e., the first HMXB formation) after $\sim$5 Myr of the starburst.
As shown in Figure \ref{fig:HeIIHb_EWHb}, the HMXB models of S19 have quantitatively explained the {\HeII}4686/{\Hb} ratios of half of our metal-poor galaxies.
However, we find that the other five metal-poor galaxies are not explained by the HMXB model, which fall in the ranges of {\EW}({\Hb})$>$100 {\AA} and $\log$({\HeII}4686/{\Hb})$>$$-2.0$.
Interestingly, three out of the five metal-poor galaxies are EMPGs (i.e., $Z<0.1 Z_{\odot}$), which are marked with red and orange circles in Figure \ref{fig:HeIIHb_EWHb}.
Especially, {\HSCEMPGd} ($0.016 Z_{\odot}$) and {\IzotovEMPG} ($0.019 Z_{\odot}$) are the representative EMPGs with the two lowest metallicities reported to date, showing high {\HeII}4686/{\Hb} ratios of $\log$({\HeII}4686/{\Hb})$\sim$$-1.6$.
Furthermore, the S19 SFG sample also include galaxies in the same ranges of {\EW}({\Hb})$>$100 {\AA} and $\log$({\HeII}4686/{\Hb})$>$$-2.0$.
S19 have argued that other X-ray sources are likely to appear fairly soon after the onset of the star formation ($\lesssim$5 Myr) in galaxies with high values of {\EW}({\Hb}) and {\HeII}4686/{\Hb}.
Although WR stars might contribute to the strong {\HeII}4686 emission, we do not find broad {\HeII}4686 emission lines typical of the WR stars \citep{Brinchmann2008b, Lopez-Sanchez2009} in our spectra.
Instead, S19 suggest that an underlying older population or shocks could also contribute to the high {\HeII}4686/{\Hb} ratios.
S19 do not discuss the metallicity in the explanation of the large {\HeII}4686/{\Hb} ratios at {\EW}({\Hb})$>$100 {\AA}.
However, the two EMPGs ({\HSCEMPGd} and {\IzotovEMPG}) in this paper emphasize that both the {\EW}({\Hb}) and metallicity may be keys to explain the {\HeII}4686/{\Hb} ratios higher than the HMXB models.

In addition to the S19 suggestions, we propose two other possibilities, for the first time, which can explain the high {\HeII}4686/{\Hb} ratios seen in the range of {\EW}({\Hb})$>$100 {\AA}.
First, we propose a possibility of super massive stars beyond 300 $M_{\odot}$.
The HMXB models of \citet{Fragos2013b, Fragos2013a} assume the Kroupa IMF \citep{Kroupa2001, Kroupa2003} with the maximum stellar mass of $M_\mathrm{max}$$=$120 $M_{\odot}$.
Thus, in the HMXB models of \citep{Fragos2013b, Fragos2013a}, the first HMXBs emerge $\sim$ 5 Myr after the start of the star formation, which corresponds to a lifetime of a star with 120 $M_{\odot}$.
On the other hand, stars more massive than 120 $M_{\odot}$ are expected to have a shorter life time than stars with 120 $M_{\odot}$.
According to the theoretical study of \citet{Yungelson2008}, super massive stars with 300 $M_{\odot}$ and 1000 $M_{\odot}$ die after 2.5 and 2.0 Myr after the onset of the star formation, respectively.
As described in Section \ref{subsubsec:FeO}, stars between 140 and 300 $M_{\odot}$ undergo thermonuclear explosions triggered by PISNe \citep{Barkat1967}, and do not leave any compact object  \citep[e.g.,][]{Heger2002}.
On the other hand, stars beyond 300 $M_{\odot}$ experience core-collapse SNe and form IMBHs \citep[e.g.,][]{Ohkubo2006}.
\citet{Ohkubo2006} estimate that BH masses become $\sim$230 and $\sim$500 $M_{\odot}$ for stars with initial masses of 500 and 1000 $M_{\odot}$, respectively.
Thus, when we assume super massive stars beyond 300 $M_{\odot}$, IMBHs appear as early as $\sim$2 Myr, and part of the IMBHs may form HMXBs.
Accretion disks of IMBHs emit very hard radiation including ionizing photons above 54.4 eV, which boosts the {\HeII}4686 intensity.
A galaxy as young as $\sim$2 Myr has {\EW}({\Hb})$\sim$300--400{\AA} according to the {\sc bpass} models.
Under the assumption of super massive stars beyond 300 $M_{\odot}$, the {\HeII}4686/{\Hb} ratio is expected to start increasing at around {\EW}({\Hb})$\sim$300--400{\AA}.
Such a model may cover the regions of {\EW}({\Hb})$>$100 {\AA} and $\log$({\HeII}4686/{\Hb})$>$$-2.0$ shown in Figure \ref{fig:HeIIHb_EWHb}.
Thus, we suggest that super massive stars beyond 300 $M_{\odot}$ would be able to explain the high ratios, $\log$({\HeII}4686/{\Hb})$>$$-2.0$ in the galaxies with {\EW}({\Hb})$>$100 {\AA}.
Note again that galaxies with $\log$({\HeII}4686/{\Hb})$>$$-2.0$ and {\EW}({\Hb})$>$100 {\AA} includes EMPGs, {\HSCEMPGd} {\SDSSEMPGc}, and {\IzotovEMPG}.
These EMPGs might form super massive star beyond 300 $M_{\odot}$ from their extremely metal-poor gas.
However, our explanation and the interpretation of S19 are based on some simple assumptions that associate the HMXB models and the {\sc bpass} stellar synthesis models. 
We propose to construct self-consistent SED models ranging from X-ray to UV with the HMXB evolution models under the assumption of $M_\mathrm{max}$$>$300 $M_{\odot}$.

Second, we also suggest a possibility of a metal-poor AGN, which can contribute to boost the {\HeII}4686 intensity of the very young galaxies.
In Paper I, we have confirmed that all of our metal-poor galaxies fall on the SFG region of the BPT diagram defined by the maximum photoionization models with stellar radiation \citep{Kewley2001}.
However, \citet{Kewley2013a} suggest that emission-line ratios calculated under the assumption of a metal-poor AGN also fall on the SFG region.
Thus, we cannot exclude the possibility of a metal-poor AGN.
\citet{Groves2004a,Groves2004b} have constructed the photo-ionization models under the assumption of AGN-like, power-row radiation.
The models of \citet{Groves2004a,Groves2004b} predict very strong {\HeII}4686 emission represented by $\log$({\HeII}4686/{\Hb})$\sim$$-1.5$ to 0.0.
On the other hand, photo-ionization models with stellar radiation \citep{Xiao2018} predict $\log$({\HeII}4686/{\Hb})$\lesssim$$-2.5$.
To explain the observed ratios of $\log$({\HeII}4686/{\Hb})$\sim$$-2.0$, the combination of AGN and stellar radiation is required.
We have checked the archival data of ROSAT and XMM, and found no detection in X-ray.
This is because the data are $\sim$2 orders of magnitudes shallower than expected X-ray luminosities ($\sim$10$^{-14}$ erg\,s$^{-1}$\,cm$^{-2}$) of our metal-poor galaxy sample, which are obtained under the assumption of $L_\mathrm{2keV}$--$M_\mathrm{UV}$ relation of AGN \citep{Lusso2010}.
Deep X-ray observations are required to constrain X-ray sources of metal-poor galaxies.

\subsection{Formation Mechanism of Super Massive Stars beyond 300 $M_{\odot}$} \label{subsec:vms}
Below, we focus only on the two representative EMPGs, {\HSCEMPGd} (0.016 $Z_{\odot}$) and {\IzotovEMPG} (0.019 $Z_{\odot}$) and discuss their high Fe/O ratios and {\HeII}4686/{\Hb} ratios.
The two representative EMPGs, {\HSCEMPGd} and {\IzotovEMPG} show the two lowest metallicities reported to date.
In Section \ref{subsubsec:FeO}, we have found that the two EMPGs, {\HSCEMPGd} and {\IzotovEMPG} show Fe/O ratios $\sim$1.0 dex higher than Galactic stars and the Fe/O evolution models at fixed metallicity.
We have concluded that the high Fe/O ratios are explained by core-collapse SNe of super massive stars beyond 300 $M_{\odot}$.
In Section \ref{subsubsec:strong_Heii}, we have also found that the two EMPGs, {\HSCEMPGd} and {\IzotovEMPG}, show both high {\HeII}4686/{\Hb} ratios ($\sim$1/40) and high {\EW}({\Hb}) ($\sim$100--300 {\AA}).
We have suggested that IMBH formed from super massive stars beyond 300 $M_{\odot}$ can explain the high {\HeII}4686/{\Hb} ratios.
Interestingly, the scenario of super massive stars beyond 300 $M_{\odot}$ explains both the high Fe/O ratios and the high {\HeII}4686/{\Hb} ratios of our EMPGs with the low metallicities ($\sim$0.02 $Z_{\odot}$), young ages ($\lesssim$50 Myr), and very low stellar mass ($\sim$10$^{5}$--10$^{6}$ $M_{\odot}$), which are undergoing early phases of the galaxy formation.
We propose, for the first time, the connection between the large {\HeII}4686/{\Hb} ratios (Section \ref{subsubsec:strong_Heii}) and the solar Fe/O ratios (Section \ref{subsubsec:FeO}).

The idea of super massive stars beyond 300 $M_{\odot}$ is not necessarily extraordinary.
\citet{Crowther2010, Crowther2016} have claimed the spectroscopic identification of super massive stars with $\sim$320 $M_{\odot}$ in the R136 star cluster of LMC.
The identification of an IMBH with $M_\mathrm{BH}>700 M_{\odot}$ in a star cluster of M82 \citep{Matsumoto2001,Ebisuzaki2001,Kaaret2001} is another indirect trace of a super massive star beyond 300 $M_{\odot}$.
This is because the SN numerical simulation \citep{Ohkubo2006} suggests that a star with initial masses beyond 300 $M_{\odot}$ leaves an IMBH with $\gtrsim$100 $M_{\odot}$.

We may wonder how such super massive stars are formed.
Below, we discuss the formation mechanisms of super massive stars beyond 300 $M_{\odot}$.
Theoretical studies \citep[e.g.,][]{Bromm2004,Omukai2003} suggest that metal-free (pop-III) stars are typically super massive ($>$300 $M_{\odot}$) because gas cooling becomes insufficient and the fragmentation mass becomes large in metal-free gas.
The critical metallicity ($Z_\mathrm{crit}$), below which super massive stars can be formed directly, is $Z_\mathrm{crit} \sim 10^{-6}$ to $10^{-4}$$Z_{\odot}$, theoretically \citep[e.g.,][]{Bromm2001b,Santoro2006,Schneider2006,Smith2007}.
The critical metallicity ($Z_\mathrm{crit} \sim 10^{-6}$--$10^{-4}$$Z_{\odot}$) is much lower than our metallicity measurements of our EMPGs, $\sim$0.02 $Z_{\odot}$.
Thus, the direct gas collapse is unlikely to be the formation mechanism of the super massive stars beyond 300 $M_{\odot}$ in our EMPGs.

We briefly discuss another possibility that EMPGs previously had a metallicity near the critical metallicity ($\sim 10^{-6}$--$10^{-4}$$Z_{\odot}$) at the time of the galaxy formation, and the metallicity has been increased by SNe within a short period.
The metallicity at the time of the galaxy formation namely depends on the metallicity of the inflow gas, which may come from a void region free from the metal contamination.
Based on the observational results of metal-poor galaxies, \citet[][]{Thuan2005b} speculate that the metallicity lower limit  (so-called ``metallicity floor'') might exist at $\sim10^{-2}$$Z_{\odot}$ at $z=0$ because the IGM has slightly been metal-enriched by the past star formation activities even in the void regions.
In addition, observational studies of Lyman alpha absorption systems \citep[e.g.,][]{Prochaska2003, Rafelski2012, Lehner2013, Quiret2016} have not yet discovered Lyman alpha absorption system below $\sim10^{-2}$$Z_{\odot}$ at $z=0$--$1$.
Hydrodynamical simulation of \citet{Martizzi2019} also demonstrates that the average IGM metallicity in the void region is $\sim10^{-2}$$Z_{\odot}$ at $z=0$.
We do not necessarily remove the possibility that EMPGs had a metallicity near the critical metallicity ($\sim 10^{-6}$--$10^{-4}$$Z_{\odot}$) at the time of the galaxy formation because part of the IGM would reach below $\sim10^{-2}$$Z_{\odot}$ \citep{Hafen2017} due to a metallicity fluctuation. 
However, the direct gas collapse may not be, at least, the main formation mechanism of the super massive stars beyond 300 $M_{\odot}$.

\citet{Ebisuzaki2001} and \citet{PortegiesZwart1999, PortegiesZwart2004, PortegiesZwart2006} has presented another formation mechanism of super massive stars beyond 300 $M_{\odot}$, where super massive stars are formed by stellar mergers under the very dense condition in a star cluster.
Numerical simulations suggest that super massive stars of 800--3000 $M_{\odot}$ have been formed by stellar mergers within $\sim$3 Myr \citep{PortegiesZwart2004}.
The super massive star formation mechanism of \citet{PortegiesZwart2004} is presented to explain an IMBH discovered in a star cluster of M82 \citep{Matsumoto2001,Ebisuzaki2001,Kaaret2001}.
The estimated BH mass of the M82 IMBH is $M_\mathrm{BH}>700 M_{\odot}$. 
This stellar merger mechanism requires the very dense star forming regions, which are typical in young, metal-poor galaxies.
Indeed, our sample EMPGs are undergoing the intensive star-formation represented by high sSFRs \citep[$\sim$300 Gyr$^{-1}$,][]{Kojima2020a} and small sizes ($\sim$ 100 pc, Paper III).
Although the stellar merger mechanism {\it itself} does not depend on metallicity, the formation mechanism of the very dense star forming regions may depend on metallicity.
A possible scenario is that a large amount of massive stars are formed within a compact region by a primordial gas (i.e., almost metal free) infall from the inter-galactic space (KS1-EMPG in Paper III).
This scenario may be associated with the top heavy IMF.

In this paper, we find that two representative EMPGs ($\sim$0.02 $Z_{\odot}$) show both the high Fe/O ratios and the high {\HeII}4686/{\Hb} ratios, which are not explained by the previous models assuming massive stars up to $100$ or $120$ $M_{\odot}$.
In the end of this section, we summarize one possible picture that we have suggested in each section of this paper.
In our picture, EMPGs are formed by a primordial gas infall, which also forms star clusters with a very high number density of massive stars (Paper III). 
In such very dense regions of star clusters, stellar mergers trigger the formation of super massive stars beyond 300 $M_{\odot}$ within $\sim$3 Myr \citep{PortegiesZwart2004}.
The super massive stars beyond 300 $M_{\odot}$ eventually eject much iron through the core-collapse SNe in $\sim$2 Myr \citep{Ohkubo2006, Yungelson2008} and form IMBHs \citep[$M_\mathrm{BH}\gtrsim 100 M_{\odot}$,][]{Ohkubo2006} boosting the {\HeII}4686 emission.
To testify this picture, spectroscopic observations are required with a high spacial resolution in multi-wavelength to identify the formation mechanism of EMPGs and to directly detect IMBH signatures in EMPGs.

\section{SUMMARY} \label{sec:summary}
We investigate element abundance ratios and ionizing radiation of 10 metal-poor galaxies at $z\lesssim0.03$, which have been discovered in the wide-field imaging data of Subaru/Hyper Sprime-Cam (HSC) and Sloan Digital Sky Survey (SDSS) by \citet{Kojima2020a}. 
The 10 metal-poor galaxies are represented by low metallicities, {\metal} =6.90--8.45, low stellar masses, {\mass}$=$4.95--7.06, and high specific star-formation rates (sSFR$\sim$300\,Gyr$^{-1}$).
These galaxies have very low masses of {\mass}$\lesssim$6, which are comparable to those of star clusters.
Such cluster-like galaxies are undergoing very early phases of the galaxy formation.
Especially, three out of the 10 galaxies are extremely metal-poor galaxies (EMPGs) defined by {\metal}$<$7.69, including {\HSCEMPGd} with the lowest metallicity (0.016 $Z_{\odot}$) reported to date. 
In addition to the 10 metal-poor galaxies, we include another EMPG from the literature \citep[{\IzotovEMPG},][]{Izotov2018a} in the sample of this paper.
{\IzotovEMPG} has the second lowest metallicity of 0.019 $Z_{\odot}$ reported to date.
The two EMPGs ({\HSCEMPGd} and {\IzotovEMPG}) are key in this paper.

\begin{itemize}

\item We estimate element abundance ratios of Ne/O, Ar/O, and N/O of our metal-poor galaxies, and compare them with local SFGs.
We find that $\alpha$-element ratios of Ne/O and Ar/O show almost constant values of $\log$(Ne/O) $\sim$$-0.8$ and $\log$(Ar/O)$\sim$$-2.5$ as a function of metallicity, respectively.
These constant Ne/O and Ar/O values are consistent with those of local SFGs.
Most of our metal-poor galaxies have N/O ratios of $\log$(N/O)$\lesssim$$-1.5$, suggesting that our metal-poor galaxies are undergoing the primary nucleosynthesis of nitrogen due to their low metallicity and young stellar population.

\item We also estimate Fe/O ratios of our metal-poor galaxies, and compare them with local SFGs.
Our metal-poor galaxy sample shows a decreasing Fe/O trend with increasing metallicity, which is consistent with the previous results of local SFGs.
We find that two EMPGs, {\HSCEMPGd} (0.016 $Z_{\odot}$) and {\IzotovEMPG} \citep[0.019 $Z_{\odot}$,][]{Izotov2018a}, show higher Fe/O ratios than observational results of Galactic stars and a model calculation of the Fe/O evolution at fixed metallicity.
Especially, {\HSCEMPGd} and {\IzotovEMPG} show the solar Fe/O ratios in spite of its very low metallicity, 0.016 and 0.019 $Z_{\odot}$, respectively. 
We discuss the three scenarios that might be able to explain the high Fe/O ratios with the extremely low metallicity: (1) the preferential dust depletion of iron, (2) a combination of metal enrichment and gas dilution caused by inflow, and (3) super massive stars beyond 300 $M_{\odot}$.
The scenario (1) is ruled out because the solar Fe/O ratios are not achieved by the dust depletion, and we do not see any correlation between the dust extinction and the Fe/O ratios.
We also exclude the scenario (2) because the observed N/O ratios are lower than the expected solar N/O ratio when the scenario (2) is true. 
Thus, we conclude that the high Fe/O ratios of the two EMPGs are attributed to super massive stars beyond 300 $M_{\odot}$, which is consistent with the young stellar ages of EMPGs ($\lesssim50$ Myr).

\item To probe ionizing radiation in our metal-poor galaxies, we inspect emission lines from various ions covering a wide range of ionization potentials.
We choose {\Hb}, {\OII}3727,3729, {\ArIII}4740, {\OIII}5007, and {\ArIV}7136 lines, which are sensitive to ionizing photon above 13.6, 13.6, 27.6, 35.1, and 40.7 eV, respectively.
Our metal-poor galaxies and local, average SFGs show sequences of {\OII}3727,3729/{\Hb}, {\ArIII}4740/{\Hb}, {\OIII}5007/{\Hb}, and {\ArIV}7136/{\Hb} as a function of metallicity, and match each other within small scatters.
The match between the two samples suggests that our metal-poor galaxies and local, average SFGs have a similar spectral shape in the energy range of 13.6--40.7 eV for a given metallicity.

\item We find that five metal-poor galaxies show both high {\HeII}4686/{\Hb} ratios ($>$1/100) and high {\EW}({\Hb}) ($>$100 {\AA}).
Interestingly, two out of the five metal-poor galaxies are the representative EMPGs, {\HSCEMPGd} (0.016 $Z_{\odot}$) and {\IzotovEMPG} (0.019 $Z_{\odot}$).
These high {\HeII}4686/{\Hb} ratios and high {\EW}({\Hb}) are not explained by the latest binary population stellar synthesis model and the latest HMXB model, where a maximum stellar mass cut, 120 $M_{\odot}$ is used.
We suggest that super massive stars beyond 300 $M_{\odot}$ can explain the high {\HeII}4686/{\Hb} ratios for galaxies of {\EW}({\Hb})$>$100 {\AA} (i.e., $\lesssim$5 Myr).
Super massive stars beyond 300 $M_{\odot}$ have very short lifetimes of $\sim$2 Myr, and form intermediate-mass black holes (IMBHs) of $\gtrsim$100 $M_{\odot}$ as early as $\sim$2 Myr after the onset of the star formation.
We do not rule out a possibility of a metal-poor AGN, which can contribute to the {\HeII}4686 boost of the very young galaxies even at $\lesssim$5 Myr.

\item Interestingly, the scenario of super massive stars beyond 300 $M_{\odot}$ explains both the high Fe/O ratios and the high {\HeII}4686/{\Hb} ratios of our EMPGs.
We also discuss a formation mechanism of super massive stars beyond 300 $M_{\odot}$.
The direct collapse of metal-poor gas is unlikely to be the formation mechanism because the critical metallicity ($Z_\mathrm{crit} \sim 10^{-6}$--$10^{-4}$$Z_{\odot}$), below which super massive stars can be formed directly, is much lower than the metallicities of the two representative EMPGs, $\sim$0.02 $Z_{\odot}$.
Instead, super massive stars beyond 300 $M_{\odot}$ would be formed by stellar mergers under the very dense condition in a star cluster.
In this picture, EMPGs are formed by a primordial gas infall, which also forms star clusters with a very high number density of massive stars. 
In such very dense regions, stellar mergers trigger the formation of super massive stars beyond 300 $M_{\odot}$ within $\sim$3 Myr.
The super massive stars beyond 300 $M_{\odot}$ eventually eject much iron through the core-collapse SNe in $\sim$2 Myr and form IMBHs ($M_\mathrm{BH}\gtrsim 100 M_{\odot}$) boosting the {\HeII}4686 emission.

\end{itemize}

\acknowledgments

\begin{acknowledgements}
We are grateful to  
Lennox Cowie,
Akio Inoue,
Taddy Kodama, 
Matthew Malkan, and
Daniel Stark
for their important and useful comments.
We thank John David Silverman and Anne Verhamme for their helpful comments on our survey name. 
We would like to express our special thanks to Daniel Kelson for his great efforts in helping us reduce and calibrate our MagE data.
We are also grateful to Yuri Izotov for his permission to show a spectrum of a representative EMPG ({\IzotovEMPG}) in this paper so that we can compare it with our galaxy spectra.

We also thank staffs of the Las Campanas observatories, the Subaru telescope, and the Keck observatories for helping us with our observations.
The observations were carried out within the framework of Subaru-Keck time exchange program, where the travel expense was supported by the Subaru Telescope, which is operated by the National Astronomical Observatory of Japan.
The authors wish to recognize and acknowledge the very significant cultural role and reverence that the summit of Maunakea has always had within the indigenous Hawaiian community.  
We are most fortunate to have the opportunity to conduct observations from this mountain.

The Hyper Suprime-Cam (HSC) collaboration includes the astronomical communities of Japan and Taiwan, and Princeton University.  The HSC instrumentation and software were developed by the National Astronomical Observatory of Japan (NAOJ), the Kavli Institute for the Physics and Mathematics of the Universe (Kavli IPMU), the University of Tokyo, the High Energy Accelerator Research Organization (KEK), the Academia Sinica Institute for Astronomy and Astrophysics in Taiwan (ASIAA), and Princeton University.  Funding was contributed by the FIRST program from the Japanese Cabinet Office, the Ministry of Education, Culture, Sports, Science and Technology (MEXT), the Japan Society for the Promotion of Science (JSPS), Japan Science and Technology Agency  (JST), the Toray Science  Foundation, NAOJ, Kavli IPMU, KEK, ASIAA, and Princeton University.

This paper makes use of software developed for the Large Synoptic Survey Telescope. We thank the LSST Project for making their code available as free software at \verb|http://dm.lsst.org|.

This paper is based on data collected at the Subaru Telescope and retrieved from the HSC data archive system, which is operated by Subaru Telescope and Astronomy Data Center (ADC) at NAOJ. Data analysis was in part carried out with the cooperation of Center for Computational Astrophysics (CfCA), NAOJ.

The Pan-STARRS1 Surveys (PS1) and the PS1 public science archive have been made possible through contributions by the Institute for Astronomy, the University of Hawaii, the Pan-STARRS Project Office, the Max Planck Society and its participating institutes, the Max Planck Institute for Astronomy, Heidelberg, and the Max Planck Institute for Extraterrestrial Physics, Garching, The Johns Hopkins University, Durham University, the University of Edinburgh, the Queen’s University Belfast, the Harvard-Smithsonian Center for Astrophysics, the Las Cumbres Observatory Global Telescope Network Incorporated, the National Central University of Taiwan, the Space Telescope Science Institute, the National Aeronautics and Space Administration under grant No. NNX08AR22G issued through the Planetary Science Division of the NASA Science Mission Directorate, the National Science Foundation grant No. AST-1238877, the University of Maryland, Eotvos Lorand University (ELTE), the Los Alamos National Laboratory, and the Gordon and Betty Moore Foundation.

This work is supported by World Premier International Research Center Initiative (WPI Initiative), MEXT, Japan, as well as KAKENHI Grant-in-Aid for Scientific Research (A) (15H02064, 17H01110, and 17H01114) through Japan Society for the Promotion of Science (JSPS).
T.K., K.Y., Y.S., and M. Onodera are supported by JSPS KAKENHI Grant Numbers, 18J12840, 18K13578, 18J12727, and 17K14257.
S.F. acknowledges support from the European Research Council (ERC) Consolidator Grant funding scheme (project ConTExt, grant No. 648179). The Cosmic Dawn Center is funded by the Danish National Research Foundation under grant No. 140.

\end{acknowledgements}

\bibliography{library}

\begin{thebibliography}{}
\expandafter\ifx\csname natexlab\endcsname\relax\def\natexlab#1{#1}\fi

\bibitem[{Aihara {et~al.}(2018)Aihara, Arimoto, Armstrong, Arnouts, Bahcall,
  Bickerton, Bosch, Bundy, Capak, Chan, Chiba, Coupon, Egami, Enoki, Finet,
  Fujimori, Fujimoto, Furusawa, Furusawa, Goto, Goulding, Greco, Greene, Gunn,
  Hamana, Harikane, Hashimoto, Hattori, Hayashi, Hayashi, He{\l}miniak,
  Higuchi, Hikage, Ho, Hsieh, Huang, Huang, Ikeda, Imanishi, Inoue, Iwasawa,
  Iwata, Jaelani, Jian, Kamata, Karoji, Kashikawa, Katayama, Kawanomoto, Kayo,
  Koda, Koike, Kojima, Komiyama, Konno, Koshida, Koyama, Kusakabe, Leauthaud,
  Lee, Lin, Lin, Lupton, Mandelbaum, Matsuoka, Medezinski, Mineo, Miyama,
  Miyatake, Miyazaki, Momose, More, More, Moritani, Moriya, Morokuma, Mukae,
  Murata, Murayama, Nagao, Nakata, Niida, Niikura, Nishizawa, Obuchi, Oguri,
  Oishi, Okabe, Okamoto, Okura, Ono, Onodera, Onoue, Osato, Ouchi, Price, Pyo,
  Sako, Sawicki, Shibuya, Shimasaku, Shimono, Shirasaki, Silverman, Simet,
  Speagle, Spergel, Strauss, Sugahara, Sugiyama, Suto, Suyu, Suzuki, Tait,
  Takada, Takata, Tamura, Tanaka, Tanaka, Tanaka, Tanaka, Terai, Terashima,
  Toba, Tominaga, Toshikawa, Turner, Uchida, Uchiyama, Umetsu, Uraguchi, Urata,
  Usuda, Utsumi, Wang, Wang, Wong, Yabe, Yamada, Yamanoi, Yasuda, Yeh,
  Yonehara, \& Yuma}]{Aihara2018a}
Aihara, H., Arimoto, N., Armstrong, R., {et~al.} 2018, PASJ, 70, 1

\bibitem[{Aihara {et~al.}(2019)Aihara, AlSayyad, Ando, Armstrong, Bosch, Egami,
  Furusawa, Furusawa, Goulding, Harikane, Hikage, Ho, Hsieh, Huang, Ikeda,
  Imanishi, Ito, Iwata, Jaelani, Kakuma, Kawana, Kikuta, Kobayashi, Koike,
  Komiyama, Li, Liang, Lin, Luo, Lupton, Lust, MacArthur, Matsuoka, Mineo,
  Miyatake, Miyazaki, More, Murata, Namiki, Nishizawa, Oguri, Okabe, Okamoto,
  Okura, Ono, Onodera, Onoue, Osato, Ouchi, Shibuya, Strauss, Sugiyama, Suto,
  Takada, Takagi, Takata, Takita, Tanaka, Terai, Toba, Uchiyama, Utsumi, Wang,
  Wang, \& Yamada}]{Aihara2019}
Aihara, H., AlSayyad, Y., Ando, M., {et~al.} 2019, PASJ, 71, 114

\bibitem[{Albareti {et~al.}(2017)Albareti, Prieto, Almeida, Anders, Anderson,
  Andrews, Arag{\'{o}}n-Salamanca, Argudo-Fern{\'{a}}ndez, Armengaud, Aubourg,
  Avila-Reese, Badenes, Bailey, Barbuy, Barger, Barrera-Ballesteros, Bartosz,
  Basu, Bates, Battaglia, Baumgarten, Baur, Bautista, Beers, Belfiore,
  Bershady, de~Lis, Bird, Bizyaev, Blanc, Blanton, Blomqvist, Bolton,
  Borissova, Bovy, Brandt, Brinkmann, Brownstein, Bundy, Burtin, Busca, Chavez,
  D{\'{i}}az, Cappellari, Carrera, Chen, Cherinka, Cheung, Chiappini,
  Chojnowski, Chuang, Chung, Cirolini, Clerc, Cohen, Comerford, Comparat,
  {Correa do Nascimento}, Cousinou, Covey, Crane, Croft, Cunha, Darling,
  Davidson, Dawson, {Da Costa}, {Da Silva Ilha}, Machado, Delubac, {De Lee},
  {De la Macorra}, {De la Torre}, Diamond-Stanic, Donor, Downes, Drory, Du, {Du
  Mas des Bourboux}, Dwelly, Ebelke, Eigenbrot, Eisenstein, Elsworth, Emsellem,
  Eracleous, Escoffier, Evans, Falc{\'{o}}n-Barroso, Fan, Favole,
  Fernandez-Alvar, Fernandez-Trincado, Feuillet, Fleming, Font-Ribera,
  Freischlad, Frinchaboy, Fu, Gao, Garcia, Garcia-Dias, Garcia-Hern{\'{a}}ndez,
  P{\'{e}}rez, Gaulme, Ge, Geisler, Gillespie, Marin, Girardi, Goddard, Chew,
  Gonzalez-Perez, Grabowski, Green, Grier, Grier, Guo, Guy, Hagen, Hall,
  Harding, Harley, Hasselquist, Hawley, Hayes, Hearty, Hekker, Toledo, Ho,
  Hogg, Holley-Bockelmann, Holtzman, Holzer, Hu, Huber, Hutchinson, Hwang,
  Ibarra-Medel, Ivans, Ivory, Jaehnig, Jensen, Johnson, Jones, Jullo,
  Kallinger, Kinemuchi, Kirkby, Klaene, Kneib, Kollmeier, Lacerna, Lane, Lang,
  Laurent, Law, Leauthaud, {Le Goff}, Li, Li, Li, Li, Liang, Liang, Lima, Lin,
  Lin, Lin, Liu, Long, Lucatello, MacDonald, MacLeod, Mackereth, Mahadevan,
  Maia, Maiolino, Majewski, Malanushenko, Malanushenko, Mallmann, Manchado,
  Maraston, Marques-Chaves, Valpuesta, Masters, Mathur, McGreer, Merloni,
  Merrifield, Mesz{\'{a}}ros, Meza, Miglio, Minchev, Molaverdikhani,
  Montero-Dorta, Mosser, Muna, Myers, Nair, Nandra, Ness, Newman, Nichol,
  Nidever, Nitschelm, O'Connell, Oravetz, Oravetz, Pace, Padilla,
  Palanque-Delabrouille, Pan, Parejko, Paris, Park, Peacock, Peirani,
  Pellejero-Ibanez, Penny, Percival, Percival, Perez-Fournon, Petitjean, Pieri,
  Pinsonneault, Pisani, Prada, Prakash, Price-Jones, Raddick, Rahman, Raichoor,
  Rembold, Reyna, Rich, Richstein, Ridl, Riffel, Riffel, Rix, Robin, Rockosi,
  Rodr{\'{i}}guez-Torres, Rodrigues, Roe, Lopes,
  Rom{\'{a}}n-Z{\'{u}}{\~{n}}iga, Ross, Rossi, Ruan, Ruggeri, Runnoe,
  Salazar-Albornoz, Salvato, Sanchez, Sanchez, Sanchez-Gallego, Santiago,
  Schiavon, Schimoia, Schlafly, Schlegel, Schneider, Sch{\"{o}}nrich,
  Schultheis, Schwope, Seo, Serenelli, Sesar, Shao, Shetrone, Shull, Aguirre,
  Skrutskie, Slosar, Smith, Smith, Sobeck, Somers, Souto, Stark, Stassun,
  Steinmetz, Stello, Bergmann, Strauss, Streblyanska, Stringfellow, Suarez,
  Sun, Taghizadeh-Popp, Tang, Tao, Tayar, Tembe, Thomas, Tinker, Tojeiro,
  Tremonti, Troup, Trump, Unda-Sanzana, Valenzuela, {Van den Bosch},
  Vargas-Maga{\~{n}}a, Vazquez, Villanova, Vivek, Vogt, Wake, Walterbos, Wang,
  Wang, Weaver, Weijmans, Weinberg, Westfall, Whelan, Wilcots, Wild, Williams,
  Wilson, Wood-Vasey, Wylezalek, Xiao, Yan, Yang, Ybarra, Yeche, Yuan,
  Zakamska, Zamora, Zasowski, Zhang, Zhao, Zhao, Zheng, Zheng, Zhou, Zhu, Zinn,
  \& Zou}]{Albareti2017}
Albareti, F.~D., Prieto, C.~A., Almeida, A., {et~al.} 2017, ApJS, 233, 25

\bibitem[{Allen {et~al.}(2008)Allen, Groves, Dopita, Sutherland, \&
  Kewley}]{Allen2008}
Allen, M.~G., Groves, B.~A., Dopita, M.~A., Sutherland, R.~S., \& Kewley, L.~J.
  2008, ApJS, 178, 20

\bibitem[{Andrews \& Martini(2013)}]{Andrews2013}
Andrews, B.~H., \& Martini, P. 2013, ApJ, 765, 140

\bibitem[{Asplund {et~al.}(2009)Asplund, Grevesse, Sauval, \&
  Scott}]{Asplund2009}
Asplund, M., Grevesse, N., Sauval, A.~J., \& Scott, P. 2009, ARA{\&}A, 47, 481

\bibitem[{Axelrod {et~al.}(2010)Axelrod, Kantor, Lupton, \&
  Pierfederici}]{Axelrod2010}
Axelrod, T., Kantor, J., Lupton, R.~H., \& Pierfederici, F. 2010, in Proc.
  SPIE, ed. N.~M. Radziwill \& A.~Bridger, Vol. 7740, 774015

\bibitem[{Baldwin {et~al.}(1981)Baldwin, Phillips, \& Terlevich}]{Baldwin1981}
Baldwin, A., Phillips, M.~M., \& Terlevich, R. 1981, PASP, 93, 817

\bibitem[{Barkat {et~al.}(1967)Barkat, Rakavy, \& Sack}]{Barkat1967}
Barkat, Z., Rakavy, G., \& Sack, N. 1967, PRL, 18, 379

\bibitem[{Bensby \& Feltzing(2006)}]{Bensby2006}
Bensby, T., \& Feltzing, S. 2006, MNRAS, 367, 1181

\bibitem[{Bensby {et~al.}(2013)Bensby, Yee, Feltzing, Johnson, Gould, Cohen,
  Asplund, Mel{\'{e}}ndez, Lucatello, Han, Thompson, Gal-Yam, Udalski, Bennett,
  Bond, Kohei, Sumi, Suzuki, Suzuki, Takino, Tristram, Yamai, \&
  Yonehara}]{Bensby2013}
Bensby, T., Yee, J.~C., Feltzing, S., {et~al.} 2013, A{\&}A, 549, A147

\bibitem[{Bosch {et~al.}(2018)Bosch, Armstrong, Bickerton, Furusawa, Ikeda,
  Koike, Lupton, Mineo, Price, Takata, Tanaka, Yasuda, Alsayyad, Becker,
  Coulton, Coupon, Garmilla, Huang, Krughoff, Lang, Leauthaud, Lim, Lust,
  Macarthur, Mandelbaum, Miyatake, Miyazaki, Murata, More, Okura, Owen,
  Swinbank, Strauss, Yamada, \& Yamanoi}]{Bosch2018}
Bosch, J., Armstrong, R., Bickerton, S., {et~al.} 2018, PASJ, 70, 1

\bibitem[{Brinchmann {et~al.}(2008)Brinchmann, Kunth, \&
  Durret}]{Brinchmann2008b}
Brinchmann, J., Kunth, D., \& Durret, F. 2008, A{\&}A, 485, 657

\bibitem[{Bromm {et~al.}(2001)Bromm, Ferrara, Coppi, \& Larson}]{Bromm2001b}
Bromm, V., Ferrara, A., Coppi, P., \& Larson, R. 2001, MNRAS, 328, 969

\bibitem[{Bromm \& Loeb(2004)}]{Bromm2004}
Bromm, V., \& Loeb, A. 2004, New Astronomy, 9, 353

\bibitem[{Campbell {et~al.}(1986)Campbell, Terlevich, \&
  Melnick}]{Campbell1986}
Campbell, A., Terlevich, R., \& Melnick, J. 1986, MNRAS, 223, 811

\bibitem[{Cardelli {et~al.}(1989)Cardelli, Clayton, \& Mathis}]{Cardelli1989}
Cardelli, J.~A., Clayton, G.~C., \& Mathis, J.~S. 1989, ApJ, 345, 245

\bibitem[{Cayrel {et~al.}(2004)Cayrel, Depagne, Spite, Hill, Spite,
  Fran{\c{c}}ois, Plez, Beers, Primas, Andersen, Barbuy, Bonifacio, Molaro, \&
  Nordstr{\"{o}}m}]{Cayrel2004}
Cayrel, R., Depagne, E., Spite, M., {et~al.} 2004, A{\&}A, 416, 1117

\bibitem[{Christensen {et~al.}(2012{\natexlab{a}})Christensen, Laursen,
  Richard, Hjorth, Milvang-Jensen, Dessauges-Zavadsky, Limousin, Grillo, \&
  Ebeling}]{Christensen2012b}
Christensen, L., Laursen, P., Richard, J., {et~al.} 2012{\natexlab{a}}, MNRAS,
  427, 1973

\bibitem[{Christensen {et~al.}(2012{\natexlab{b}})Christensen, Richard, Hjorth,
  Milvang-Jensen, Laursen, Limousin, Dessauges-Zavadsky, Grillo, \&
  Ebeling}]{Christensen2012a}
Christensen, L., Richard, J., Hjorth, J., {et~al.} 2012{\natexlab{b}}, MNRAS,
  427, 1953

\bibitem[{Coupon {et~al.}(2018)Coupon, Czakon, Bosch, Komiyama, Medezinski,
  Miyazaki, \& Oguri}]{Coupon2018}
Coupon, J., Czakon, N., Bosch, J., {et~al.} 2018, PASJ, 70, 1

\bibitem[{Crowther \& Hadfield(2006)}]{Crowther2006}
Crowther, P.~A., \& Hadfield, L.~J. 2006, A{\&}A, 449, 711

\bibitem[{Crowther {et~al.}(2010)Crowther, Schnurr, Hirschi, Yusof, Parker,
  Goodwin, \& Kassim}]{Crowther2010}
Crowther, P.~A., Schnurr, O., Hirschi, R., {et~al.} 2010, MNRAS, 408, 731

\bibitem[{Crowther {et~al.}(2016)Crowther, Caballero-Nieves, Bostroem,
  {Ma{\'{i}}z Apell{\'{a}}niz}, Schneider, Walborn, Angus, Brott, Bonanos, {De
  Koter}, {De Mink}, Evans, Gr{\"{a}}fener, Herrero, Howarth, Langer, Lennon,
  Puls, Sana, \& Vink}]{Crowther2016}
Crowther, P.~A., Caballero-Nieves, S.~M., Bostroem, K.~A., {et~al.} 2016,
  MNRAS, 458, 624

\bibitem[{Ebisuzaki {et~al.}(2001)Ebisuzaki, Makino, Tsuru, Funato, {Portegies
  Zwart}, Hut, McMillan, Matsushita, Matsumoto, \& Kawabe}]{Ebisuzaki2001}
Ebisuzaki, T., Makino, J., Tsuru, T.~G., {et~al.} 2001, ApJ, 562, L19

\bibitem[{Eldridge {et~al.}(2017)Eldridge, Stanway, Xiao, McClelland, Taylor,
  Ng, Greis, \& Bray}]{Eldridge2017}
Eldridge, J.~J., Stanway, E.~R., Xiao, L., {et~al.} 2017, PASA, 34, e058

\bibitem[{Elmegreen \& Elmegreen(2017)}]{Elmegreen2017b}
Elmegreen, D.~M., \& Elmegreen, B.~G. 2017, ApJ, 851, L44

\bibitem[{Ferland {et~al.}(2013)Ferland, Porter, {Van Hoof}, Williams, Abel,
  Lykins, Shaw, Henney, \& Stancil}]{Ferland2013}
Ferland, G.~J., Porter, R.~L., {Van Hoof}, P. A.~M., {et~al.} 2013, RMxAA, 49,
  137

\bibitem[{Fischer \& Tachiev(2004)}]{Fischer2004}
Fischer, C.~F., \& Tachiev, G. 2004, Atomic Data and Nuclear Data Tables, 87,
  doi:10.1016/j.adt.2004.02.001

\bibitem[{Fragos {et~al.}(2013{\natexlab{a}})Fragos, Lehmer, Naoz, Zezas, \&
  Basu-Zych}]{Fragos2013b}
Fragos, T., Lehmer, B.~D., Naoz, S., Zezas, A., \& Basu-Zych, A.
  2013{\natexlab{a}}, ApJL, 776, L31

\bibitem[{Fragos {et~al.}(2013{\natexlab{b}})Fragos, Lehmer, Tremmel,
  Tzanavaris, Basu-Zych, Belczynski, Hornschemeier, Jenkins, Kalogera, Ptak, \&
  Zezas}]{Fragos2013a}
Fragos, T., Lehmer, B., Tremmel, M., {et~al.} 2013{\natexlab{b}}, ApJ, 764, 41

\bibitem[{Furusawa {et~al.}(2018)Furusawa, Koike, Takata, Okura, Miyatake,
  Lupton, Bickerton, Price, Bosch, Yasuda, Mineo, Yamada, Miyazaki, Nakata,
  Koshida, Komiyama, Utsumi, Kawanomoto, Jeschke, Noumaru, Schubert, Iwata,
  Finet, Fujiyoshi, Tajitsu, Terai, \& Lee}]{Furusawa2018}
Furusawa, H., Koike, M., Takata, T., {et~al.} 2018, PASJ, 70, 1

\bibitem[{Galav{\'{i}}s {et~al.}(1997)Galav{\'{i}}s, Mendoza, \&
  Zeippen}]{Galavis1997}
Galav{\'{i}}s, M.~E., Mendoza, C., \& Zeippen, C.~J. 1997, A{\&}AS, 123, 159

\bibitem[{Garnett(1992)}]{Garnett1992}
Garnett, D.~R. 1992, AJ, 103, 1330

\bibitem[{Gratton {et~al.}(2003)Gratton, Carretta, Claudi, Lucatello, \&
  Barbieri}]{Gratton2003}
Gratton, R.~G., Carretta, E., Claudi, R., Lucatello, S., \& Barbieri, M. 2003,
  A{\&}A, 404, 187

\bibitem[{Groves {et~al.}(2004{\natexlab{a}})Groves, Dopita, \&
  Sutherland}]{Groves2004a}
Groves, B.~A., Dopita, M.~A., \& Sutherland, R.~S. 2004{\natexlab{a}}, ApJS,
  153, 9

\bibitem[{Groves {et~al.}(2004{\natexlab{b}})Groves, Dopita, \&
  Sutherland}]{Groves2004b}
---. 2004{\natexlab{b}}, ApJS, 153, 75

\bibitem[{Guseva {et~al.}(2017)Guseva, Izotov, Fricke, \& Henkel}]{Guseva2017}
Guseva, N.~G., Izotov, Y.~I., Fricke, K.~J., \& Henkel, C. 2017, A{\&}A, 599,
  A65

\bibitem[{Hafen {et~al.}(2017)Hafen, Faucher-Gigu{\`{e}}re,
  Angl{\'{e}}s-Alc{\'{a}}zar, Kere{\v{s}}, Feldmann, Chan, Quataert, Murray, \&
  Hopkins}]{Hafen2017}
Hafen, Z., Faucher-Gigu{\`{e}}re, C.-A., Angl{\'{e}}s-Alc{\'{a}}zar, D.,
  {et~al.} 2017, MNRAS, 469, 2292

\bibitem[{Hanuschik(2003)}]{Hanuschik2003}
Hanuschik, R.~W. 2003, A{\&}A, 407, 1157

\bibitem[{Harikane {et~al.}(2018)Harikane, Ouchi, Shibuya, Kojima, Zhang, Itoh,
  Ono, Higuchi, Inoue, Chevallard, Capak, Nagao, Onodera, Faisst, Martin,
  Rauch, Bruzual, Charlot, Davidzon, Fujimoto, Hilmi, Ilbert, Lee, Matsuoka,
  Silverman, \& Toft}]{Harikane2018b}
Harikane, Y., Ouchi, M., Shibuya, T., {et~al.} 2018, ApJ, 859, 84

\bibitem[{Heger \& Woosley(2002)}]{Heger2002}
Heger, A., \& Woosley, S.~E. 2002, ApJ, 567, 532

\bibitem[{Isobe {et~al.}(2020)Isobe, Ouchi, Kojima, Shibuya, Hayashi, Rauch,
  Kikuchihara, Zhang, Ono, Fujimoto, Harikane, Kim, Komiyama, Kusakabe, Lee,
  Mawatari, Onodera, Sugahara, \& Yabe}]{Isobe2020}
Isobe, Y., Ouchi, M., Kojima, T., {et~al.} 2020, arXiv e-prints,
  arXiv:2004.11444

\bibitem[{Ivezi{\'{c}} {et~al.}(2019{\natexlab{a}})Ivezi{\'{c}}, Kahn, Tyson,
  Abel, Acosta, Allsman, Alonso, AlSayyad, Anderson, Andrew, {P. Angel},
  Angeli, Ansari, Antilogus, Araujo, Armstrong, Arndt, Astier, Aubourg, Auza,
  Axelrod, Bard, Barr, Barrau, Bartlett, Bauer, Bauman, Baumont, Bechtol,
  Bechtol, Becker, Becla, Beldica, Bellavia, Bianco, Biswas, Blanc, Blazek,
  Blandford, Bloom, Bogart, Bond, Booth, Borgland, Borne, Bosch, Boutigny,
  Brackett, Bradshaw, Brandt, Brown, Bullock, Burchat, Burke, Cagnoli,
  Calabrese, Callahan, Callen, Carlin, Carlson, Chandrasekharan,
  Charles-Emerson, Chesley, Cheu, Chiang, Chiang, Chirino, Chow, Ciardi,
  Claver, Cohen-Tanugi, Cockrum, Coles, Connolly, Cook, Cooray, Covey, Cribbs,
  Cui, Cutri, Daly, Daniel, Daruich, Daubard, Daues, Dawson, Delgado,
  Dellapenna, de~Peyster, de~Val-Borro, Digel, Doherty, Dubois,
  Dubois-Felsmann, Durech, Economou, Eifler, Eracleous, Emmons, Neto, Ferguson,
  Figueroa, Fisher-Levine, Focke, Foss, Frank, Freemon, Gangler, Gawiser,
  Geary, Gee, Geha, Gessner, Gibson, Gilmore, Glanzman, Glick, Goldina,
  Goldstein, Goodenow, Graham, Gressler, Gris, Guy, Guyonnet, Haller, Harris,
  Hascall, Haupt, Hernandez, Herrmann, Hileman, Hoblitt, Hodgson, Hogan,
  Howard, Huang, Huffer, Ingraham, Innes, Jacoby, Jain, Jammes, Jee, Jenness,
  Jernigan, Jevremovi{\'{c}}, Johns, Johnson, Johnson, Jones, Juramy-Gilles,
  Juri{\'{c}}, Kalirai, Kallivayalil, Kalmbach, Kantor, Karst, Kasliwal, Kelly,
  Kessler, Kinnison, Kirkby, Knox, Kotov, Krabbendam, Krughoff, Kub{\'{a}}nek,
  Kuczewski, Kulkarni, Ku, Kurita, Lage, Lambert, Lange, Langton, Guillou,
  Levine, Liang, Lim, Lintott, Long, Lopez, Lotz, Lupton, Lust, MacArthur,
  Mahabal, Mandelbaum, Markiewicz, Marsh, Marshall, Marshall, May, McKercher,
  McQueen, Meyers, Migliore, Miller, Mills, Miraval, Moeyens, Moolekamp, Monet,
  Moniez, Monkewitz, Montgomery, Morrison, Mueller, Muller, Arancibia, Neill,
  Newbry, Nief, Nomerotski, Nordby, O'Connor, Oliver, Olivier, Olsen,
  O'Mullane, Ortiz, Osier, Owen, Pain, Palecek, Parejko, Parsons, Pease,
  Peterson, Peterson, Petravick, Petrick, Petry, Pierfederici, Pietrowicz,
  Pike, Pinto, Plante, Plate, Plutchak, Price, Prouza, Radeka, Rajagopal,
  Rasmussen, Regnault, Reil, Reiss, Reuter, Ridgway, Riot, Ritz, Robinson,
  Roby, Roodman, Rosing, Roucelle, Rumore, Russo, Saha, Sassolas, Schalk,
  Schellart, Schindler, Schmidt, Schneider, Schneider, Schoening, Schumacher,
  Schwamb, Sebag, Selvy, Sembroski, Seppala, Serio, Serrano, Shaw, Shipsey,
  Sick, Silvestri, Slater, Smith, Smith, Sobhani, Soldahl, Storrie-Lombardi,
  Stover, Strauss, Street, Stubbs, Sullivan, Sweeney, Swinbank, Szalay, Takacs,
  Tether, Thaler, Thayer, Thomas, Thornton, Thukral, Tice, Trilling, Turri,
  Berg, Berk, Vetter, Virieux, Vucina, Wahl, Walkowicz, Walsh, Walter, Wang,
  Wang, Warner, Wiecha, Willman, Winters, Wittman, Wolff, Wood-Vasey, Wu, Xin,
  Yoachim, \& Zhan}]{Ivezic2008}
Ivezi{\'{c}}, {\v{Z}}., Kahn, S.~M., Tyson, J.~A., {et~al.} 2019{\natexlab{a}},
  ApJ, 873, 111

\bibitem[{Ivezi{\'{c}} {et~al.}(2019{\natexlab{b}})Ivezi{\'{c}}, Kahn, Tyson,
  Abel, Acosta, Allsman, Alonso, AlSayyad, Anderson, Andrew, {P. Angel},
  Angeli, Ansari, Antilogus, Araujo, Armstrong, Arndt, Astier, Aubourg, Auza,
  Axelrod, Bard, Barr, Barrau, Bartlett, Bauer, Bauman, Baumont, Bechtol,
  Bechtol, Becker, Becla, Beldica, Bellavia, Bianco, Biswas, Blanc, Blazek,
  Blandford, Bloom, Bogart, Bond, Booth, Borgland, Borne, Bosch, Boutigny,
  Brackett, Bradshaw, Brandt, Brown, Bullock, Burchat, Burke, Cagnoli,
  Calabrese, Callahan, Callen, Carlin, Carlson, Chandrasekharan,
  Charles-Emerson, Chesley, Cheu, Chiang, Chiang, Chirino, Chow, Ciardi,
  Claver, Cohen-Tanugi, Cockrum, Coles, Connolly, Cook, Cooray, Covey, Cribbs,
  Cui, Cutri, Daly, Daniel, Daruich, Daubard, Daues, Dawson, Delgado,
  Dellapenna, de~Peyster, de~Val-Borro, Digel, Doherty, Dubois,
  Dubois-Felsmann, Durech, Economou, Eifler, Eracleous, Emmons, Neto, Ferguson,
  Figueroa, Fisher-Levine, Focke, Foss, Frank, Freemon, Gangler, Gawiser,
  Geary, Gee, Geha, Gessner, Gibson, Gilmore, Glanzman, Glick, Goldina,
  Goldstein, Goodenow, Graham, Gressler, Gris, Guy, Guyonnet, Haller, Harris,
  Hascall, Haupt, Hernandez, Herrmann, Hileman, Hoblitt, Hodgson, Hogan,
  Howard, Huang, Huffer, Ingraham, Innes, Jacoby, Jain, Jammes, Jee, Jenness,
  Jernigan, Jevremovi{\'{c}}, Johns, Johnson, Johnson, Jones, Juramy-Gilles,
  Juri{\'{c}}, Kalirai, Kallivayalil, Kalmbach, Kantor, Karst, Kasliwal, Kelly,
  Kessler, Kinnison, Kirkby, Knox, Kotov, Krabbendam, Krughoff, Kub{\'{a}}nek,
  Kuczewski, Kulkarni, Ku, Kurita, Lage, Lambert, Lange, Langton, Guillou,
  Levine, Liang, Lim, Lintott, Long, Lopez, Lotz, Lupton, Lust, MacArthur,
  Mahabal, Mandelbaum, Markiewicz, Marsh, Marshall, Marshall, May, McKercher,
  McQueen, Meyers, Migliore, Miller, Mills, Miraval, Moeyens, Moolekamp, Monet,
  Moniez, Monkewitz, Montgomery, Morrison, Mueller, Muller, Arancibia, Neill,
  Newbry, Nief, Nomerotski, Nordby, O'Connor, Oliver, Olivier, Olsen,
  O'Mullane, Ortiz, Osier, Owen, Pain, Palecek, Parejko, Parsons, Pease,
  Peterson, Peterson, Petravick, Petrick, Petry, Pierfederici, Pietrowicz,
  Pike, Pinto, Plante, Plate, Plutchak, Price, Prouza, Radeka, Rajagopal,
  Rasmussen, Regnault, Reil, Reiss, Reuter, Ridgway, Riot, Ritz, Robinson,
  Roby, Roodman, Rosing, Roucelle, Rumore, Russo, Saha, Sassolas, Schalk,
  Schellart, Schindler, Schmidt, Schneider, Schneider, Schoening, Schumacher,
  Schwamb, Sebag, Selvy, Sembroski, Seppala, Serio, Serrano, Shaw, Shipsey,
  Sick, Silvestri, Slater, Smith, Smith, Sobhani, Soldahl, Storrie-Lombardi,
  Stover, Strauss, Street, Stubbs, Sullivan, Sweeney, Swinbank, Szalay, Takacs,
  Tether, Thaler, Thayer, Thomas, Thornton, Thukral, Tice, Trilling, Turri,
  Berg, Berk, Vetter, Virieux, Vucina, Wahl, Walkowicz, Walsh, Walter, Wang,
  Wang, Warner, Wiecha, Willman, Winters, Wittman, Wolff, Wood-Vasey, Wu, Xin,
  Yoachim, \& Zhan}]{Ivezic2019}
---. 2019{\natexlab{b}}, ApJ, 873, 111

\bibitem[{Iwamoto {et~al.}(1999)Iwamoto, Brachwitz, Nomoto, Kishimoto, Umeda,
  Hix, \& Thielemann}]{Iwamoto1999}
Iwamoto, K., Brachwitz, F., Nomoto, K., {et~al.} 1999, ApJS, 125, 439

\bibitem[{Izotov {et~al.}(2009)Izotov, Guseva, Fricke, \&
  Papaderos}]{Izotov2009}
Izotov, Y.~I., Guseva, N.~G., Fricke, K.~J., \& Papaderos, P. 2009, A{\&}A,
  503, 61

\bibitem[{Izotov {et~al.}(2006{\natexlab{a}})Izotov, Schaerer, Blecha, Royer,
  Guseva, \& North}]{Izotov2006b}
Izotov, Y.~I., Schaerer, D., Blecha, A., {et~al.} 2006{\natexlab{a}}, A{\&}A,
  459, 71

\bibitem[{Izotov {et~al.}(2006{\natexlab{b}})Izotov, Stasi{\'{n}}ska, Meynet,
  Guseva, \& Thuan}]{Izotov2006}
Izotov, Y.~I., Stasi{\'{n}}ska, G., Meynet, G., Guseva, N.~G., \& Thuan, T.~X.
  2006{\natexlab{b}}, A{\&}A, 448, 955

\bibitem[{Izotov \& Thuan(1998)}]{Izotov1998a}
Izotov, Y.~I., \& Thuan, T.~X. 1998, ApJ, 497, 227

\bibitem[{Izotov \& Thuan(1999)}]{Izotov1999}
---. 1999, ApJ, 511, 639

\bibitem[{Izotov {et~al.}(2012)Izotov, Thuan, \& Guseva}]{Izotov2012}
Izotov, Y.~I., Thuan, T.~X., \& Guseva, N.~G. 2012, A{\&}A, 546, A122

\bibitem[{Izotov {et~al.}(2019)Izotov, Thuan, \& Guseva}]{Izotov2019a}
---. 2019, MNRAS, 483, 5491

\bibitem[{Izotov {et~al.}(2018)Izotov, Thuan, Guseva, \& Liss}]{Izotov2018a}
Izotov, Y.~I., Thuan, T.~X., Guseva, N.~G., \& Liss, S.~E. 2018, MNRAS, 473,
  1956

\bibitem[{Johansson {et~al.}(2000)Johansson, Zethson, Hartman, Ekberg,
  Ishibashi, Davidson, \& Gull}]{Johansson2000}
Johansson, S., Zethson, T., Hartman, H., {et~al.} 2000, A{\&}A, 361, 977

\bibitem[{Juri{\'{c}} {et~al.}(2015)Juri{\'{c}}, Kantor, Lim, Lupton,
  Dubois-Felsmann, Jenness, Axelrod, Aleksi{\'{c}}, Allsman, AlSayyad, Alt,
  Armstrong, Basney, Becker, Becla, Bickerton, Biswas, Bosch, Boutigny, Kind,
  Ciardi, Connolly, Daniel, Daues, Economou, Chiang, Fausti, Fisher-Levine,
  Freemon, Gee, Gris, Hernandez, Hoblitt, Ivezi{\'{c}}, Jammes,
  Jevremovi{\'{c}}, Jones, Kalmbach, Kasliwal, Krughoff, Lang, Lurie, Lust,
  Mullally, MacArthur, Melchior, Moeyens, Nidever, Owen, Parejko, Peterson,
  Petravick, Pietrowicz, Price, Reiss, Shaw, Sick, Slater, Strauss, Sullivan,
  Swinbank, {Van Dyk}, Vuj{\v{c}}i{\'{c}}, Withers, \& Yoachim}]{Juric2015}
Juri{\'{c}}, M., Kantor, J., Lim, K.-T., {et~al.} 2015, arXiv e-prints,
  arXiv:1512.07914

\bibitem[{Kaaret {et~al.}(2001)Kaaret, Prestwich, Zezas, Murray, Kim, Kilgard,
  Schlegel, \& Ward}]{Kaaret2001}
Kaaret, P., Prestwich, A.~H., Zezas, A., {et~al.} 2001, MNRAS, 321, L29

\bibitem[{Kawanomoto {et~al.}(2018)Kawanomoto, Uraguchi, Komiyama, Miyazaki,
  Furusawa, Finet, Hattori, Wang, Yasuda, \& Suzuki}]{Kawanomoto2018}
Kawanomoto, S., Uraguchi, F., Komiyama, Y., {et~al.} 2018, PASJ, 70, 1

\bibitem[{Kelson(2003)}]{Kelson2003}
Kelson, D. 2003, PASP, 115, 688

\bibitem[{Kelson {et~al.}(2000)Kelson, Illingworth, van Dokkum, \&
  Franx}]{Kelson2000}
Kelson, D.~D., Illingworth, G.~D., van Dokkum, P.~G., \& Franx, M. 2000, ApJ,
  531, 159

\bibitem[{Kennicutt(1998)}]{Kennicutt1998}
Kennicutt, R.~C. 1998, ARA{\&}A, 36, 189

\bibitem[{Kewley {et~al.}(2013)Kewley, Dopita, Leitherer, Dav{\'{e}}, Yuan,
  Allen, Groves, \& Sutherland}]{Kewley2013a}
Kewley, L.~J., Dopita, M.~A., Leitherer, C., {et~al.} 2013, ApJ, 774, 100

\bibitem[{Kewley {et~al.}(2001)Kewley, Dopita, Sutherland, Heisler, \&
  Trevena}]{Kewley2001}
Kewley, L.~J., Dopita, M.~A., Sutherland, R.~S., Heisler, C.~A., \& Trevena, J.
  2001, ApJ, 556, 121

\bibitem[{Kisielius {et~al.}(2009)Kisielius, Storey, Ferland, \&
  Keenan}]{Kisielius2009}
Kisielius, R., Storey, P.~J., Ferland, G.~J., \& Keenan, F.~P. 2009, MNRAS,
  397, 903

\bibitem[{Kojima {et~al.}(2017)Kojima, Ouchi, Nakajima, Shibuya, Harikane, \&
  Ono}]{Kojima2017}
Kojima, T., Ouchi, M., Nakajima, K., {et~al.} 2017, PASJ, 69, 1

\bibitem[{Kojima {et~al.}(2020)Kojima, Ouchi, Rauch, Ono, Nakajima, Isobe,
  Fujimoto, Harikane, Hashimoto, Hayashi, Komiyama, Kusakabe, Kim, Lee, Mukae,
  Nagao, Onodera, Shibuya, Sugahara, Umemura, \& Yabe}]{Kojima2020a}
Kojima, T., Ouchi, M., Rauch, M., {et~al.} 2020, ApJ, 898, 142

\bibitem[{Komiyama {et~al.}(2018)Komiyama, Obuchi, Nakaya, Kamata, Kawanomoto,
  Utsumi, Miyazaki, Uraguchi, Furusawa, Morokuma, Uchida, Miyatake, Mineo,
  Fujimori, Aihara, Karoji, Gunn, \& Wang}]{Komiyama2018}
Komiyama, Y., Obuchi, Y., Nakaya, H., {et~al.} 2018, PASJ, 70, 1

\bibitem[{Kroupa(2001)}]{Kroupa2001}
Kroupa, P. 2001, MNRAS, 322, 231

\bibitem[{Kroupa \& Weidner(2003)}]{Kroupa2003}
Kroupa, P., \& Weidner, C. 2003, ApJ, 598, 1076

\bibitem[{Kunth \& {\"{O}}stlin(2000)}]{Kunth2000}
Kunth, D., \& {\"{O}}stlin, G. 2000, A{\&}A, 10, 1

\bibitem[{Lecureur {et~al.}(2007)Lecureur, Hill, Zoccali, Barbuy, G{\'{o}}mez,
  Minniti, Ortolani, \& Renzini}]{Lecureur2007}
Lecureur, A., Hill, V., Zoccali, M., {et~al.} 2007, A{\&}A, 465, 799

\bibitem[{Legrand {et~al.}(1997)Legrand, Kunth, Roy, Mas-Hesse, \&
  Walsh}]{Legrand1997}
Legrand, F., Kunth, D., Roy, J.~R., Mas-Hesse, J.~M., \& Walsh, J.~R. 1997,
  A{\&}A, 326, 18

\bibitem[{Lehner {et~al.}(2013)Lehner, Howk, Tripp, Tumlinson, Prochaska,
  O'Meara, Thom, Werk, Fox, \& Ribaudo}]{Lehner2013}
Lehner, N., Howk, J.~C., Tripp, T.~M., {et~al.} 2013, ApJ, 770,
  doi:10.1088/0004-637X/770/2/138

\bibitem[{L{\'{o}}pez-S{\'{a}}nchez \& Esteban(2009)}]{Lopez-Sanchez2009}
L{\'{o}}pez-S{\'{a}}nchez, A.~R., \& Esteban, C. 2009, A{\&}A, 508, 615

\bibitem[{L{\'{o}}pez-S{\'{a}}nchez \& Esteban(2010)}]{Lopez-Sanchez2010a}
L{\'{o}}pez-S{\'{a}}nchez, {\'{A}}.~R., \& Esteban, C. 2010, A{\&}A, 516, A104

\bibitem[{Luridiana {et~al.}(2015)Luridiana, Morisset, \& Shaw}]{Luridiana2015}
Luridiana, V., Morisset, C., \& Shaw, R.~A. 2015, A{\&}A, 573, A42

\bibitem[{Lusso {et~al.}(2010)Lusso, Comastri, Vignali, Zamorani, Brusa, Gilli,
  Iwasawa, Salvato, Civano, Elvis, Merloni, Bongiorno, Trump, Koekemoer,
  Schinnerer, {Le Floc'H}, Cappelluti, Jahnke, Sargent, Silverman, Mainieri,
  Fiore, Bolzonella, {Le F{\`{e}}vre}, Garilli, Iovino, Kneib, Lamareille,
  Lilly, Mignoli, Scodeggio, \& Vergani}]{Lusso2010}
Lusso, E., Comastri, A., Vignali, C., {et~al.} 2010, A{\&}A, 512, A34

\bibitem[{Ly {et~al.}(2014)Ly, Malkan, Nagao, Kashikawa, Shimasaku, \&
  Hayashi}]{Ly2014}
Ly, C., Malkan, M.~A., Nagao, T., {et~al.} 2014, ApJ, 780, 122

\bibitem[{Mainali {et~al.}(2017)Mainali, Kollmeier, Stark, Simcoe, Walth,
  Newman, \& Miller}]{Mainali2017}
Mainali, R., Kollmeier, J.~A., Stark, D.~P., {et~al.} 2017, ApJ, 836, L14

\bibitem[{Martizzi {et~al.}(2019)Martizzi, Vogelsberger, Artale, Haider,
  Torrey, Marinacci, Nelson, Pillepich, Weinberger, Hernquist, Naiman, \&
  Springel}]{Martizzi2019}
Martizzi, D., Vogelsberger, M., Artale, M.~C., {et~al.} 2019, MNRAS, 486, 3766

\bibitem[{Matsumoto {et~al.}(2001)Matsumoto, Tsuru, Koyama, Awaki, Canizares,
  Kawai, Matsushita, \& Kawabe}]{Matsumoto2001}
Matsumoto, H., Tsuru, T.~G., Koyama, K., {et~al.} 2001, ApJ, 547, L25

\bibitem[{McLaughlin \& Bell(2000)}]{McLaughlin2000}
McLaughlin, B.~M., \& Bell, K.~L. 2000, Journal of Physics B: Atomic, Molecular
  and Optical Physics, 33, 597

\bibitem[{Mendoza \& Zeippen(1982)}]{Mendoza1982}
Mendoza, C., \& Zeippen, C.~J. 1982, MNRAS, 198, 127

\bibitem[{Miyazaki {et~al.}(2012)Miyazaki, Komiyama, Nakaya, Kamata, Doi,
  Hamana, Karoji, Furusawa, Kawanomoto, Morokuma, Ishizuka, Nariai, Tanaka,
  Uraguchi, Utsumi, Obuchi, Okura, Oguri, Takata, Tomono, Kurakami, Namikawa,
  Usuda, Yamanoi, Terai, Uekiyo, Yamada, Koike, Aihara, Fujimori, Mineo,
  Miyatake, Yasuda, Nishizawa, Saito, Tanaka, Uchida, Katayama, Wang, Chen,
  Lupton, Loomis, Bickerton, Price, Gunn, Suzuki, Miyazaki, Muramatsu,
  Yamamoto, Endo, Ezaki, Itoh, Miwa, Yokota, Matsuda, Ebinuma, \&
  Takeshi}]{Miyazaki2012}
Miyazaki, S., Komiyama, Y., Nakaya, H., {et~al.} 2012, in Proc. SPIE, ed. I.~S.
  McLean, S.~K. Ramsay, \& H.~Takami, 84460Z

\bibitem[{Miyazaki {et~al.}(2018)Miyazaki, Komiyama, Kawanomoto, Doi, Furusawa,
  Hamana, Hayashi, Ikeda, Kamata, Karoji, Koike, Kurakami, Miyama, Morokuma,
  Nakata, Namikawa, Nakaya, Nariai, Obuchi, Oishi, Okada, Okura, Tait, Takata,
  Tanaka, Tanaka, Terai, Tomono, Uraguchi, Usuda, Utsumi, Yamada, Yamanoi,
  Aihara, Fujimori, Mineo, Miyatake, Oguri, Uchida, Tanaka, Yasuda, Takada,
  Murayama, Nishizawa, Sugiyama, Chiba, Futamase, Wang, Chen, Ho, Liaw, Chiu,
  Ho, Lai, Lee, Jeng, Iwamura, Armstrong, Bickerton, Bosch, Gunn, Lupton,
  Loomis, Price, Smith, Strauss, Turner, Suzuki, Miyazaki, Muramatsu, Yamamoto,
  Endo, Ezaki, Ito, Kawaguchi, Sofuku, Taniike, Akutsu, Dojo, Kasumi, Matsuda,
  Imoto, Miwa, Suzuki, Takeshi, \& Yokota}]{Miyazaki2018}
Miyazaki, S., Komiyama, Y., Kawanomoto, S., {et~al.} 2018, PASJ, 70, 1

\bibitem[{{Munoz Burgos} {et~al.}(2009){Munoz Burgos}, Loch, Ballance, \&
  Boivin}]{MunozBurgos2009}
{Munoz Burgos}, J.~M., Loch, S.~D., Ballance, C.~P., \& Boivin, R.~F. 2009,
  A{\&}A, 500, 1253

\bibitem[{Nakajima {et~al.}(2016)Nakajima, Ellis, Iwata, Inoue, Kusakabe,
  Ouchi, \& Robertson}]{Nakajima2016}
Nakajima, K., Ellis, R.~S., Iwata, I., {et~al.} 2016, ApJ, 831, L9

\bibitem[{Nakajima \& Ouchi(2014)}]{Nakajima2014}
Nakajima, K., \& Ouchi, M. 2014, MNRAS, 442, 900

\bibitem[{Nomoto {et~al.}(2013)Nomoto, Kobayashi, \& Tominaga}]{Nomoto2013}
Nomoto, K., Kobayashi, C., \& Tominaga, N. 2013, ARA{\&}A, 51, 457

\bibitem[{Nugis \& Lamers(2000)}]{Nugis2000}
Nugis, T., \& Lamers, H.~J. 2000, A{\&}A, 360, 227

\bibitem[{Ohkubo {et~al.}(2006)Ohkubo, Umeda, Maeda, Nomoto, Suzuki, Tsuruta,
  \& Rees}]{Ohkubo2006}
Ohkubo, T., Umeda, H., Maeda, K., {et~al.} 2006, ApJ, 645, 1352

\bibitem[{Oke \& Gunn(1983)}]{Oke1983}
Oke, J.~B., \& Gunn, J.~E. 1983, ApJ, 266, 713

\bibitem[{Omukai \& Palla(2003)}]{Omukai2003}
Omukai, K., \& Palla, F. 2003, ApJ, 589, 677

\bibitem[{Ono {et~al.}(2010)Ono, Ouchi, Shimasaku, Dunlop, Farrah, McLure, \&
  Okamura}]{Ono2010b}
Ono, Y., Ouchi, M., Shimasaku, K., {et~al.} 2010, ApJ, 724, 1524

\bibitem[{Ono {et~al.}(2018)Ono, Ouchi, Harikane, Toshikawa, Rauch, Yuma,
  Sawicki, Shibuya, Shimasaku, Oguri, Willott, Akhlaghi, Akiyama, Coupon,
  Kashikawa, Komiyama, Konno, Lin, Matsuoka, Miyazaki, Nagao, Nakajima,
  Silverman, Tanaka, Taniguchi, \& Wang}]{Ono2018}
Ono, Y., Ouchi, M., Harikane, Y., {et~al.} 2018, PASJ, 70, 2

\bibitem[{P{\'{e}}rez-Montero \& Contini(2009)}]{Perez-Montero2009}
P{\'{e}}rez-Montero, E., \& Contini, T. 2009, MNRAS, 398, 949

\bibitem[{P{\'{e}}rez-Montero {et~al.}(2013)P{\'{e}}rez-Montero, Contini,
  Lamareille, Maier, Carollo, Kneib, {Le F{\`{e}}vre}, Lilly, Mainieri,
  Renzini, Scodeggio, Zamorani, Bardelli, Bolzonella, Bongiorno, Caputi,
  Cucciati, {De La Torre}, {De Ravel}, Franzetti, Garilli, Iovino, Kampczyk,
  Knobel, Kova{\v{c}}, {Le Borgne}, {Le Brun}, Mignoli, Pell{\`{o}}, Peng,
  Presotto, Ricciardelli, Silverman, Tanaka, Tasca, Tresse, Vergani, \&
  Zucca}]{Perez-Montero2013}
P{\'{e}}rez-Montero, E., Contini, T., Lamareille, F., {et~al.} 2013, A{\&}A,
  549, A25

\bibitem[{{Portegies Zwart} {et~al.}(2004){Portegies Zwart}, Baumgardt, Hut,
  Makino, \& McMillan}]{PortegiesZwart2004}
{Portegies Zwart}, S.~F., Baumgardt, H., Hut, P., Makino, J., \& McMillan,
  S.~L. 2004, Nature, 428, 724

\bibitem[{{Portegies Zwart} {et~al.}(2006){Portegies Zwart}, Baumgardt,
  McMillan, Makino, Hut, \& Ebisuzaki}]{PortegiesZwart2006}
{Portegies Zwart}, S.~F., Baumgardt, H., McMillan, S. L.~W., {et~al.} 2006,
  ApJ, 641, 319

\bibitem[{{Portegies Zwart} {et~al.}(1999){Portegies Zwart}, Makino, McMillan,
  \& Hut}]{PortegiesZwart1999}
{Portegies Zwart}, S.~F., Makino, J., McMillan, S.~L., \& Hut, P. 1999, A{\&}A,
  348, 117

\bibitem[{Press {et~al.}(2007)Press, Teukolsky, Vetterling, \&
  Flannery}]{Press2007}
Press, W.~H., Teukolsky, S.~A., Vetterling, T.~W., \& Flannery, B.~P. 2007,
  {Numerical Recipies, The Art of Scientific Computing} (Cambridge University
  Press)

\bibitem[{Prochaska {et~al.}(2003)Prochaska, Gawiser, Wolfe, Castro, \&
  Djorgovski}]{Prochaska2003}
Prochaska, J.~X., Gawiser, E., Wolfe, A.~M., Castro, S., \& Djorgovski, S.~G.
  2003, ApJ, 595, L9

\bibitem[{Quinet(1996)}]{Quinet1996}
Quinet, P. 1996, A{\&}AS, 116, 573

\bibitem[{Quiret {et~al.}(2016)Quiret, P{\'{e}}roux, Zafar, Kulkarni, Jenkins,
  Milliard, Rahmani, Popping, Rao, Turnshek, \& Monier}]{Quiret2016}
Quiret, S., P{\'{e}}roux, C., Zafar, T., {et~al.} 2016, MNRAS, 458, 4074

\bibitem[{Rafelski {et~al.}(2012)Rafelski, Wolfe, Prochaska, Neeleman, \&
  Mendez}]{Rafelski2012}
Rafelski, M., Wolfe, A.~M., Prochaska, J.~X., Neeleman, M., \& Mendez, A.~J.
  2012, ApJ, 755, 89

\bibitem[{Ramsbottom \& Bell(1997)}]{Ramsbottom1997}
Ramsbottom, C., \& Bell, K. 1997, Atomic Data and Nuclear Data Tables, 66, 65

\bibitem[{Rodriguez(2003)}]{Rodriguez2003}
Rodriguez, M. 2003, ApJ, 590, 296

\bibitem[{Rodriguez \& Rubin(2005)}]{Rodriguez2005}
Rodriguez, M., \& Rubin, R.~H. 2005, ApJ, 626, 900

\bibitem[{Santoro \& Shull(2006)}]{Santoro2006}
Santoro, F., \& Shull, J.~M. 2006, ApJ, 643, 26

\bibitem[{Saxena {et~al.}(2020{\natexlab{a}})Saxena, Pentericci, Mirabelli,
  Schaerer, Schneider, Cullen, Amorin, Bolzonella, Bongiorno, Carnall,
  Castellano, Cucciati, Fontana, Fynbo, Garilli, Gargiulo, Guaita, Hathi,
  Hutchison, Koekemoer, Marchi, Mcleod, Mclure, Papovich, Pozzetti, Talia, \&
  Zamorani}]{Saxena2020a}
Saxena, A., Pentericci, L., Mirabelli, M., {et~al.} 2020{\natexlab{a}}, A{\&}A,
  636, arXiv:1911.09999

\bibitem[{Saxena {et~al.}(2020{\natexlab{b}})Saxena, Pentericci, Schaerer,
  Schneider, Amorin, Bongiorno, Calabr{\`{o}}, Castellano, Cimatti, Cullen,
  Fontana, Fynbo, Hathi, McLeod, Talia, \& Zamorani}]{Saxena2020b}
Saxena, A., Pentericci, L., Schaerer, D., {et~al.} 2020{\natexlab{b}}, MNRAS,
  496, 3796

\bibitem[{Schaerer {et~al.}(2019)Schaerer, Fragos, \& Izotov}]{Schaerer2019}
Schaerer, D., Fragos, T., \& Izotov, Y.~I. 2019, A{\&}A, 622, L10

\bibitem[{Schlegel {et~al.}(1998)Schlegel, Finkbeiner, \& Davis}]{Schlegel1998}
Schlegel, D.~J., Finkbeiner, D.~P., \& Davis, M. 1998, ApJ, 500, 525

\bibitem[{Schneider {et~al.}(2006)Schneider, Omukai, Inoue, \&
  Ferrara}]{Schneider2006}
Schneider, R., Omukai, K., Inoue, A.~K., \& Ferrara, A. 2006, MNRAS, 369, 1437

\bibitem[{Senchyna {et~al.}(2019)Senchyna, Stark, Chevallard, Charlot, Jones,
  \& Vidal-Garc{\'{i}}a}]{Senchyna2019b}
Senchyna, P., Stark, D.~P., Chevallard, J., {et~al.} 2019, MNRAS, 488, 3492

\bibitem[{Senchyna {et~al.}(2020)Senchyna, Stark, Mirocha, Reines, Charlot,
  Jones, \& Mulchaey}]{Senchyna2020}
Senchyna, P., Stark, D.~P., Mirocha, J., {et~al.} 2020, MNRAS, 494, 941

\bibitem[{Senchyna {et~al.}(2017)Senchyna, Stark, Vidal-Garc{\'{i}}a,
  Chevallard, Charlot, Mainali, Jones, Wofford, Feltre, \&
  Gutkin}]{Senchyna2017}
Senchyna, P., Stark, D.~P., Vidal-Garc{\'{i}}a, A., {et~al.} 2017, MNRAS, 472,
  2608

\bibitem[{Shapley {et~al.}(2017)Shapley, Sanders, Reddy, Kriek, Freeman,
  Mobasher, Siana, Coil, Leung, DeGroot, Shivaei, Price, Azadi, \&
  Aird}]{Shapley2017}
Shapley, A.~E., Sanders, R.~L., Reddy, N.~A., {et~al.} 2017, ApJ, 846, L30

\bibitem[{Shirazi \& Brinchmann(2012)}]{Shirazi2012}
Shirazi, M., \& Brinchmann, J. 2012, MNRAS, 421, 1043

\bibitem[{Skillman(1989)}]{Skillman1989c}
Skillman, E.~D. 1989, ApJ, 347, 883

\bibitem[{Smith \& Sigurdsson(2007)}]{Smith2007}
Smith, B.~D., \& Sigurdsson, S. 2007, ApJ, 661, L5

\bibitem[{Stanway {et~al.}(2016)Stanway, Eldridge, \& Becker}]{Stanway2016}
Stanway, E.~R., Eldridge, J.~J., \& Becker, G.~D. 2016, MNRAS, 456, 485

\bibitem[{Stark {et~al.}(2014)Stark, Richard, Siana, Charlot, Freeman, Gutkin,
  Wofford, Robertson, Amanullah, Watson, \& Milvang-jensen}]{Stark2014a}
Stark, D.~P., Richard, J., Siana, B., {et~al.} 2014, MNRAS, 445, 3200

\bibitem[{Stark {et~al.}(2015)Stark, Walth, Charlot, Cl{\'{e}}ment, Feltre,
  Gutkin, Richard, Mainali, Robertson, Siana, Tang, \& Schenker}]{Stark2015b}
Stark, D.~P., Walth, G., Charlot, S., {et~al.} 2015, MNRAS, 454, 1393

\bibitem[{Stark {et~al.}(2017)Stark, Ellis, Charlot, Chevallard, Tang, Belli,
  Zitrin, Mainali, Gutkin, Vidal-Garc{\'{i}}a, Bouwens, \& Oesch}]{Stark2017}
Stark, D.~P., Ellis, R.~S., Charlot, S., {et~al.} 2017, MNRAS, 464, 469

\bibitem[{Stasi{\'{n}}ska \& Izotov(2003)}]{Stasinska2003}
Stasi{\'{n}}ska, G., \& Izotov, Y. 2003, A{\&}A, 397, 71

\bibitem[{Stasi{\'{n}}ska {et~al.}(2015)Stasi{\'{n}}ska, Izotov, Morisset, \&
  Guseva}]{Stasinska2015}
Stasi{\'{n}}ska, G., Izotov, Y., Morisset, C., \& Guseva, N. 2015, A{\&}A, 576,
  A83

\bibitem[{Steidel {et~al.}(2016)Steidel, Strom, Pettini, Rudie, Reddy, \&
  Trainor}]{Steidel2016}
Steidel, C.~C., Strom, A.~L., Pettini, M., {et~al.} 2016, ApJ, 826, 159

\bibitem[{Storey \& Hummer(1995)}]{Storey1995}
Storey, P.~J., \& Hummer, D.~G. 1995, MNRAS, 272, 41

\bibitem[{Storey {et~al.}(2014)Storey, Sochi, \& Badnell}]{Storey2014}
Storey, P.~J., Sochi, T., \& Badnell, N.~R. 2014, MNRAS, 441, 3028

\bibitem[{Storey \& Zeippen(2000)}]{Storey2000}
Storey, P.~J., \& Zeippen, C.~J. 2000, MNRAS, 312, 813

\bibitem[{Suzuki \& Maeda(2018)}]{Suzuki2018}
Suzuki, A., \& Maeda, K. 2018, ApJ, 852, 101

\bibitem[{Tayal(2011)}]{Tayal2011}
Tayal, S.~S. 2011, ApJS, 195, 12

\bibitem[{Thuan {et~al.}(2005)Thuan, {Lecavelier des Etangs}, \&
  Izotov}]{Thuan2005b}
Thuan, T.~X., {Lecavelier des Etangs}, A., \& Izotov, Y.~I. 2005, ApJ, 621, 269

\bibitem[{Tominaga {et~al.}(2007)Tominaga, Umeda, \& Nomoto}]{Tominaga2007b}
Tominaga, N., Umeda, H., \& Nomoto, K. 2007, ApJ, 660, 516

\bibitem[{Vanzella {et~al.}(2017)Vanzella, Balestra, Gronke, Karman, Caminha,
  Dijkstra, Rosati, {De Barros}, Caputi, Grillo, Tozzi, Meneghetti, Mercurio,
  \& Gilli}]{Vanzella2016a}
Vanzella, E., Balestra, I., Gronke, M., {et~al.} 2017, MNRAS, 465, 3803

\bibitem[{Vincenzo {et~al.}(2016)Vincenzo, Belfiore, Maiolino, Matteucci, \&
  Ventura}]{Vincenzo2016}
Vincenzo, F., Belfiore, F., Maiolino, R., Matteucci, F., \& Ventura, P. 2016,
  MNRAS, 458, 3466

\bibitem[{Wise {et~al.}(2012)Wise, Turk, Norman, \& Abel}]{Wise2012}
Wise, J.~H., Turk, M.~J., Norman, M.~L., \& Abel, T. 2012, ApJ, 745, 50

\bibitem[{Xiao {et~al.}(2018)Xiao, Stanway, \& Eldridge}]{Xiao2018}
Xiao, L., Stanway, E.~R., \& Eldridge, J.~J. 2018, MNRAS, 477, 904

\bibitem[{York {et~al.}(2000)York, Adelman, {Anderson, Jr.}, Anderson, Annis,
  Bahcall, Bakken, Barkhouser, Bastian, Berman, Boroski, Bracker, Briegel,
  Briggs, Brinkmann, Brunner, Burles, Carey, Carr, Castander, Chen, Colestock,
  Connolly, Crocker, Csabai, Czarapata, Davis, Doi, Dombeck, Eisenstein,
  Ellman, Elms, Evans, Fan, Federwitz, Fiscelli, Friedman, Frieman, Fukugita,
  Gillespie, Gunn, Gurbani, de~Haas, Haldeman, Harris, Hayes, Heckman,
  Hennessy, Hindsley, Holm, Holmgren, Huang, Hull, Husby, Ichikawa, Ichikawa,
  Ivezi{\'{c}}, Kent, Kim, Kinney, Klaene, Kleinman, Kleinman, Knapp, Korienek,
  Kron, Kunszt, Lamb, Lee, Leger, Limmongkol, Lindenmeyer, Long, Loomis,
  Loveday, Lucinio, Lupton, MacKinnon, Mannery, Mantsch, Margon, McGehee,
  McKay, Meiksin, Merelli, Monet, Munn, Narayanan, Nash, Neilsen, Neswold,
  Newberg, Nichol, Nicinski, Nonino, Okada, Okamura, Ostriker, Owen, Pauls,
  Peoples, Peterson, Petravick, Pier, Pope, Pordes, Prosapio, Rechenmacher,
  Quinn, Richards, Richmond, Rivetta, Rockosi, Ruthmansdorfer, Sandford,
  Schlegel, Schneider, Sekiguchi, Sergey, Shimasaku, Siegmund, Smee, Smith,
  Snedden, Stone, Stoughton, Strauss, Stubbs, SubbaRao, Szalay, Szapudi,
  Szokoly, Thakar, Tremonti, Tucker, Uomoto, {Vanden Berk}, Vogeley, Waddell,
  Wang, Watanabe, Weinberg, Yanny, \& Yasuda}]{York2000}
York, D.~G., Adelman, J., {Anderson, Jr.}, J.~E., {et~al.} 2000, AJ, 120, 1579

\bibitem[{Yungelson {et~al.}(2008)Yungelson, {Van Den Heuvel}, Vink, Zwart, \&
  {De Koter}}]{Yungelson2008}
Yungelson, L.~R., {Van Den Heuvel}, E.~P., Vink, J.~S., Zwart, S.~F., \& {De
  Koter}, A. 2008, A{\&}A, 477, 223

\end{thebibliography}

\end{document}